\documentclass[twocolumn]{aa}  
\usepackage{xcolor}
\usepackage{listings}
\usepackage{natbib}
\usepackage{subcaption}
\bibpunct{(}{)}{;}{a}{}{,}
\usepackage{booktabs,caption}
\usepackage[flushleft]{threeparttable}
\usepackage{adjustbox}

\newcommand\gaia{\textit{Gaia}}

\definecolor{codegreen}{rgb}{0,0.6,0}
\definecolor{codegray}{rgb}{0.5,0.5,0.5}
\definecolor{codepurple}{rgb}{0.58,0,0.82}
\definecolor{backcolour}{rgb}{0.95,0.95,0.92}

\defcitealias{massari23}{Paper I}

\lstdefinestyle{mystyle}{
    commentstyle=\color{codegreen},
    keywordstyle=\color{magenta},
    numberstyle=\tiny\color{codegray},
    stringstyle=\color{codepurple},
    basicstyle=\ttfamily,
    breakatwhitespace=false,         
    breaklines=true,                 
    captionpos=b,                    
    keepspaces=true,                 
    numbers=none,                    
    showspaces=false,                
    showstringspaces=false,
    showtabs=false,                  
    tabsize=2
}
\title{Cluster Ages to Reconstruct the Milky Way Assembly (CARMA)}
\subtitle{II. The age-metallicity relation of Gaia-Sausage-Enceladus globular clusters}

\titlerunning{CARMA II.}
\authorrunning{Aguado-Agelet et al.}
\author{Fernando Aguado-Agelet \inst{1,2}, 
Davide Massari \inst{3}, 
Matteo Monelli \inst{4,5}, 
Santi Cassisi \inst{5,6}, 
Carme Gallart \inst{7}, 
Edoardo Ceccarelli \inst{3,8},
Yllari Kay González Koda \inst{9},
Tom\'as Ruiz-Lara \inst{9}, 
Elena Pancino \inst{10}, 
Sara Saracino \inst{10, 11}, 
Maurizio Salaris \inst{11}
}

\institute{              atlanTTic, Universidade de Vigo, Escola de Enxeñar\'ia de Telecomunicaci\'on, 36310, Vigo, Spain\\ \email{faguado@uvigo.gal}
             \and
             Universidad de La Laguna, Avda. Astrof\'isico Fco. S\'anchez, E-38205 La Laguna, Tenerife, Spain
             \and
             INAF - Osservatorio di Astrofisica e Scienza dello Spazio di Bologna, Via Gobetti 93/3, I-40129 Bologna, Italy
             \and 
             INAF – Osservatorio Astronomico di Roma, Via Frascati 33, 00078 Monte Porzio Catone, Roma, Italy
             \and
             INAF – Osservatorio Astronomico di Abruzzo, Via M. Maggini, 64100 Teramo, Italy
             \and
             INFN - Sezione di Pisa, Universit\'a di Pisa, Largo Pontecorvo 3, 56127 Pisa, Italy
             \and
             Instituto de Astrof\'isica de Canarias, Calle V\'ia L\'actea s/n, E-38206 La Laguna, Tenerife, Spain
             \and
             Dipartimento di Fisica e Astronomia, Universit\`a degli Studi di Bologna, Via Piero Gobetti 93/2, 40129 Bologna, Italy
             \and
             Universidad de Granada, Departamento de Física Teórica y del Cosmos, Campus Fuente Nueva, Edificio Mecenas, 18071 Granada, Spain
             \and
             INAF - Osservatorio Astrofisico di Arcetri, Largo E. Fermi 5, I-50125 Firenze, Italy
             \and
             Astrophysics Research Institute, Liverpool John Moores University, 146 Brownlow Hill, Liverpool L3 5RF, UK
}

\abstract{We present the age determination of 13 globular clusters that are dynamically associated with the Gaia-Sausage-Enceladus (GSE) merger event, as part of the CARMA project's effort to trace the Milky Way assembly history. We used deep and homogeneous archival {\it Hubble Space Telescope} data, and applied isochrone fitting to derive homogeneous age estimates.
We find that the majority of the selected clusters form a well-defined age-metallicity relation, with a few outliers. Among these, NGC~288 and NGC~6205 are more than 2 Gyr older than the other GSE globular clusters at a similar metallicity, and are therefore interpreted as probably having originated in situ. Moreover, NGC~7099 is somewhat younger than the average GSE trend, which suggests a possible alternative dwarf galaxy progenitor, while NGC~5286 is slightly older, as if its progenitor was characterised by greater star-formation efficiency. Another remarkable feature of the resulting age-metallicity relation is the presence of two epochs of globular cluster formation, with a duration of $\sim0.3$ Gyr each and separated by $\sim2$ Gyr. 
These findings are in excellent agreement with the age-metallicity relation recently found for halo field stars, which clearly hints at episodic star-formation in GSE. The age of the two formation epochs is similar to the mean age of the two groups of in-situ globular clusters previously studied by CARMA. These epochs might therefore precisely pinpoint two important dynamical events that Gaia-Sausage-Enceladus had with the Milky Way during its evolutionary history. Finally, we discuss the correlation between the recent spectroscopic determination of Si and Eu, and the clusters age and origin.}


\begin{document}
\flushbottom
\maketitle
\thispagestyle{empty}

\section{Introduction}

The quest to reconstruct the Milky Way (MW) assembly history is experiencing a revolution that began with data release 2 of the {\it Gaia} mission \citep[][]{gaiadr2}. In fact, with the availability of 5D phase-space information, complemented by the line-of-sight velocities provided by large spectroscopic surveys such as APOGEE \citep{apogee}, GALAH \citep{galah}, the Gaia ESO Survey \citep{gaiaeso}, H3 \citep{h3}, and LAMOST \citep{lamost}, the possibility to study the orbital properties of nearby halo stars has led to the discovery and characterisation of past merger events that have shaped the MW into its current appearance.
It is now clear that the nearby halo is dominated by the remnant of the latest significant merger with the Gaia-Sausage-Enceladus (GSE) dwarf galaxy \citep[][]{helmi18, belokurov18}. Additional contributions come from several less massive mergers like Sagittarius \citep[][]{ibata94}, Sequoia \citep[][]{myeong19}, the Helmi streams \citep[][]{helmi99}, Antaeus \citep[][]{oria22, ceccarelli24a}, and other candidates associated to coherent sub-structures in the integrals of motion space \citep[see e.g.][]{massari19, koppelman19, horta21, malhan22, dodd23, mikkola23}.

However, it has become increasingly evident that the interpretation of sub-structures in the dynamical space is complicated by the fact that the debris of merger events naturally overlaps \citep[see e.g.][]{jeanbaptiste17, koppelman20, chen24, mori24}, and that the same merger can create different coherent dynamical sub-structures, just like in-situ disc stars can do when perturbed by a massive merger \citep[see e.g.][]{amarante22, khoperskov22, belokurov23}. Ultimately, a purely dynamical selection produces samples of merger debris that are neither pure nor complete \cite[][]{buder22, rey23}, and should be complemented by additional information on other conserved properties such as chemistry \citep[see e.g.][]{naidu20, horta20, ceccarelli24a} or age \cite[see e.g.][]{montalban21, xiang22, queiroz23}.

In this sense, globular clusters (GCs) play an important role, as they are the tracers of the MW assembly history whose age can be measured in the most precise way \cite[see e.g.][]{vandenberg13, massari23}. With such a precise measurement, the age-metallicity relation (AMR) of MW GCs has proved to be a powerful tool to assess the origin of GCs as in-situ or accreted stellar systems \citep[see e.g.][]{marinfranch09, forbes10, leaman13, kruijssen19, massari19, callingham22}. Unfortunately, precise GC age measurements are limited to relatively small samples. Different age indicators, photometric systems, assumptions on distance, reddening and metallicity, and theoretical models are just some of the many sources of systematic uncertainties that affect different compilations of GC ages and that might add up to $\sim2$ Gyr \citep[][]{massari19}. To overcome these limitations, the CARMA project \citep[][hereafter Paper I]{massari23} has started an effort to derive accurate ages for the entire MW GC system, based on the isochrone fitting of each GC colour magnitude diagram (CMD) built by using homogeneous optical photometry taken with the {\it Hubble Space Telescope} (HST), and adopting the most up-to-date set of homogeneous theoretical models provided by the BaSTI database \citep[][]{hidalgo18, pietrinferni21}.

In the first proof-of-concept paper in the series, we show the power of our method applied to the pair of GCs NGC~6388 and NGC~6441, for which neither dynamics \citep{massari19, callingham22} nor chemistry \citep{horta20, minelli21, carretta22} could unambiguously determine the origin. By comparing their relative age with that a further four GCs with a clear in-situ origin, and with the AMR of halo field stars in the solar neighbourhood, we demonstrated that both GCs were born in-situ in the MW.
In this second paper of the series, we focus on the GC population of the GSE merger \citep{helmi18, belokurov18}. GSE is the latest significant accretion event experienced by the MW about 9-11 Gyr ago \citep[see e.g.][]{dimatteo19, gallart19, kruijssen19, naidu21}. Its stellar mass estimates range from a few$\times10^8$ M$_{\odot}$ \citep[e.g.][]{lane23} to some $\times10^9$ M$_{\odot}$ \citep[e.g.][]{fattahi19, das20}. Its accretion seems to have had a dramatic impact on the evolution of the MW, as it led, for example, to the appearance of the thick disc as we currently know it \citep{helmi18, gallart19, ciuca24}. Due to its high mass, GSE likely hosted a large system of GCs, the current number of candidate members varying between 20 and 26 \citep[][]{massari19, forbes20, callingham22, chen24}. Starting from a sub-sample of likely GSE members with publicly available HST photometry, we aim to improve the uncertainty on the individual GC associations, and to assess some of the properties of the GSE dwarf galaxy from the findings on the AMR of its GC system.

The paper is organised as follows. Section~\ref{data_methodology} summarises the methods and the data used in this work. Section~\ref{results} presents the results of our GC age-dating and interprets the age-metallicity relation (AMR) of GSE GCs. Finally, a discussion on the most relevant findings is offered in Section ~\ref{conclusion}. The results of the isochrone fit of each individual CMD are shown in the Appendix.

\section{Data and methodology} \label{data_methodology}

The approach adopted in this work strictly follows the analysis presented in \citetalias{massari23}, and extends it to a sample of bona fide GSE clusters. One of the most important keys of the CARMA project is homogeneity, both in the data (source, data reduction, calibration), theoretical models (complete and self-consistent framework of the BaSTI library), and methods. We briefly summarise the specific aspects relevant for the current work.

\subsection{Gaia-Enceladus globular cluster data} \label{subsec:data_set}

The sample includes 13 out of the 17 GCs associated with GSE according to both \citet{massari19}
and \citet{callingham22}. The four remaining clusters are ESO-SC06, NGC~6235 and NGC~6864 (which are excluded because no publicly available deep HST/ACS photometry exists) and NGC~5139 ($\omega$ Cen, which will be analysed in a separate work due to the complexity of its stellar populations). The sample is listed in Table \ref{tab:sample}, which summarises the name of the target and the derived values of metallicity, reddening, distance modulus, and age.

The photometry used in this work comes from the public database of the HST UV Globular Cluster Survey (HUGS) project \citep{piotto15}. In particular, we used the $F606W$ and $F814W$ catalogues labelled as 'method-2' \citep[see also][]{anderson08, nardiello18}. The reasons for this choice are numerous and described in detail in \citetalias{massari23}, but in brief, this photometric system is the only one common to almost all MW GCs, and these catalogues typically offer the deepest CMD. Additionally, this band combination is the least sensitive to the presence of multiple evolutionary sequences caused by chemically peculiar stellar populations. The original HUGS catalogues were processed in order to have the cleanest possible sample of stars to perform a reliable comparison with theoretical isochrones: \emph{i)} only stars with a membership probability larger than 90\% were retained; \emph{ii)} the apparent magnitudes were corrected for differential reddening using the method described in \citet{milone12}; \emph{iii)} sources in the innermost regions were removed to avoid poor photometric measurements due to crowding (20 to 60\arcsec depending on the cluster); \emph{iv)} highly peculiar populations (such as those significantly enriched in C+N+O or He) were removed by identifying them in the multi-band chromosome maps \citep{milone17}.

\subsection{Age estimates} \label{subsec:age_derivation}

The age derivation was performed using an isochrones fitting approach, following the procedures described in detail in \citetalias{massari23}. Briefly, we adopted theoretical models from the latest release of the BaSTI stellar evolution library \citep{hidalgo18, pietrinferni21} that covered a fine grid in age and metallicity. Age spans a wide range from 6 to 15 Gyr in steps of 100 Myr, and metallicity spans from -2.5 dex to 0.0 dex in steps of 0.01 dex. Solar-scaled models including diffusion were consistently used. We note that the choice of solar-scaled models has been purposely made to avoid making any assumptions on the $\alpha$-element abundance of GCs, which information is either prone to large systematic errors due to very different literature sources or entirely missing. We effectively absorb the term on the $\alpha$-element abundance by working in global metallicity [M/H], rather than in iron abundance [Fe/H]. This is particularly justified by the fact that the photometry we analyse is in optical bands. In these bands, the equivalency between solar-scaled and $\alpha$-enhanced models {\it at the same global metallicity} has been demonstrated by \cite{salaris93, cassisi04} through the relation
\begin{equation}\label{eq1}
    [M/H] = [Fe/H]+\log(0.694\times10^{[\alpha/Fe]}+0.301).
\end{equation}  
Finally, since some of the clusters are affected by large extinction, we applied a temperature-dependent reddening correction to the models for clusters with E(B--V) $>$ 0.1 mag, for which the dependency of E(B-V) on the stellar temperature is non-negligible \citep[see][]{girardi08}.

The values of the gaussian priors used by the code for the distance modulus (DM), metallicity, and colour excess (E(B-V)) are adopted from the Harris catalogue \citep[][2010 edition]{harris96}. The dispersion associated with each parameter is assumed as follows: $\sigma_{DM} = 0.1$ mag, $\sigma_{[M/H]} = 0.1$ dex,  $\sigma_{E(B-V)} = 0.05$ mag. The adoption of the same dispersion for all GCs is justified by the fact that these values are conservative numbers, and are typically associated as uncertainties to the Harris compilation over the entire MW GC sample, whereas the GCs analysed here are typically nearby, poorly extinct, and with existing, high-precision photometric and spectroscopic investigations. For each individual GC, two separate isochrone fitting solutions are obtained for the (m$_{F606W}$, m$_{F606W}$-m$_{F814W}$) and (m$_{F814W}$, m$_{F606W}$-m$_{F814W}$) CMDs. These are determined by minimising the likelihood function, which is composed of two terms. The first evaluates the consistency of each solution with the initial priors, while the second quantifies the distance of each individual star from the theoretical model under scrutiny. As the final result, we use the average value of the two solutions (which are shown for each individual GC in the Appendix), while the overall uncertainties are computed such as to encompass the upper and lower limits of both runs combined. In fact, the uncertainties that come from the single runs are purely intrinsic, and as such they likely underestimate the actual errors. Our final uncertainties, instead, alleviate this problem, and at the same time take into account possible systematics related to uncertainties in the computation of the theoretical models. We finally remark that the age values presented here should be interpreted strictly in a relative sense.

\begin{table*}[!htbp]
\centering
\caption{\label{tab:results} Results of the isochrone fitting. }
\label{tab:sample}
\begin{tabular}{lcccc}
Name & [M/H] & E(B-V) & DM & age  \\
& & [mag] & [mag] & [Gyr]  \\
\hline
\\
NGC~288 & -1.12 \(^{+0.08}_{-0.08}\) & 0.02 \(^{+0.01}_{-0.01}\) & 14.77 \(^{+0.01}_{-0.01}\) & 13.75 \(^{+0.28}_{-0.22}\) \\\\ \vspace{0.1cm} 
NGC~362 & -1.16 \(^{+0.03}_{-0.04}\) & 0.03 \(^{+0.01}_{-0.01}\) & 14.79 \(^{+0.01}_{-0.01}\) & 11.47 \(^{+0.11}_{-0.10}\) \\\\ \vspace{0.1cm} 
NGC~1261 & -1.20 \(^{+0.05}_{-0.02}\) & 0.01 \(^{+0.01}_{-0.01}\) & 16.06 \(^{+0.01}_{-0.01}\) & 11.81 \(^{+0.06}_{-0.13}\) \\\\ \vspace{0.1cm} 
NGC~1851 & -1.05 \(^{+0.01}_{-0.01}\) & 0.04 \(^{+0.01}_{-0.01}\) & 15.39 \(^{+0.02}_{-0.01}\) & 11.41 \(^{+0.05}_{-0.06}\) \\\\ \vspace{0.1cm} 
NGC~2298 & -1.64 \(^{+0.11}_{-0.03}\) & 0.22 \(^{+0.01}_{-0.01}\) & 14.93 \(^{+0.02}_{-0.02}\) & 13.09 \(^{+0.41}_{-0.35}\) \\\\ \vspace{0.1cm} 
NGC~2808 & -1.04 \(^{+0.01}_{-0.01}\) & 0.22 \(^{+0.01}_{-0.01}\) & 15.01 \(^{+0.01}_{-0.01}\) & 11.09 \(^{+0.11}_{-0.10}\) \\\\ \vspace{0.1cm} 
NGC~5286 & -1.47 \(^{+0.04}_{-0.04}\) & 0.25 \(^{+0.01}_{-0.01}\) & 15.20 \(^{+0.01}_{-0.01}\) & 13.47 \(^{+0.17}_{-0.05}\) \\\\ \vspace{0.1cm} 
NGC~5897 & -1.58 \(^{+0.09}_{-0.08}\) & 0.12 \(^{+0.01}_{-0.00}\) & 15.52 \(^{+0.02}_{-0.02}\) & 13.10 \(^{+0.22}_{-0.46}\) \\\\ \vspace{0.1cm} 
NGC~6205 & -1.29 \(^{+0.07}_{-0.08}\) & 0.01 \(^{+0.01}_{-0.01}\) & 14.34 \(^{+0.03}_{-0.03}\) & 14.06 \(^{+0.35}_{-0.41}\) \\\\ \vspace{0.1cm} 
NGC~6341 & -1.77 \(^{+0.08}_{-0.09}\) & 0.01 \(^{+0.01}_{-0.01}\) & 14.59 \(^{+0.02}_{-0.02}\) & 13.58 \(^{+0.22}_{-0.21}\) \\\\ \vspace{0.1cm} 
NGC~6779 & -1.67 \(^{+0.04}_{-0.02}\) & 0.24 \(^{+0.01}_{-0.01}\) & 15.07 \(^{+0.01}_{-0.01}\) & 13.52 \(^{+0.22}_{-0.20}\) \\\\ \vspace{0.1cm} 
NGC~7089 & -1.41 \(^{+0.02}_{-0.02}\) & 0.05 \(^{+0.01}_{-0.01}\) & 15.32 \(^{+0.01}_{-0.01}\) & 12.75 \(^{+0.29}_{-0.09}\) \\\\ \vspace{0.1cm} 
NGC~7099 & -1.83 \(^{+0.02}_{-0.02}\) & 0.04 \(^{+0.01}_{-0.01}\) & 14.61 \(^{+0.01}_{-0.01}\) & 13.05 \(^{+0.12}_{-0.09}\) \\\\ \vspace{0.1cm} 

\end{tabular}
\tablefoot{The CMD fits and the corner plots are shown in Figure \ref{fig:ngc288} and in the Appendix. These values are collected together with the other CARMA results at {\tt https://www.oas.inaf.it/en/research/m2-en/carma-en/}}
\end{table*}

\newcommand{\NGC}{NGC288}

\begin{figure*}
        \centering
        \begin{subfigure}{0.45\textwidth}
            \includegraphics[width=\textwidth]{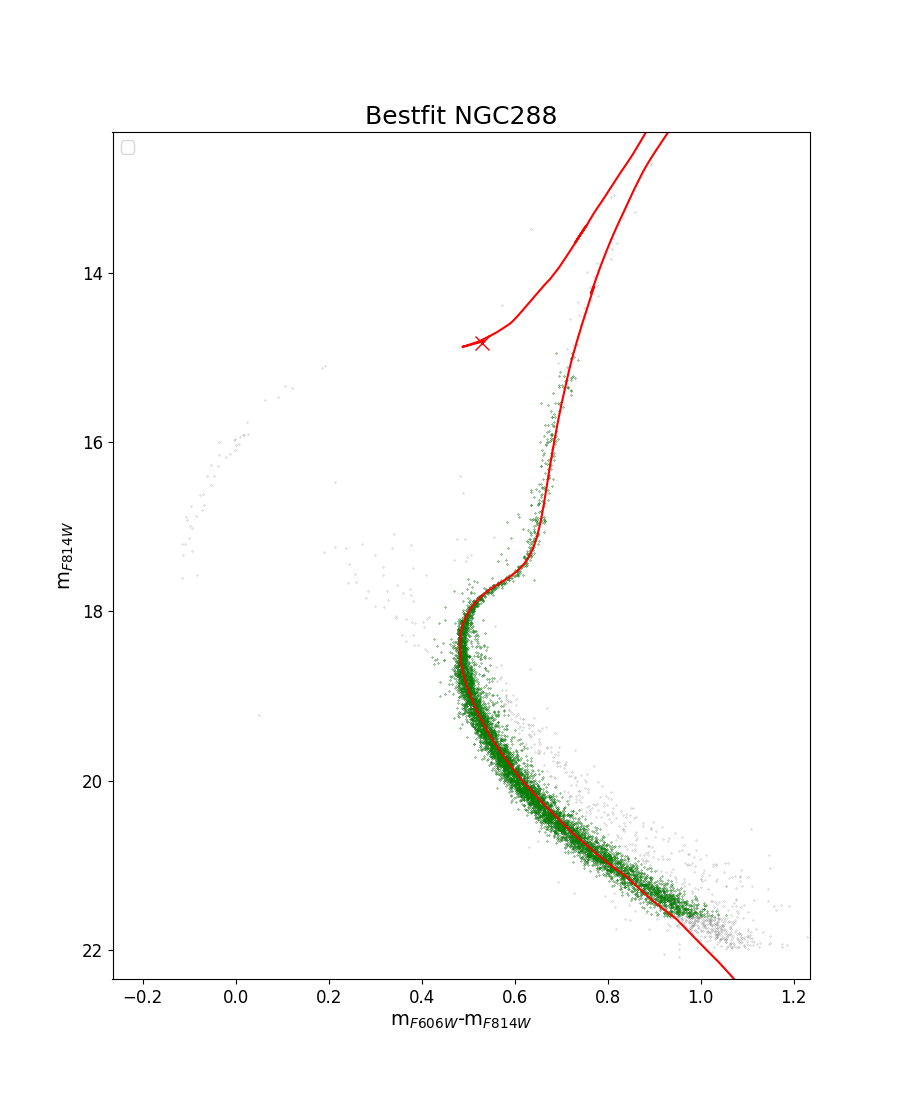}
            \caption[]
            {{\small }}    
        \end{subfigure}
        \begin{subfigure}{0.45\textwidth}  
            \includegraphics[width=\textwidth]{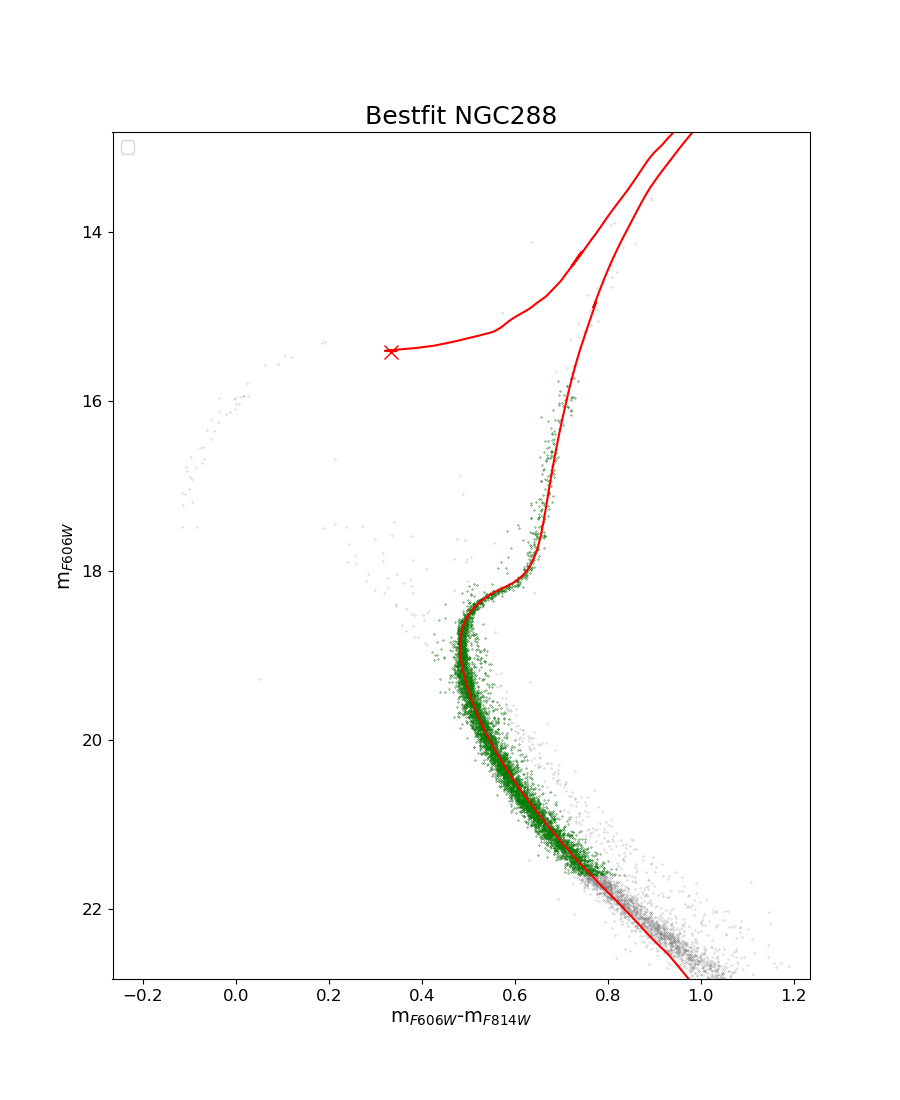}
            \caption[]
            {{\small }}    
        \end{subfigure}
        \vskip\baselineskip
        \begin{subfigure}{0.45\textwidth}   
            \includegraphics[width=\textwidth]{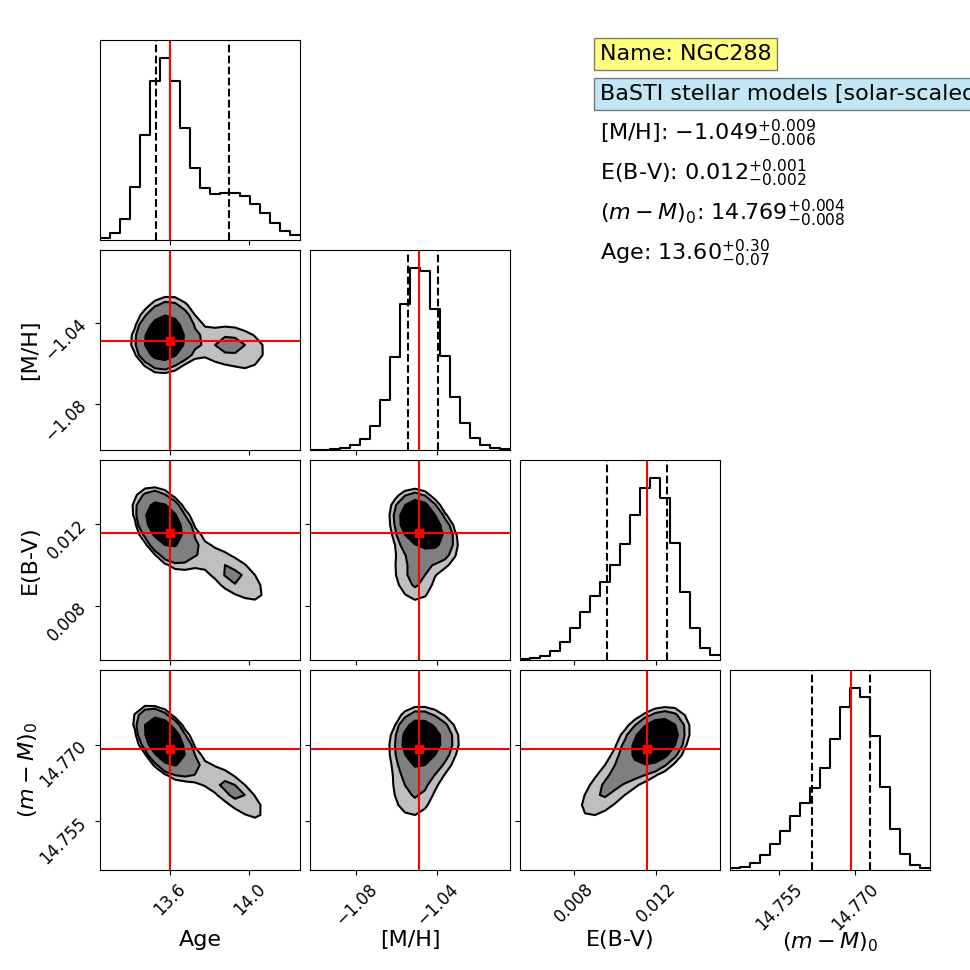}
            \caption[]
            {{\small }}    
        \end{subfigure}
        \begin{subfigure}{0.45\textwidth}   
            \includegraphics[width=\textwidth]{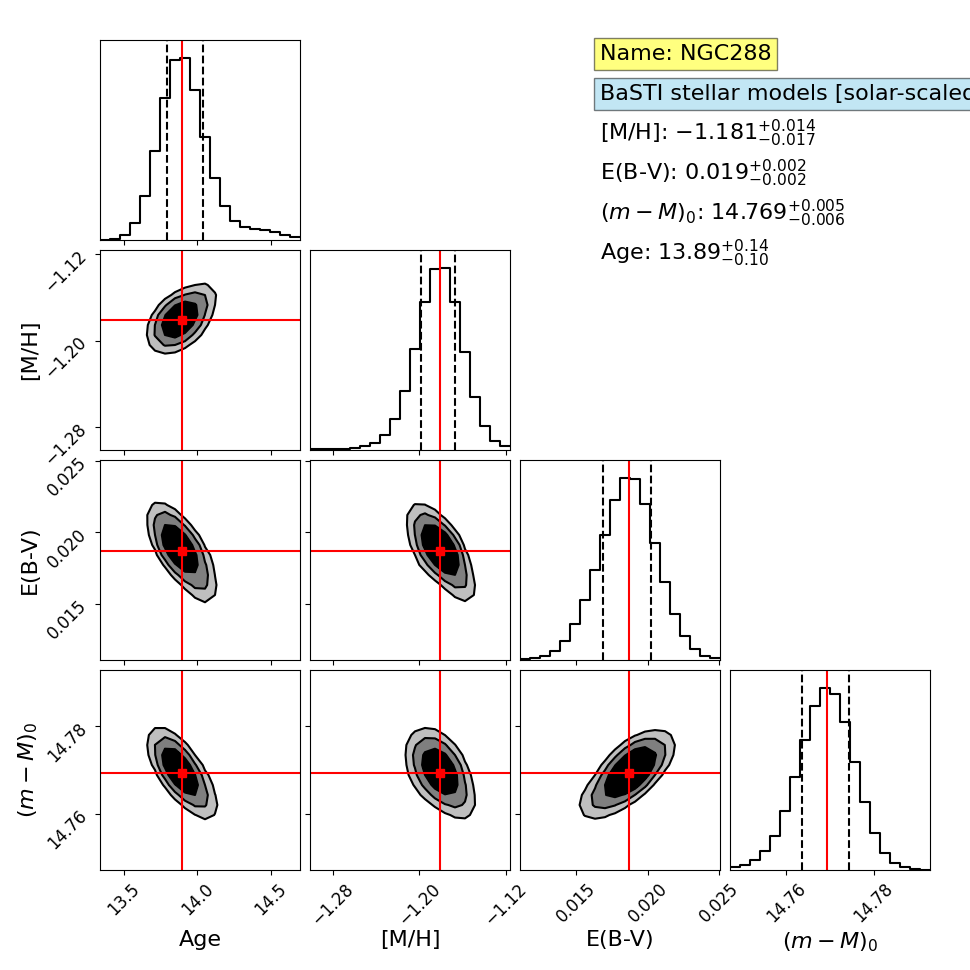}
            \caption[]
            {{\small }}    
        \end{subfigure}
        \caption[]
        {\small Results for NGC~288. Panel (a): Best-fit model in the (m$_{F814W}$, m$_{F606W}$-m$_{F814W}$) CMD. Panel (b): Best-fit model in the (m$_{F606W}$, m$_{F606W}$-m$_{F814W}$) CMD. Panel (c): Posterior distributions for the output parameters and the best-fit solution, quoted in the labels, in the (m$_{F814W}$, m$_{F606W}$-m$_{F814W}$) CMD. Panel (d): Posterior distributions for the output parameters and the best-fit solution, quoted in the labels, in the (m$_{F606W}$, m$_{F606W}$-m$_{F814W}$) CMD.} 
       \label{fig:ngc288}
    \end{figure*}

\section{Results}\label{results}

Figure \ref{fig:ngc288} presents the results of the isochrone fit for NGC~288 as an example. Similar plots for the other clusters are presented in the Appendix. The plot displays the solution derived in both CMDs: (m$_{F814W}$, m$_{F606W}$-m$_{F814W}$) and  (m$_{F606W}$, m$_{F606W}$-m$_{F814W}$) in the left and right columns, respectively. The top panels present the best-fitting isochrone superposed to each CMDs, while the bottom panels show the posterior distributions of the parameters of the model, including their correlation. The corner plots report the best-fit values for the metallicity, reddening, DM, and age.

In addition to the quality selection described in Sect.~\ref{subsec:data_set}, for each CMD the stars used to determine the best fit have been selected within a colour range from the mean-ridge line, thus to exclude residual outliers, obvious binaries, and most importantly blue straggler stars that mimic younger populations and would bias the age estimate. Grey and green points in the CMD show the sources retained after the quality cuts and those used in the fit, respectively. The lower and upper magnitude cuts were optimised on a cluster-by-cluster basis, to reach a balance between the number of red giant branch (RGB, driving the solution in [M/H]) and main sequence (MS, most sensitive to age) stars. In fact, we found that for the most massive GCs, a brighter cut in the MS helped the code to reach a solution in [M/H] more consistent with spectroscopic values available in the literature. 
The plots for NGC~288 (and similarly for all the other targets) display a good isochrone fit from the tip of the RGB down to the faint MS stars. The corner plot typically shows a well-defined minimum for the four output parameters.

\begin{figure}[ht!]
\center{
\includegraphics[width=\columnwidth]{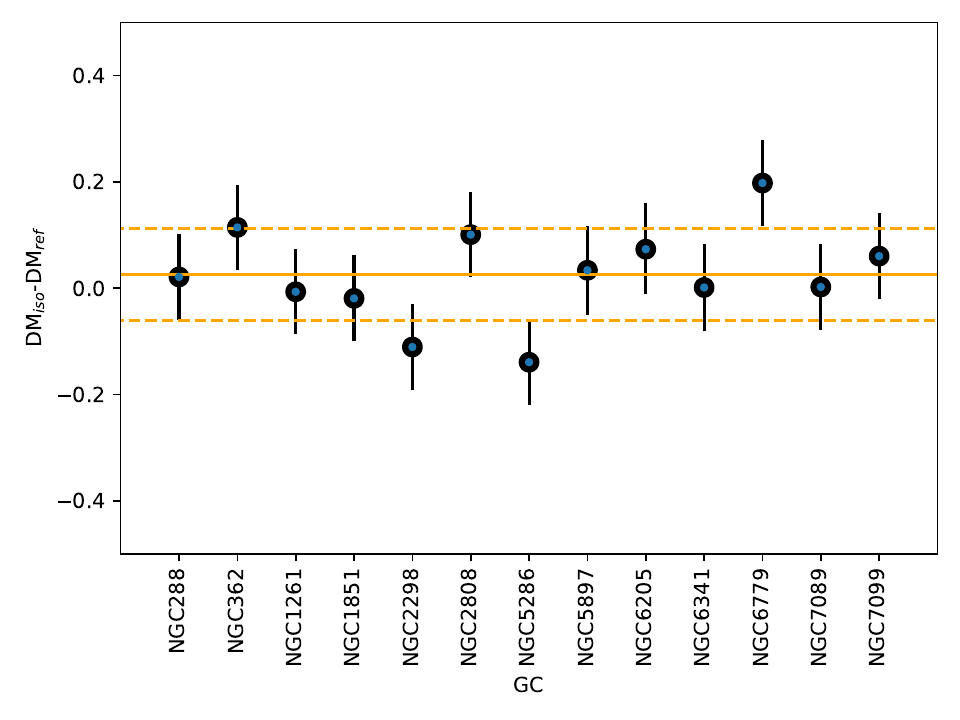}
\includegraphics[width=\columnwidth]{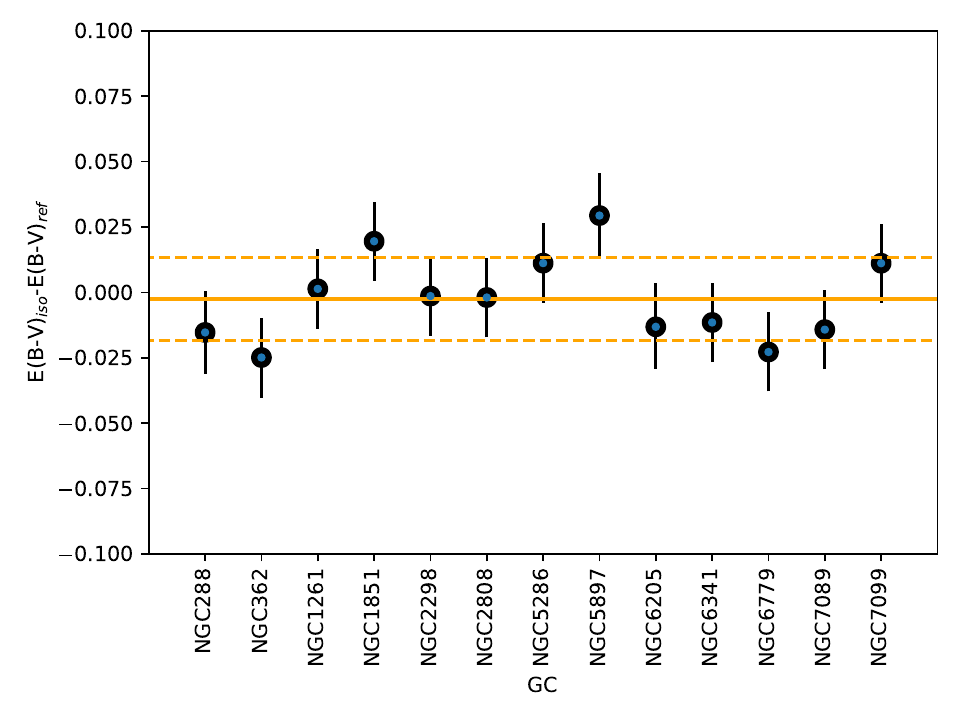}
\includegraphics[width=\columnwidth]{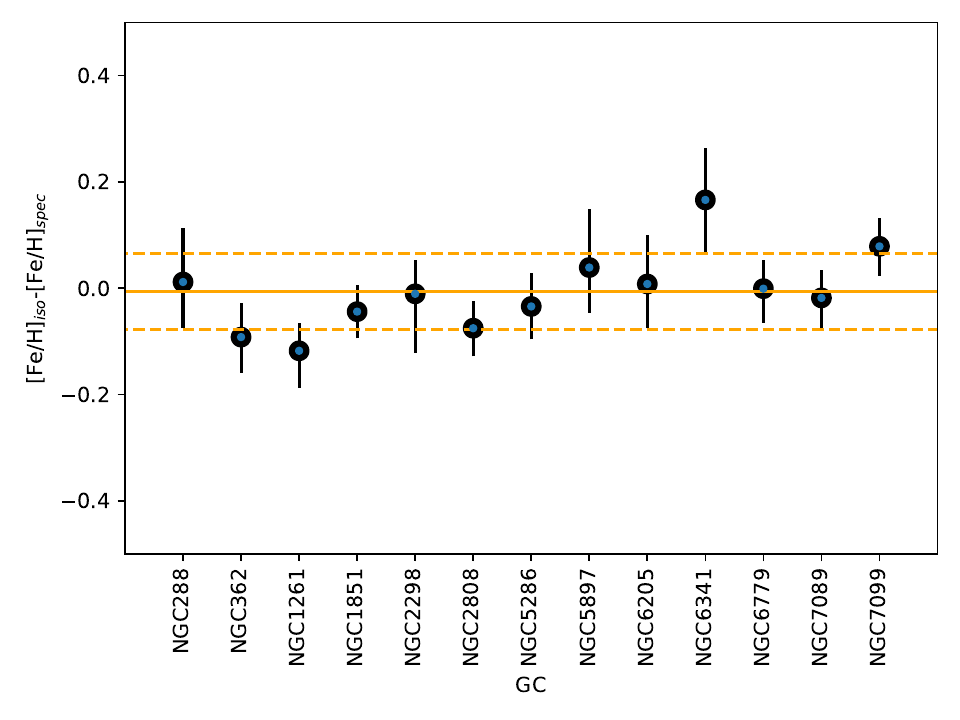}
}
\caption{Difference between literature values and the parameters obtained as output of the isochrone fitting for the true distance modulus (top panel), colour excess (middle panel), and metallicity (bottom panel). The solid and dashed lines mark the mean value and the dispersion around the mean of each distribution.}
\label{fig:diff}
\end{figure}

The results of the isochrone fitting in terms of [M/H], E($B-V$), ($m-M$)$_0$, and age for the 13 GCs under analysis are summarised in Table \ref{tab:sample}. 
Figure \ref{fig:diff} shows the difference between the literature values from the Harris catalogue\footnote{We refer to the 2010 edition, which provides GC properties with reasonable homogeneity, at least for the sample of GCs analysed here.} and the outcome of the isochrone fitting for distance, reddening, and metallicity for the 13 clusters. Note that the global metallicity [M/H] that we find as the output of the fit has been translated into iron abundance [Fe/H] according to Equation~\ref{eq1} and by using the [$\alpha$/Fe] versus [Fe/H] trend observed in GSE stars\footnote{The choice to adopt a relation determined from field stars to study GCs is well justified by homogeneous comparisons like in \cite{horta20}} \citep[see][]{helmi18}, well described by the relation [$\alpha$/Fe]$=-0.2\times$[Fe/H] for $-2.5<$[Fe/H]$<-0.5$. The solid and dashed lines represent the mean difference and the 1-$\sigma$ dispersion. For all three parameters, the mean difference is always consistent with zero within 1-$\sigma$. This is also true when comparing to the DM values found in \citep[][]{baumgardt21}, in which case we find a mean difference of $\Delta=-0.01$ with $\sigma=0.03$. In general, no particularly pathologic cases are evident, which supports the validity of the solutions found.
Moreover, the dispersion around the mean of the three distributions matches the typical uncertainties associated to each data point. This is an indication that these uncertainties, computed as the sum in quadrature between those provided in Table~\ref{tab:results} and those from the \cite{harris96} catalogue \citepalias[see also][]{massari23}, are a fair representation of the actual ones.

\subsection{The GSE age-metallicity relation}\label{sec:amr}

\begin{figure}
    \centering
        \includegraphics[scale=0.65]{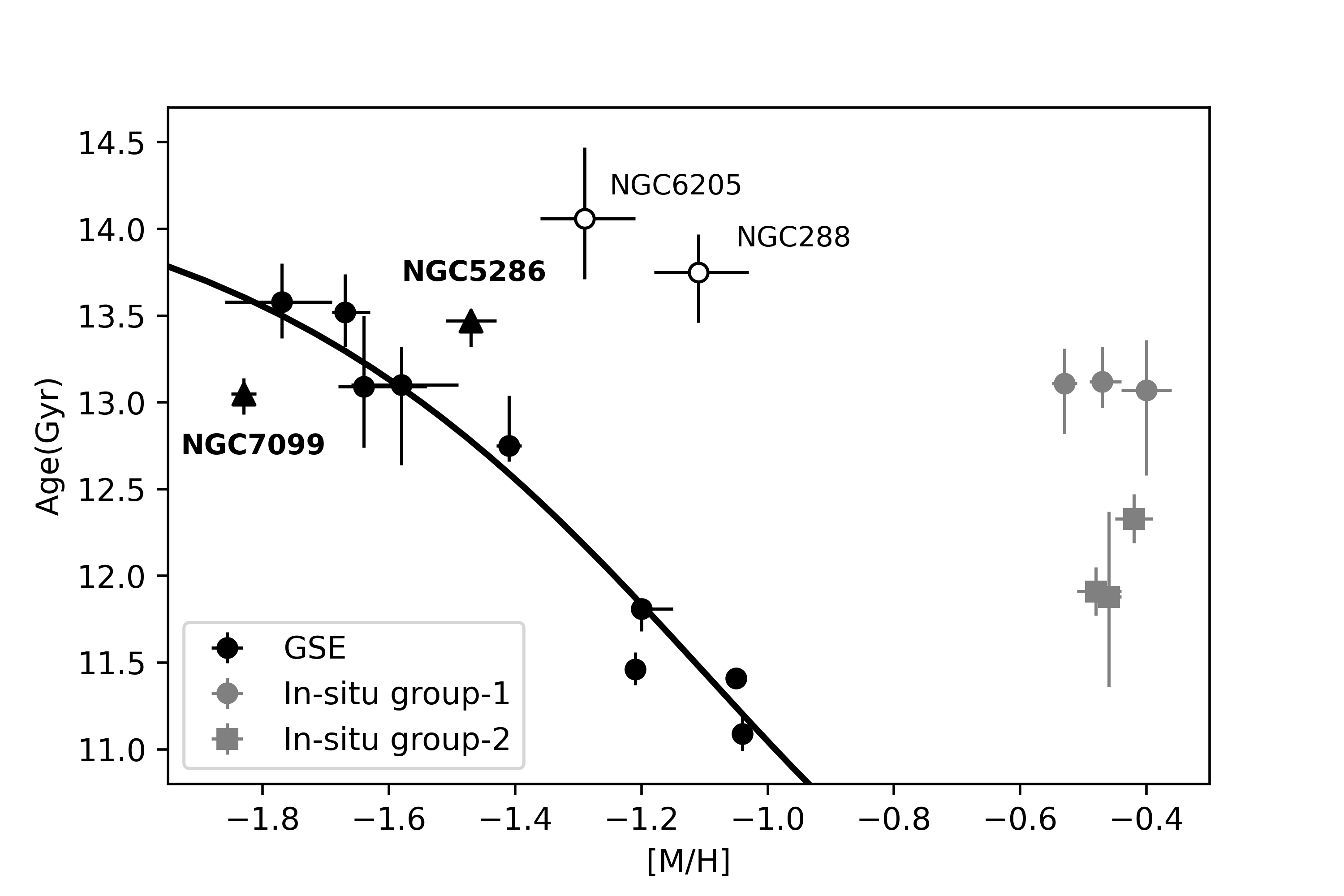}
    \caption{\small AMR of our sample of dynamically selected GSE GCs. Candidate in-situ clusters (namely NGC~288 and NGC~6205) are shown as open circles. GCs with uncertain membership to GSE (namely NGC~5286 and NGC~7099) are shown as filled black triangles. The metal-rich, in-situ GCs analysed in \citetalias{massari23} are shown in grey for the sake of comparison, with the younger and older groups marked with different symbols. A representative AMR model for the GSE genuine members is also shown as a solid black line.}
    \label{fig:amr}
\end{figure}

Fig.~\ref{fig:amr} shows the age-metallicity relation for our sample of GSE GCs. For reference, the location of the in-situ GC analysed by \citepalias{massari23} is also indicated with grey symbols.
This AMR clearly shows some remarkable features:\\
$i$) The youngest GSE GCs are significantly more metal-poor compared to the in-situ GCs of a similar age. This is an expected property, as dwarf galaxies like GSE form stars with a lower efficiency compared to the MW, and lose a larger fraction of metals due to the shallower gravitational potential. As a result of these combined effects, the chemical evolution of GSE reaches a lower metallicity compared to the MW in the same amount of time \citep[this manifests in the mass-metallicity relation, see e.g.][]{kirby13, kruijssen19}. \\
$ii$) Most of the GSE GCs describe a very tight AMR that covers a total lifespan of $2.5$ Gyr. When we consider  the one used in \cite{massari19} for GSE as a reference trend, simply rescaled to the different absolute age scale of this work (see the solid black line in Fig.~\ref{fig:amr}), the scatter around this trend is only $0.07$ dex along metallicity, and $0.16$ Gyr along age, with a handful of possible outliers;\\
$iii$) The most obvious outliers are NGC~288 and NGC~6205 (M~13), which we find to be more than 2 Gyr older compared to the other GSE GCs at the same metallicity. The old age of NGC~288 is not a novel finding, but has been debated since the 1990s. Both \cite{green90} and \cite{sarajedini90} found it to be older than NGC~362 by 2-3 Gyr. This finding was later challenged by \cite{stetson96}, who found no age difference between the two GCs. Later investigations \citep[including e.g.][]{bellazzini01, marinfranch09, vandenberg13} resulted in a range of age values, which we settle in this work. An age difference of $>2.5$ Gyr compared to the four younger GSE GCs (which include NGC~362) clearly advocates for a different origin for NGC~288, likely an in-situ one. A more detailed and conclusive analysis of this hypothesis, based on additional high-resolution spectroscopic data, is presented in a forthcoming paper \citep{ceccarelli25}. In the case of NGC~6205 as well, previous homogeneous relative age determinations resulted in conflicting conclusions. \cite{vandenberg13} found it to be about as old as NGC~288 and more than 1 Gyr older than other GSE GCs at similar metallicities like NGC~1261 and NGC~362, which is qualitatively consistent with our findings. On the other hand, \cite{marinfranch09} classified it among the young GCs, even if for these authors too NGC~6205 is $\sim10$\% older than NGC~1261 and NGC~362. Interestingly, the classifications of \cite{belokurov24} and \cite{chen24} (that use [Al/Fe] as a criterion)  provide opposite results for the in-situ or accreted origin of NGC~6205, whereas our findings indicate a likely in-situ origin. For these reasons, hereafter we exclude both NGC~288 and NGC~6205 from the analysis of purely GSE GCs;\\
$iv$) The other possible outliers are NGC~5286, which is older compared to the trend defined by the majority of the GSE sample, and NGC~7099, which is instead younger. In particular, these two clusters are off the trend by $3.5\sigma$ and $2.2\sigma$, respectively. One possibility is that these two GCs are in fact genuine members of the GSE population, in which case their location in the AMR might be indicative of a somewhat inhomogeneous metallicity distribution, or of a metallicity gradient, in the primordial GSE dwarf. Alternatively, these two GCs might be contaminating the sample of pure GSE GCs. If the two GCs are contaminants, NGC~7099 should belong to a merger event that formed with a lower star-formation efficiency compared to GSE \citep[see the AMR models in e.g.][]{massari19, callingham22, souza24}, like Sequoia or the Helmi streams. Given its retrograde orbit\footnote{These values are computed based on the proper motion values in \cite{vasiliev21} and distances in \cite{baumgardt21}, and following the same prescriptions as in \cite{massari19}.} at E$=-173969$ km$^2$/s$^{2}$ and L$_{z}=-252$ km/s~kpc, Sequoia seems the most likely candidate. On the other hand, NGC~5286 should belong to a galaxy that formed stars more efficiently than GSE, and the only possibilities are either Kraken \citep{kruijssen19, massari19} or the MW itself. Chemistry does not yet provide a conclusive way of resolving these doubts \citep[e.g.][]{monty24}, as no homogeneous derivation of elemental abundances sensitive to the GC origin at this intermediate metallicity \citep[like Mg, Ca, Ti, Zn, Eu, see][]{ceccarelli24b} exist. Future additions by the CARMA project of the age of GCs from different progenitors will shed light on the actual origin of NGC~7099 and NGC~5286;\\
$v$) Finally, the AMR of the remaining nine genuine members of GSE clearly shows evidence for two events of GC formation, separated by about 2 Gyr. Such a bimodal distribution in age is statistically confirmed by the application of the gaussian mixture model algorithm developed in \cite{muratov10}, which rejects the hypothesis of a unimodal fit with a probability P > 96\%. This is also in agreement with the findings of \cite{valenzuela24}. The earlier episode happened at $t\simeq13.2$ Gyr ($\sigma$=0.3 Gyr), and is responsible for the formation of NGC~2298, NGC~5897, NGC~6341, NGC~6779, and NGC~7089. The later one took place at $t\simeq11.4$ Gyr ($\sigma$=0.3 Gyr), and gave rise to NGC~362, NGC~1261, NGC~1851, and NGC~2808. It is interesting to note that these young GSE GCs formed at about the same time as the young in-situ GCs, which formed at t$\simeq12$ Gyr. This might be evidence for these almost simultaneous events of GC formation to have been caused by the same trigger. We discuss these features further in light of a comparison with the AMR of GSE stars in the next Section \ref{GSEstars}.

\subsection{Comparison with field stars}\label{GSEstars}

\begin{figure}[ht!]
\center{
\includegraphics[width=\columnwidth]{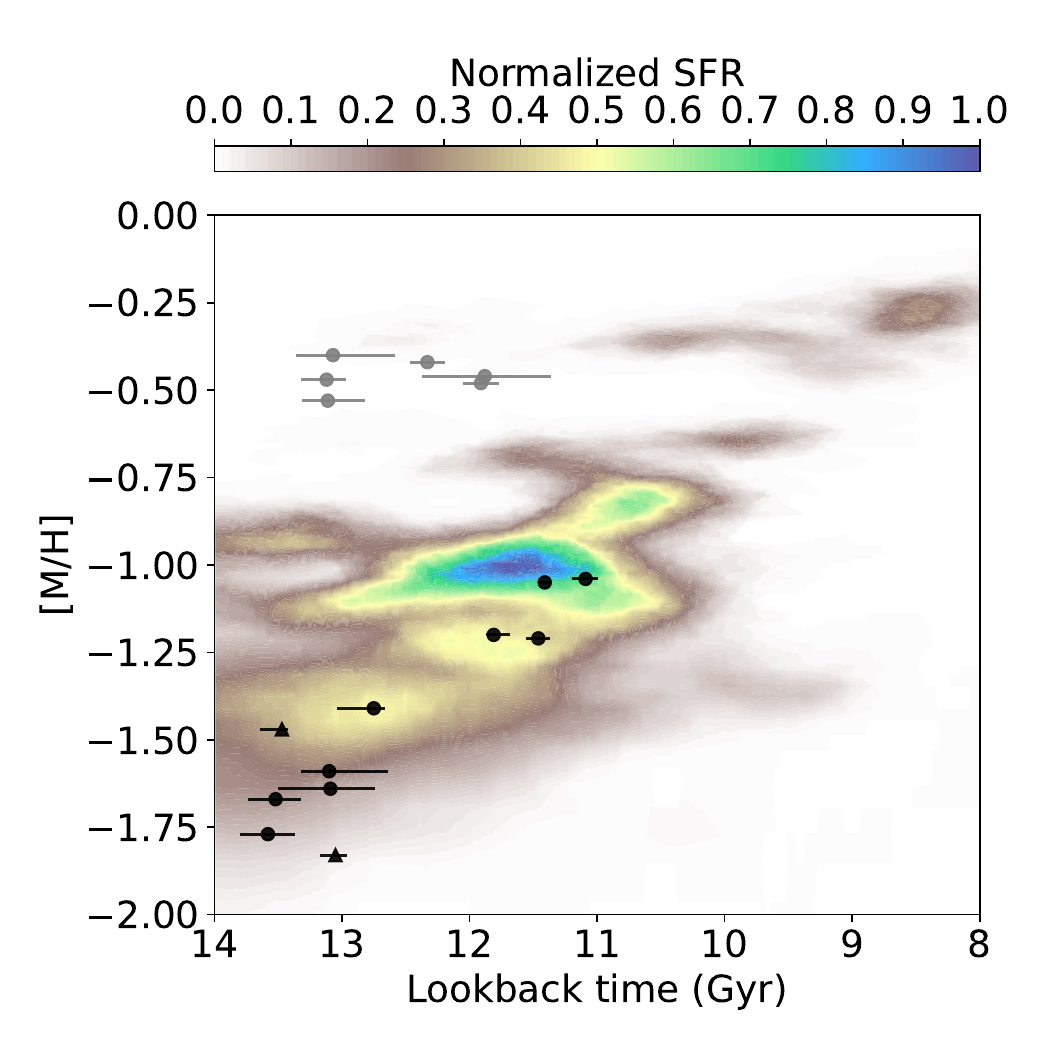}
}
\caption{Comparison between the AMRs of the most likely GSE GCs (black symbols) and stars (density map, \citealt{gk2025}). The location of the six in-situ GCs studied in \citetalias{massari23} is also shown with gray symbols for reference.}
\label{fig:age}
\end{figure}

\citet[][hereafter GK2025]{gk2025} have analyzed the age-metallicity distribution of GSE field stars within a volume of 1.2 kpc from the Sun. They dynamically selected three samples of GSE stars within this volume with different criteria and possible levels of contamination, either in action-angle coordinates \citep[following][]{feuillet21, horta24}, or using a single-linkage clustering algorithm as in \cite{dodd23}. In Fig.~\ref{fig:age} we show the superposition of the eleven likely GSE GCs' ages and metallicities and the stellar age-metallicity distribution derived for the field stars of the sample named 'GSE-group' in GK2025, which is the intermediate of their three samples in terms of number of stars and possible degree of contamination.

The two AMRs are remarkably similar in terms of their slope, and there is evidence for similarly bursty distributions (even though GK2025 warn about the fact that the bursty behaviour of the age-metallicity distribution has to be taken with caution due to the low number statistics that can affect the solution). However, providing a direct comparison between the location of the star-formation peaks is tricky, because both the methods and the data used to determine the two AMRs are entirely different. The size of the possible systematic offsets in age and metallicity due to these differences have been estimated in GK2025 under different assumptions (see their Appendix C for details). Briefly, the possible offsets are quantified by applying their CMD fitting technique (CMDft.Gaia) to GSE GC stars using purely {\it Gaia} data and a similar approach to that followed for the analysis of GSE field stars. However, we remark that it is basically impossible to perfectly replicate the CMD fitting technique on GC photometry, since for example the distance of the relatively nearby field stars used in that work can be determined robustly from {\it Gaia} parallaxes, while both crowding issues and the larger distance of the GCs make the same {\it Gaia} parallaxes very inaccurate for GC stars \citep[see e.g.][]{pancino17}. Additionally, the reddening of the field stars is determined from 3D reddening maps \citep{lallement18, green19} that do not reach the distance of the clusters and, therefore, all-sky 2D maps are used to estimate their reddening, which may thus be non-homogeneous with the reddening of field stars. Therefore, such an investigation should be considered as evidence that systematic differences might be at play but that they are difficult to quantify precisely. 
In spite of these possible offsets (found to be of a maximum of $\simeq$ 1 Gyr in age and $\simeq$ 0.2 dex in metallicity), the agreement between the ages and metallicities of the clusters and those of the main features in the AMR of field stars is remarkable. There is marginal evidence that the old peak of GC formation might be related to the most metal-poor signal in the stellar AMR, located at t$\gtrsim12.5$ and [M/H]$\lesssim-1.4$, whereas the young peak of GC formation could be related to the most evident period of enhanced star formation at t$\sim11.8$ and [M/H]$\sim-1.0$. The combination of these complementary but totally independent studies provides further evidence for the existence of episodic star formation in GSE. These kinds of events are predicted to be related to the dynamical interactions between the incoming merging galaxy and the host \citep{dicintio21,Orkney2022_bursts}. The exact correlation between a burst in star formation and a dynamical event like the pericentric passage, a fly-by, or the final coalesce phase is not straightforward \citep[though see, e.g.][]{dimatteo08}. It might depend on the orbital properties of the merger, as well as on its gas content and mass \citep[][]{dicintio21}. Simulations of dwarf galaxy mergers \citep[e.g.][]{walker96, dimatteo08, villalobos08} show that the typical separation between the first and second pericentric passages is of $0.5$-$1$ Gyr, whereas the separation between the first pericentric passage and the final coalescence of the dwarf galaxy into the host can reach up to $\sim2$ Gyr. 

We are thus inclined to interpret the two observed peaks in the GC age distribution in terms of two possibilities:\\
$i$) The first peak of GC formation happened when GSE was experiencing its very early evolution in isolation, while the second peak coincides with one particular dynamical event of interaction with the MW, this being either the first pericentric passage or the phase of final coalescence. This interpretation finds support in the results by GK2025, where the population called GSE0 by the authors shares a similar location in the AMR with the most metal-poor GCs of the first peak;\\
$ii$) Both peaks are originated by the dynamical interaction with the MW, and given the separation of 2 Gyr, we favour the scenario according to which the first burst coincides with the first pericentric passage and the second one with the coalescence phase. 

In both cases, it is interesting to note that the duration of the two bursts we found is $\sim300$ Myr. This number is similar to our typical age uncertainties, and as such it is probably an upper limit for the duration of the formation episodes. However, it matches well the average duration for star-forming bursts found by \cite{dimatteo08} in their suite of simulations. Moreover, the age of the two peaks of GSE GC formation is rather similar to the mean age of the two groups of in-situ GCs discussed in \citetalias{massari23} and shown in Fig.~\ref{fig:amr} as grey symbols. This could be indicative that GC formation in both the host (the MW) and the accreting dwarf (GSE) might have been triggered by mutual dynamical interaction, in agreement with predictions for stars from simulations \citep{dicintio21}. If this purely tentative interpretation were confirmed, our age estimates would provide a very precise and dynamically independent measurement of the timescale related to the dynamical evolution of the GSE merger.

\subsection{The [Eu/Si] trend of GSE GCs}\label{sec:discussion}

\begin{figure}[ht!]
\center{
\includegraphics[width=\columnwidth]{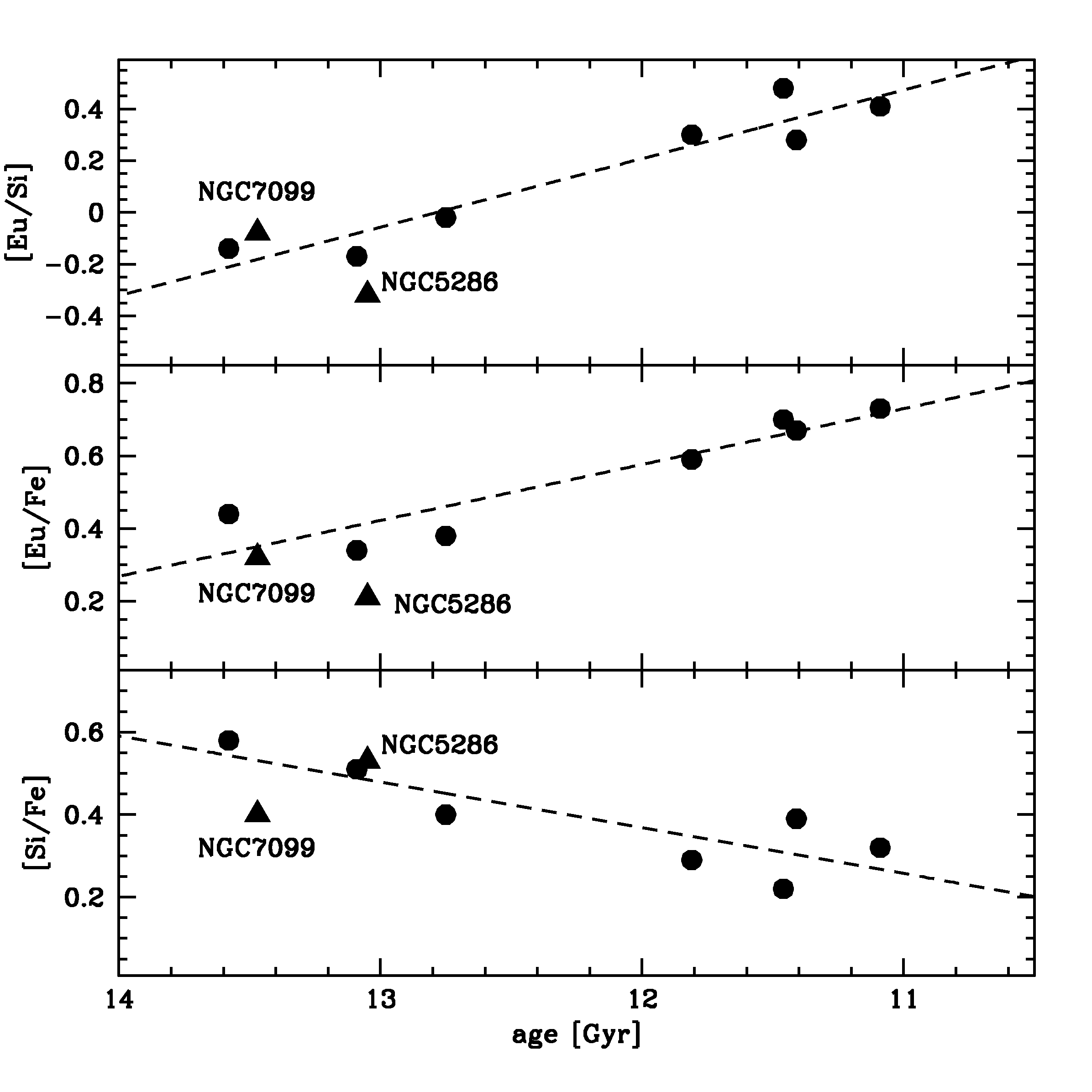}
}
\caption{Observed trends of [Si/Fe] (bottom panel), [Eu/Fe] (middle panel), and [Eu/Si] (top panel) as a function of age for our sample of accreted GCs. The black triangles indicate the two GCs for which the association to GSE is uncertain given their age (see the corresponding labels). Chemical abundances are taken from the compilation in \citet{monty24}.}
\label{fig:eusi_age}
\end{figure}

The AMR described by our sample of GCs has allowed us to select nine GCs genuinely associated with GSE (namely NGC~362, NGC~1261, NGC~1851, NGC~2298, NGC~2808, NGC~5897, NGC~6341, NGC~6779, and NGC~7089) and two additional ones (NGC~7099 and NGC~5286) whose association is more uncertain, but that very likely belong to the accreted MW GC population given their age and dynamical properties.
\cite{monty24} used the [Eu/Si] abundance ratio as a chemical tool to distinguish between accreted and in-situ MW GCs. Moreover by linking the [Eu/Si] trend for the sample of GSE GCs defined in \cite{myeong19} with the age estimates by \cite{vandenberg13}, they found an observational way to time the star-formation history of the GSE dwarf galaxy at high, sub-Gigayear resolution. We are now in a position to leverage on our findings to further improve upon these results by investigating a very pure sample of nine GSE GCs (those among the 21 candidates by \citealt{myeong19} for which we could confirm the membership to GSE based on their location in the AMR plane) and using an extremely high-precision relative age. The results on the [Eu/Fe], [Si/Fe], and [Eu/Si] trends as a function of age are shown in Fig.~\ref{fig:eusi_age}, where the abundances are the same as those used by \cite{monty24}. 

The first interesting feature is that NGC~5286, which, as described in Sect.~\ref{sec:amr}, could be a contaminant of the GSE GC sample belonging to a galaxy that formed stars with higher star-formation efficiency, is the GC that shows the lowest value of [Eu/Si] (see the top panel of Fig.~\ref{fig:eusi_age}). As described by the chemical evolutionary models in \cite{monty24}, a low value of [Eu/Si] is exactly what should be expected for such a case. This evidence therefore supports the idea that NGC~5286 might not be part of the GSE GC system. On the other hand, NGC~7099 (the other candidate contaminant) has a [Eu/Si] value consistent with that of the genuine GSE GCs at a similar age. Not much can be drawn on its actual origin from this chemical space.

The second clear feature is that all the three chemical spaces show very tight abundance trends with age when considering genuine GSE GCs.
In particular, in good agreement with \cite{monty24}, our [Eu/Si] describes an increasing trend, which ranges from [Eu/Si]$\sim-0.2$ at t=13.6 Gyr, to [Eu/Si]$\sim0.4$ at t=11.1 Gyr. A linear fit to the distribution of values leads to an increase rate of $\delta$[Eu/Fe]$=0.26$ yr$^{-1}$. The other two additional panels are best fit by rates of and $\delta$[Eu/Fe]$=-0.15$ yr$^{-1}$ (see the dashed black line in the middle panel of Fig.~\ref{fig:eusi_age}), and $\delta$[Si/Fe]$=-0.11$ yr$^{-1}$ (see the dashed black line in the bottom panel), which can be useful constraints for chemical evolution models.

\section{Summary and conclusions}\label{conclusion}

In this second paper of the CARMA series, we have further demonstrated the power of precise age measurements in unravelling the complex history of the Milky Way’s assembly through its system of GCs. Expanding on the methods and findings presented in \citetalias{massari23}, we applied the refined isochrone fitting technique to a sample of thirteen GCs that are dynamically associated to the GSE merger event \citep[][]{massari19, callingham22}, with the aim of accurately determining their ages and, by extension, confirming their origin and reconstructing some of the properties of the progenitor galaxy.

Our results show that the AMR of the GCs analysed in this study span a 3 Gyr-wide range of ages across $\sim1$ dex in metallicity. The unprecedented precision in the age determination allows us to clearly identify a few GCs that, despite their dynamical properties, are probably not members of the GSE GC system. The most obvious ones are NGC~288 and NGC~6205, which we find to be significantly older (by 2 Gyr or more) than the other GSE GCs at a similar metallicity. This is a clear indication \citep[e.g.][]{leaman13, kruijssen19} that these two systems might have been born in the Milky Way, rather than in an accreted dwarf galaxy.
Moreover, out of the remaining eleven likely accreted GCs, nine describe a very tight AMR, whereas two, namely NGC~7099 and NGC~5286, appear as outliers. While the location in the AMR plane, the retrograde orbit, and the [Eu/Si] abundance \citep{monty24} of NGC~7099 still indicate a likely accreted origin, possibly from the Sequoia dwarf galaxy, the properties of NGC~5286 leave its association quite uncertain, with its possible progenitors being GSE itself, Kraken, or the MW. In this respect, \cite{belokurov24} find the [Al/Fe] content of NGC~5286 to be consistent with an in-situ origin.

The remaining nine genuine members of GSE describe a well-defined AMR that shows a remarkably clear bimodal distribution. The two peaks of GC formation are separated by $\sim2$ Gyr, have a duration (computed as the 1$\sigma$ age dispersion) of 0.3 Gyr, and their mean age is similar to that of the two groups of in-situ GCs studied in \citetalias{massari23}. The same slope and possible bursty features have been detected in the AMR of GSE field stars by \cite{gk2025} using CMD fitting. Our interpretation is that these two bursts of star formation might be linked to the orbital properties of GSE, with the first pericentric passage and/or its final phase of coalescence with the MW being primary candidates responsible for the star-formation triggers \citep[][]{dimatteo08}. The earliest epoch of GC formation, whose location in the AMR matches remarkably well that of GSE stars \citep[][]{horta24} as measured by \cite{xiang22}, might even correspond to the initial phase during which GSE evolved in isolation. In any case, our age estimates would provide a robust and independent measurement of the dynamical timescales of the GSE merger event. Finally, these nine GSE GCs describe tight trends in the chemical evolution of elements like Eu and Si, which we quantify in an attempt to provide tight constraints in chemical evolution models.

Our findings highlight the critical role of high-precision, homogeneous photometry, theoretical models and methods in enhancing the precision of GC relative age estimates, which are crucial in resolving ambiguities concerning the origin of the MW GCs and therefore in depicting a detailed picture of our Galaxy assembly. All the CARMA results will be publicly available and kept up-to-date at {\tt https://www.oas.inaf.it/en/research/m2-en/carma-en/}.

\begin{acknowledgements}

DM, SC, and EP acknowledge financial support from PRIN-MIUR-22: CHRONOS: adjusting the clock(s) to unveil the CHRONO-chemo-dynamical Structure of the Galaxy” (PI: S. Cassisi). SS aknowledges funding from the European Union under the grant ERC-2022-AdG, {\em "StarDance: the non-canonical evolution of stars in clusters"}, Grant Agreement 101093572, PI: E. Pancino.
CG, TRL and SC acknowledge support from the Agencia Estatal de Investigación del Ministerio de Ciencia e Innovación (AEI-MCINN) under grant "At the forefront of Galactic Archaeology: evolution of the luminous and dark matter components of the Milky Way and Local Group dwarf galaxies in the Gaia era" with reference PID2020-118778GB-I00/10.13039/501100011033 and PID2023-150319NB-C21/C22/10.13039/501100011033. CG also acknowledge support from the Severo Ochoa program through CEX2019-000920-S.
TRL acknowledges support from Juan de la Cierva fellowship (IJC2020-043742-I) and Ram\'on y Cajal fellowship (RYC2023-043063-I, financed by MCIU/AEI/10.13039/501100011033 and by the FSE+).
M.M. acknowledges support from the Agencia Estatal de Investigaci\'on del Ministerio de Ciencia e Innovaci\'on (MCIN/AEI) under the grant "RR Lyrae stars, a lighthouse to distant galaxies and early galaxy evolution" and the European Regional Development Fun (ERDF) with reference PID2021-127042OB-I00.
Co-funded by the European Union (ERC-2022-AdG, "StarDance: the non-canonical evolution of stars in clusters", Grant Agreement 101093572, PI: E. Pancino). Views and opinions expressed are however those of the author(s) only and do not necessarily reflect those of the European Union or the European Research Council. Neither the European Union nor the granting authority can be held responsible for them

Based on observations with the NASA/ESA HST, obtained
at the Space Telescope Science Institute, which is operated by
AURA, Inc., under NASA contract NAS 5-26555. This
research made use of emcee \citep{foreman-mackey2013emcee}.
This work has made use of data from the European Space Agency (ESA) mission
\gaia\ (\url{https://www.cosmos.esa.int/gaia}), processed by the \gaia\
Data Processing and Analysis Consortium (DPAC,
\url{https://www.cosmos.esa.int/web/gaia/dpac/consortium}). Funding for the DPAC
has been provided by national institutions, in particular the institutions
participating in the \gaia\ Multilateral Agreement.
This project has received funding from the European Research Council (ERC) under the European Union’s Horizon 2020 research and innovation programme (grant agreement No. 804240) for S.S. and Á.S. M.M. acknowledges support from the Agencia Estatal de Investigaci\'on del Ministerio de Ciencia e Innovaci\'on (MCIN/AEI) under the grant "RR Lyrae stars, a lighthouse to distant galaxies and early galaxy evolution" and the European Regional Development Fun (ERDF) with reference PID2021-127042OB-I00, and from the Spanish Ministry of Science and Innovation (MICINN) through the Spanish State Research Agency, under Severo Ochoa Programe 2020-2023 (CEX2019-000920-S).

\end{acknowledgements}

\bibliographystyle{aa}
\bibliography{aa54262-25corr}

\begin{appendix}
\onecolumn
\section{Isochrone fitting results}
In this Appendix we show the results of our isochrone fitting algorithm applied to the 13 GCs under study in this work, with the exception of NGC~288 that is already shown in Fig.~\ref{fig:ngc288} in the main paper. Each figure is made up of four panels. The lower ones show the posterior distribution of the parameters of the model, including their correlation, resulting from the fit of the (m$_{F814W}$, m$_{F606W}$-m$_{F814W}$) CMD and of the (m$_{F606W}$, m$_{F606W}$-m$_{F814W}$) CMD. The upper panels show the isochrone corresponding to the best-fitting solution to the observed CMDs, where the green symbols mark the stars that were actually selected for the fit. We refer the reader to \citetalias{massari23} for further details about the method.

\renewcommand{\NGC}{NGC362}
\begin{figure*}[h!]
        \centering
        \begin{subfigure}{0.45\textwidth}
            \includegraphics[width=\textwidth]{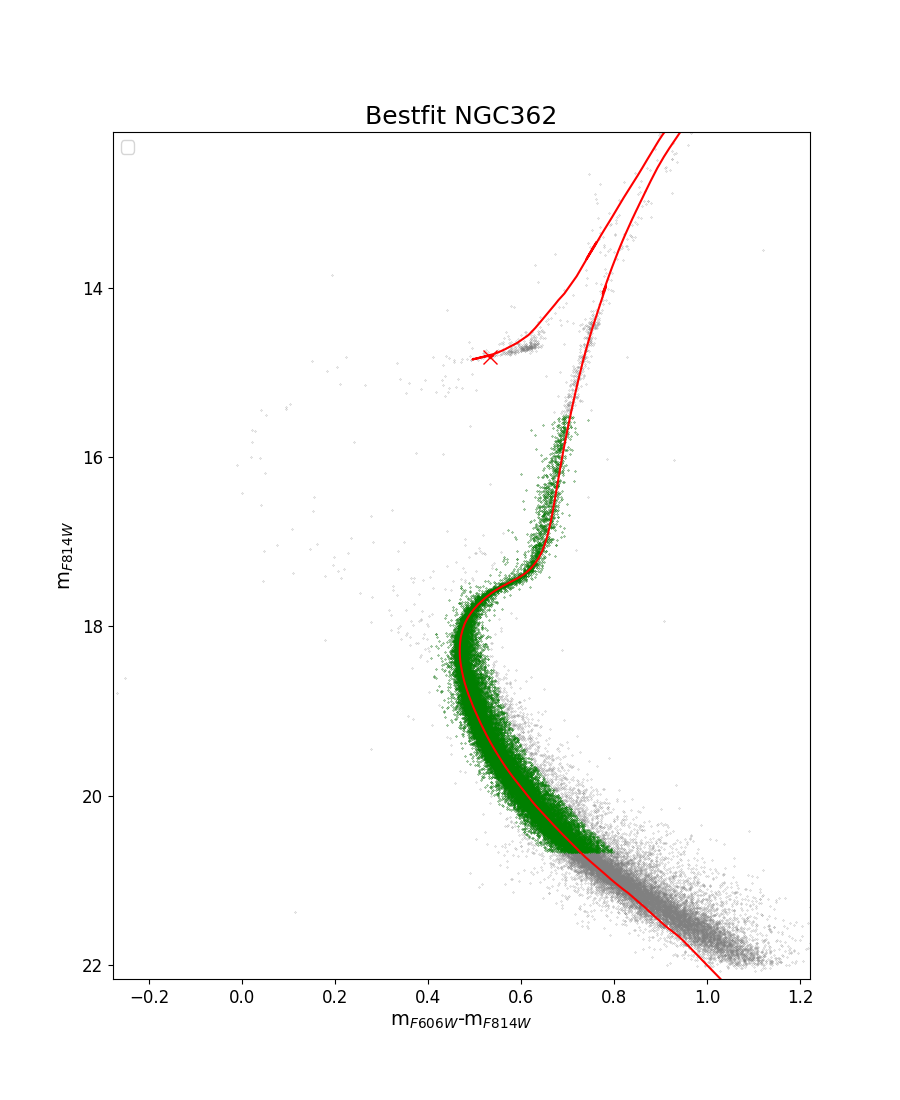}
            \caption[]
            {{\small }}    
        \end{subfigure}
        \begin{subfigure}{0.45\textwidth}  
            \includegraphics[width=\textwidth]{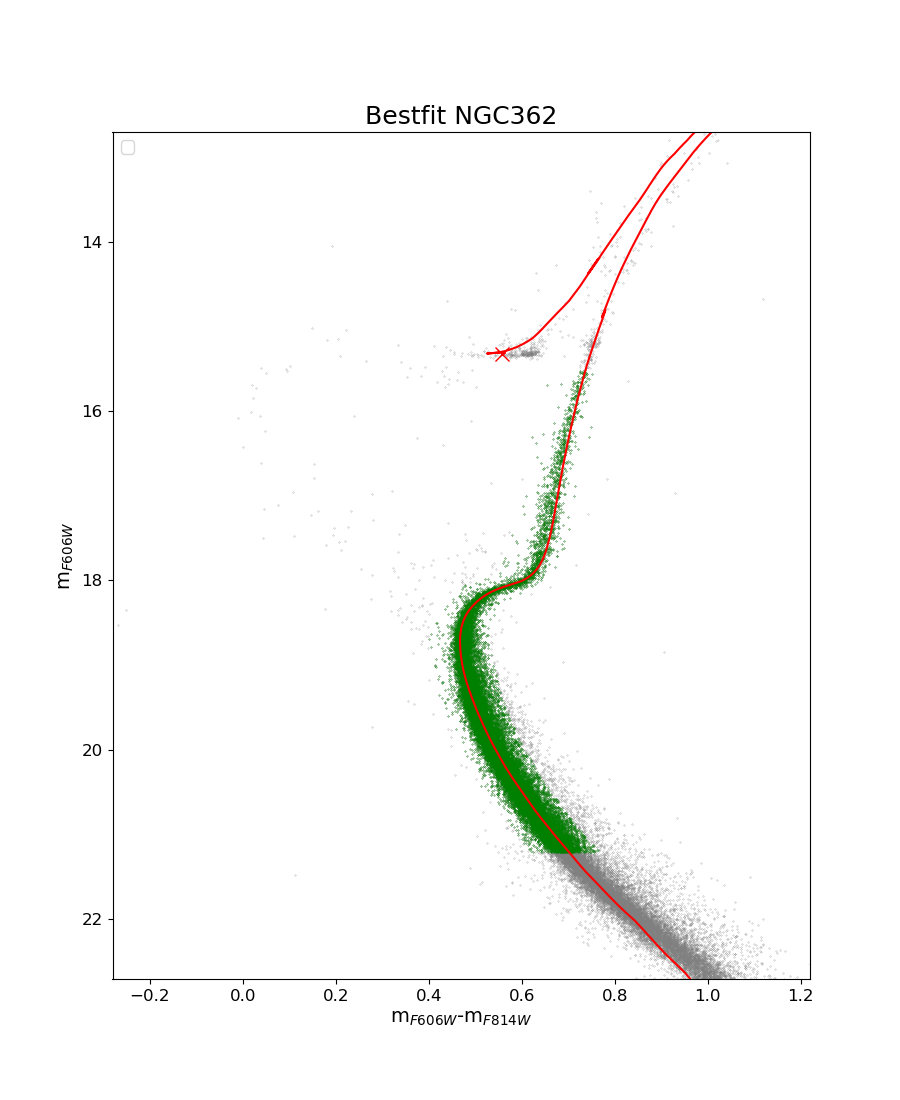}
            \caption[]
            {{\small }}    
        \end{subfigure}
        \vskip\baselineskip
        \begin{subfigure}{0.45\textwidth}   
            \includegraphics[width=\textwidth]{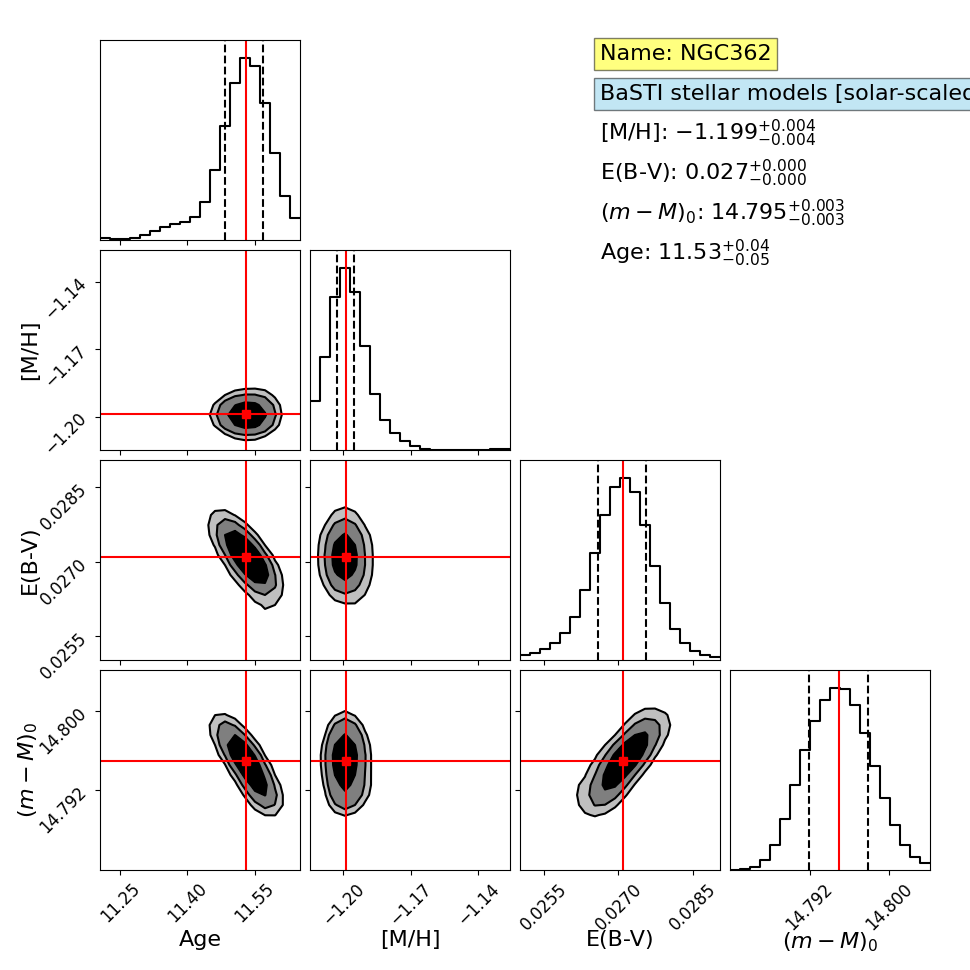}
            \caption[]
            {{\small }}    
        \end{subfigure}
        \begin{subfigure}{0.45\textwidth}   
            \includegraphics[width=\textwidth]{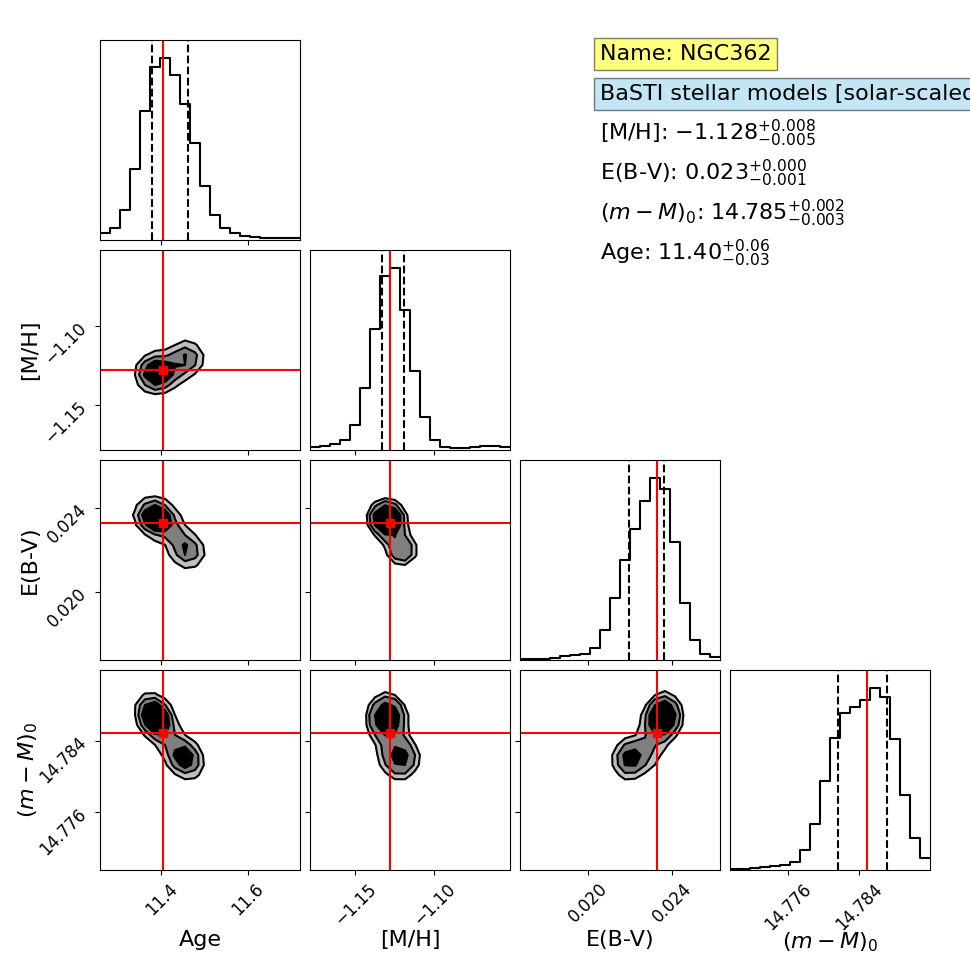}
            \caption[]
            {{\small }}    
        \end{subfigure}
        \caption[]
        {\small Results for \NGC. The meaning of the panels is the same as in Fig.~A.1} 
    \end{figure*}

\renewcommand{\NGC}{NGC1261}
\begin{figure*}
        \centering
        \begin{subfigure}{0.45\textwidth}
            \includegraphics[width=\textwidth]{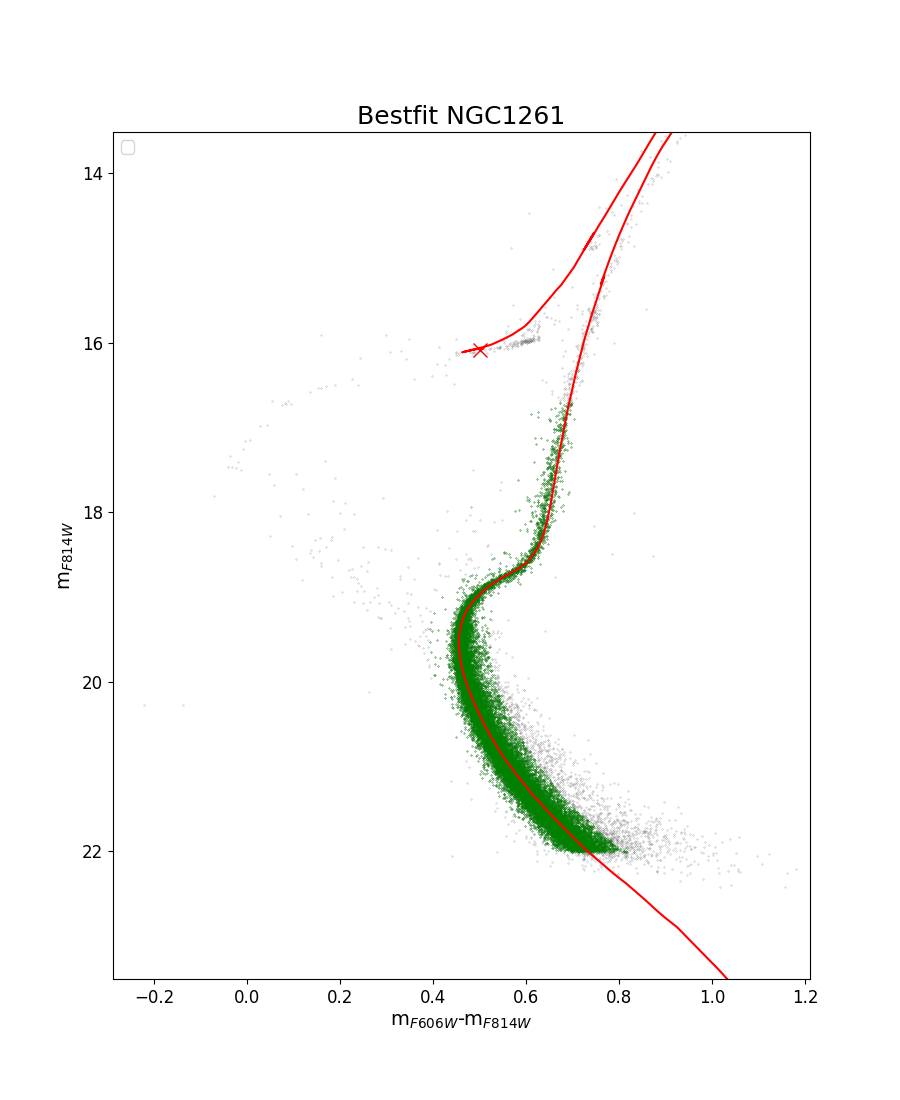}
            \caption[]
            {{\small }}    
        \end{subfigure}
        \begin{subfigure}{0.45\textwidth}  
            \includegraphics[width=\textwidth]{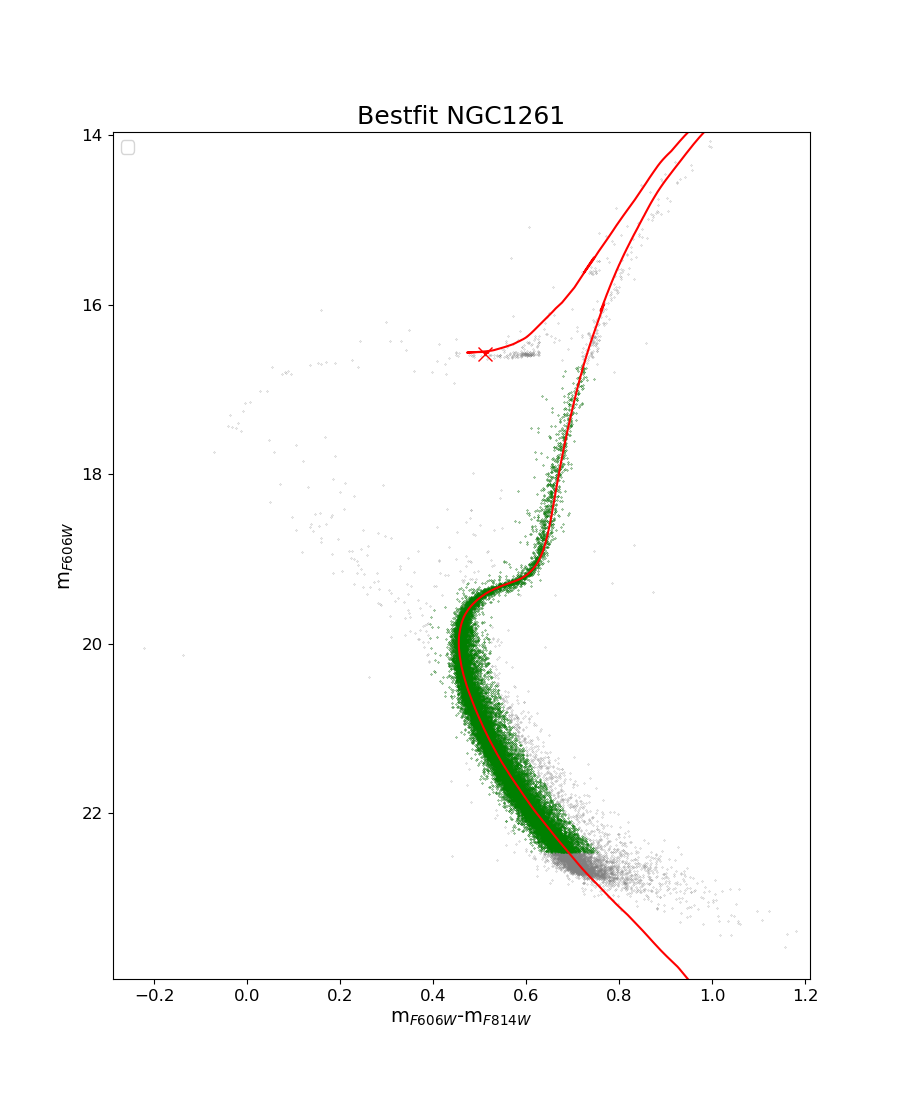}
            \caption[]
            {{\small }}    
        \end{subfigure}
        \vskip\baselineskip
        \begin{subfigure}{0.45\textwidth}   
            \includegraphics[width=\textwidth]{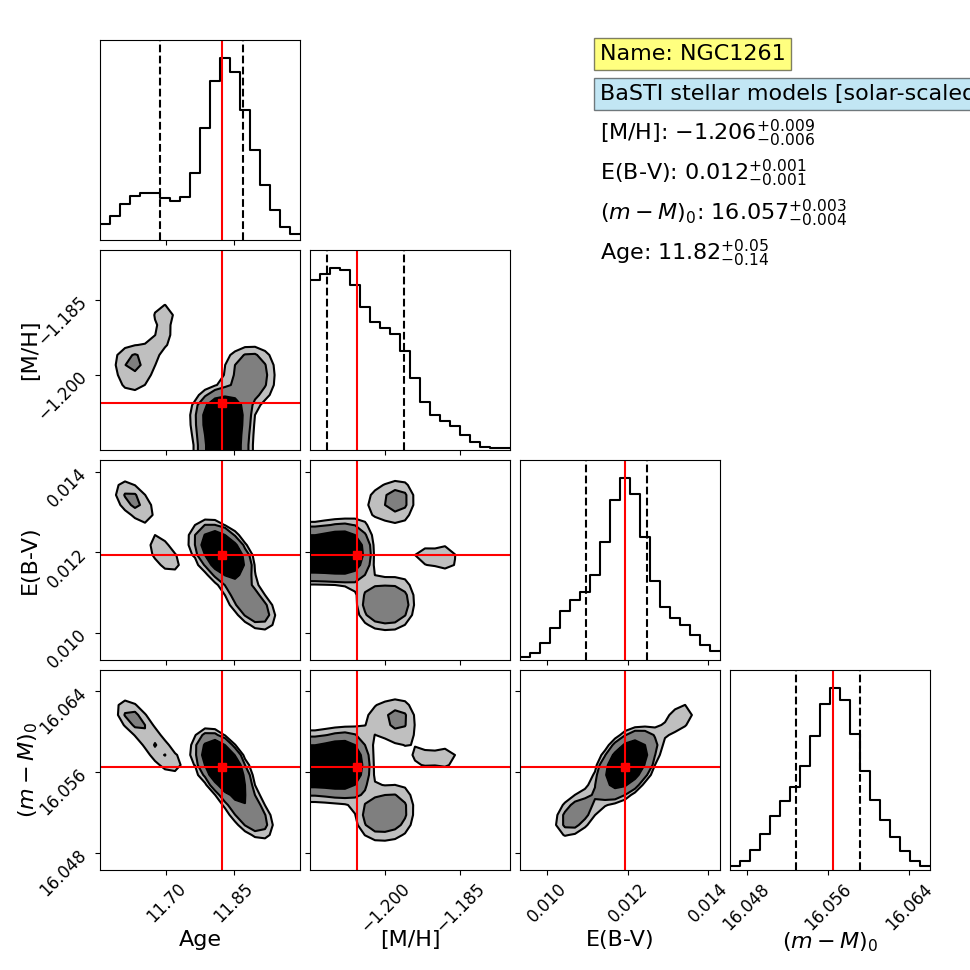}
            \caption[]
            {{\small }}    
        \end{subfigure}
        \begin{subfigure}{0.45\textwidth}   
            \includegraphics[width=\textwidth]{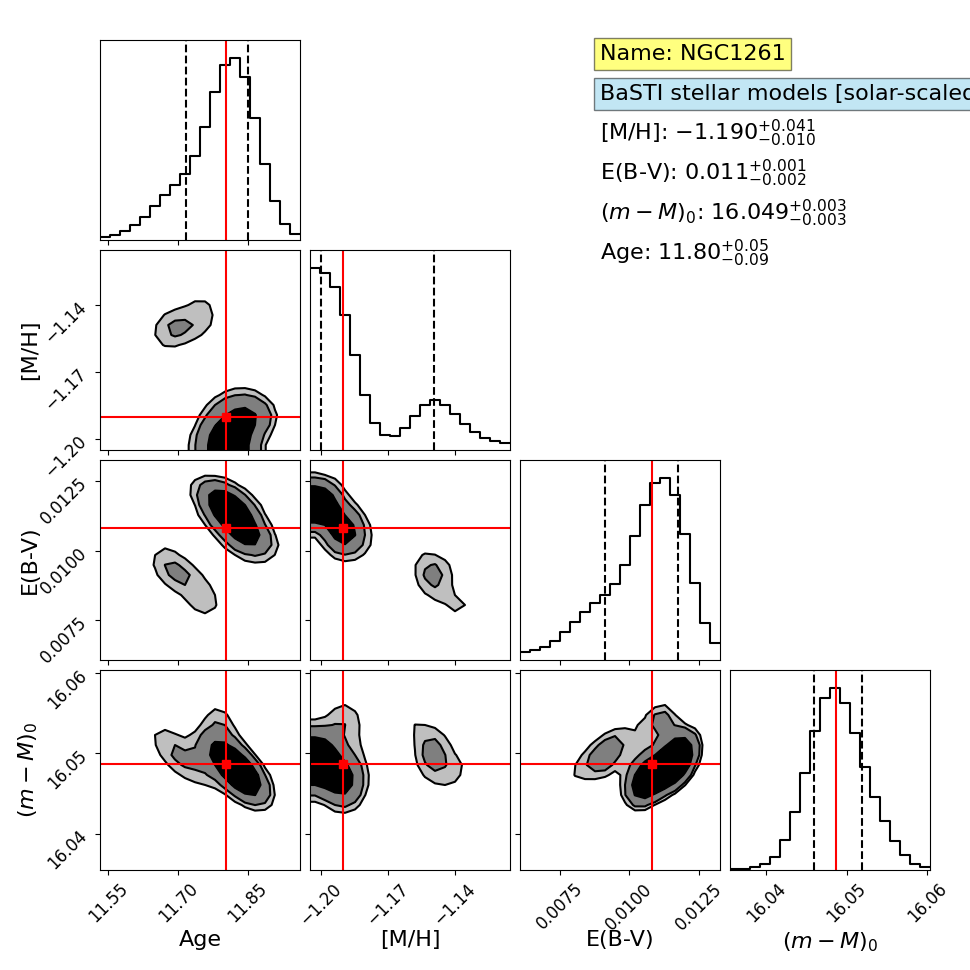}
            \caption[]
            {{\small }}    
        \end{subfigure}
        \caption[]
        {\small Results for \NGC. The meaning of the panels is the same as in Fig.~A.1} 
    \end{figure*}
    
\renewcommand{\NGC}{NGC1851}
\begin{figure*}
        \centering
        \begin{subfigure}{0.45\textwidth}
            \includegraphics[width=\textwidth]{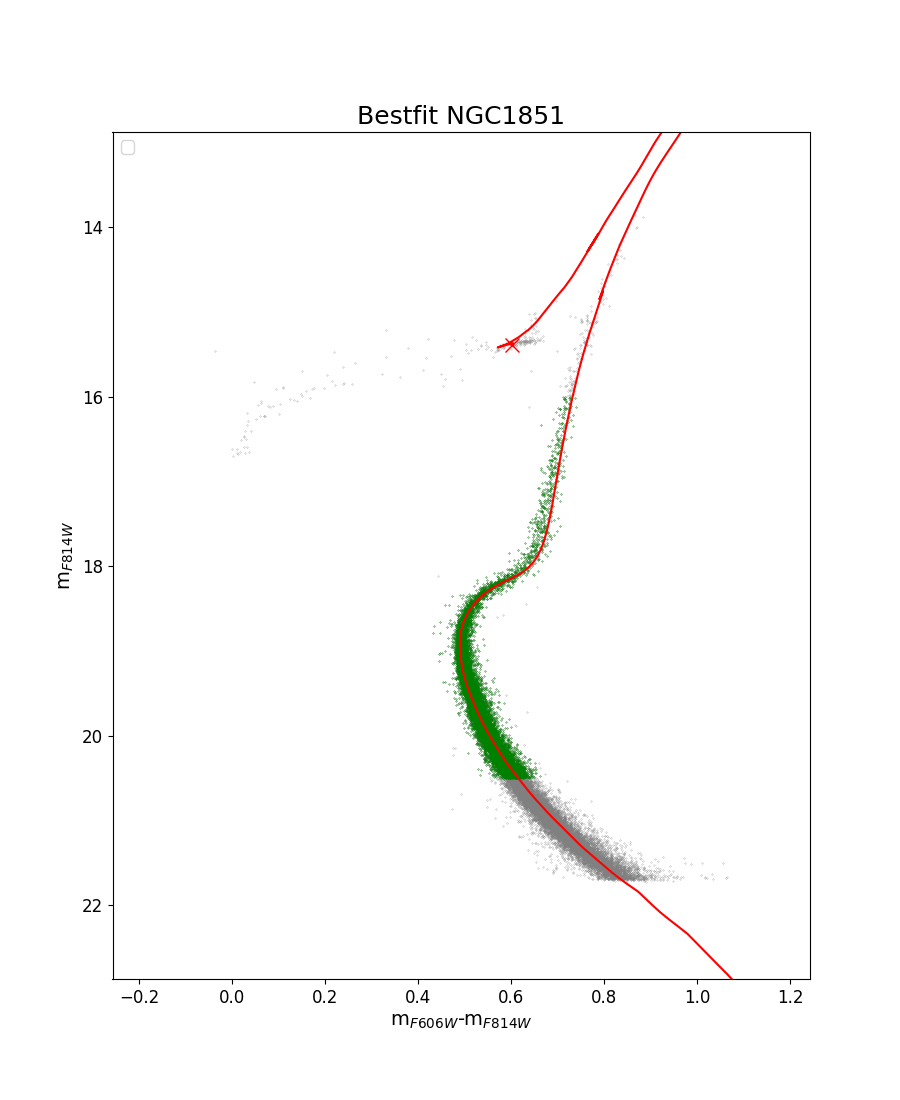}
            \caption[]
            {{\small }}    
        \end{subfigure}
        \begin{subfigure}{0.45\textwidth}  
            \includegraphics[width=\textwidth]{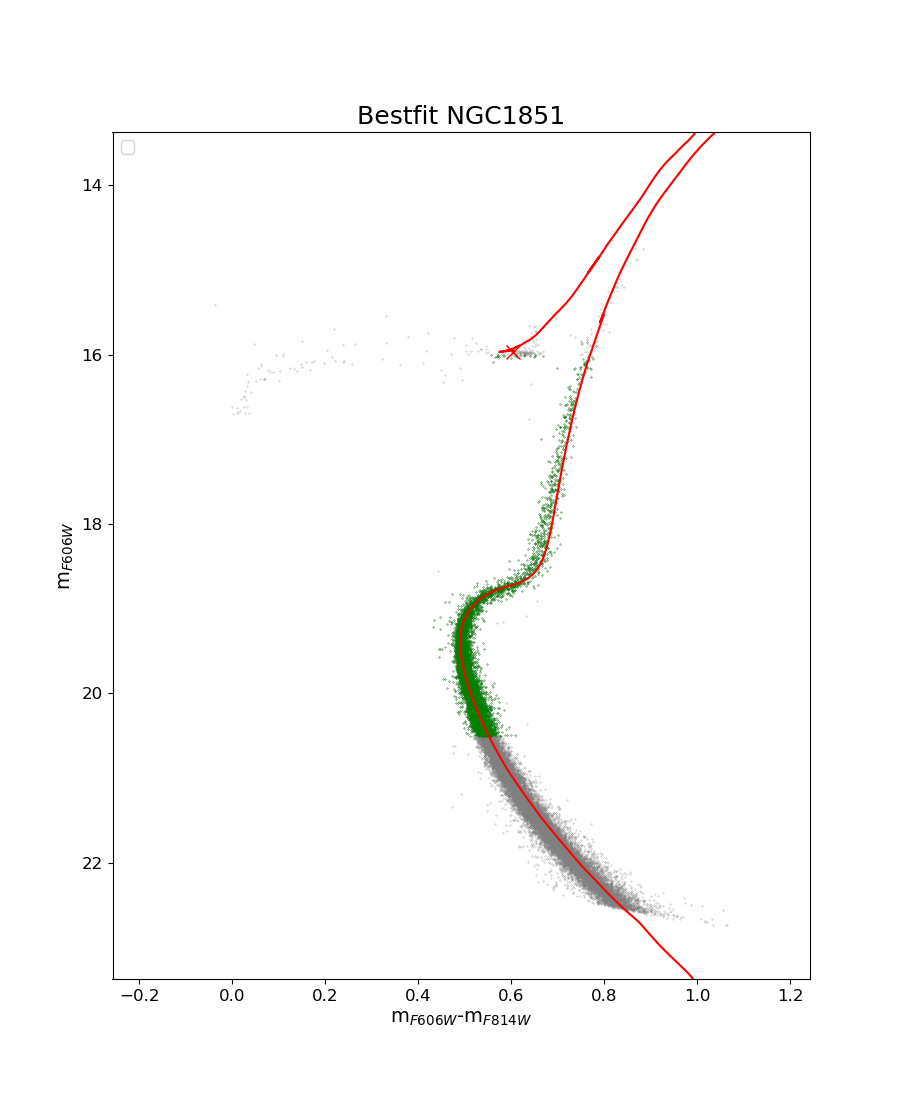}
            \caption[]
            {{\small }}    
        \end{subfigure}
        \vskip\baselineskip
        \begin{subfigure}{0.45\textwidth}   
            \includegraphics[width=\textwidth]{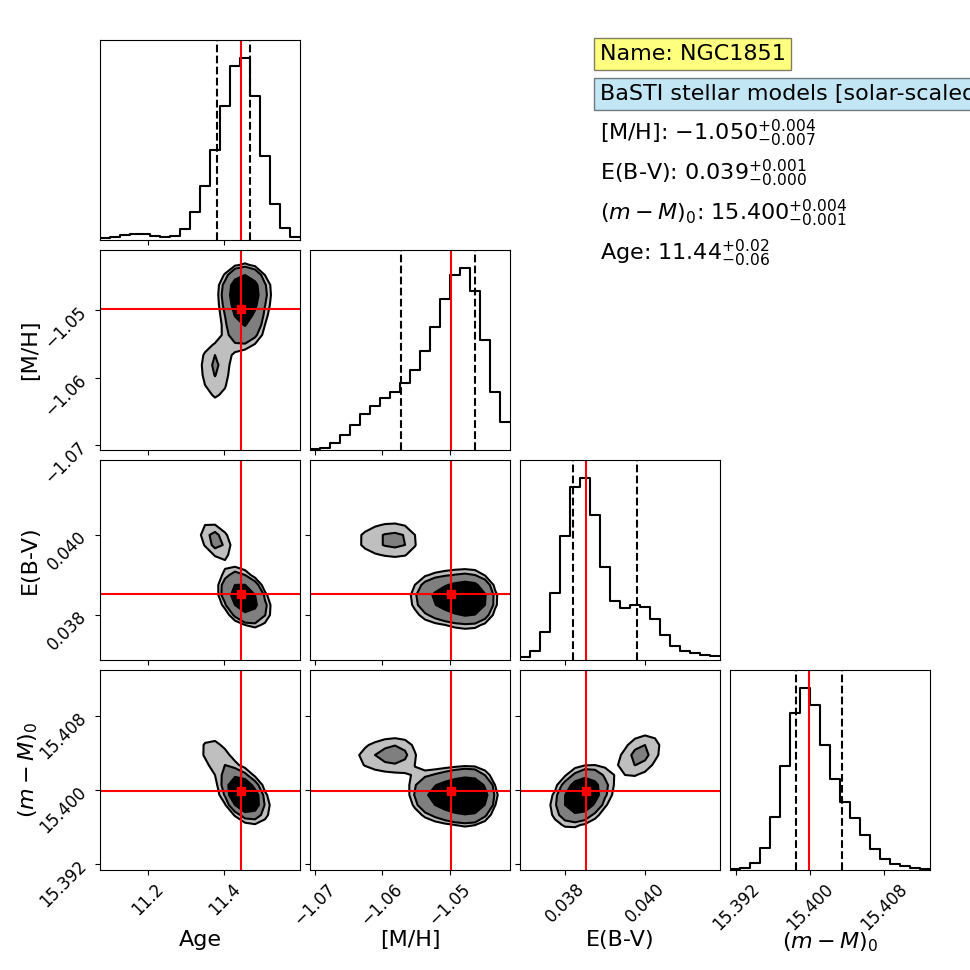}
            \caption[]
            {{\small }}    
        \end{subfigure}
        \begin{subfigure}{0.45\textwidth}   
            \includegraphics[width=\textwidth]{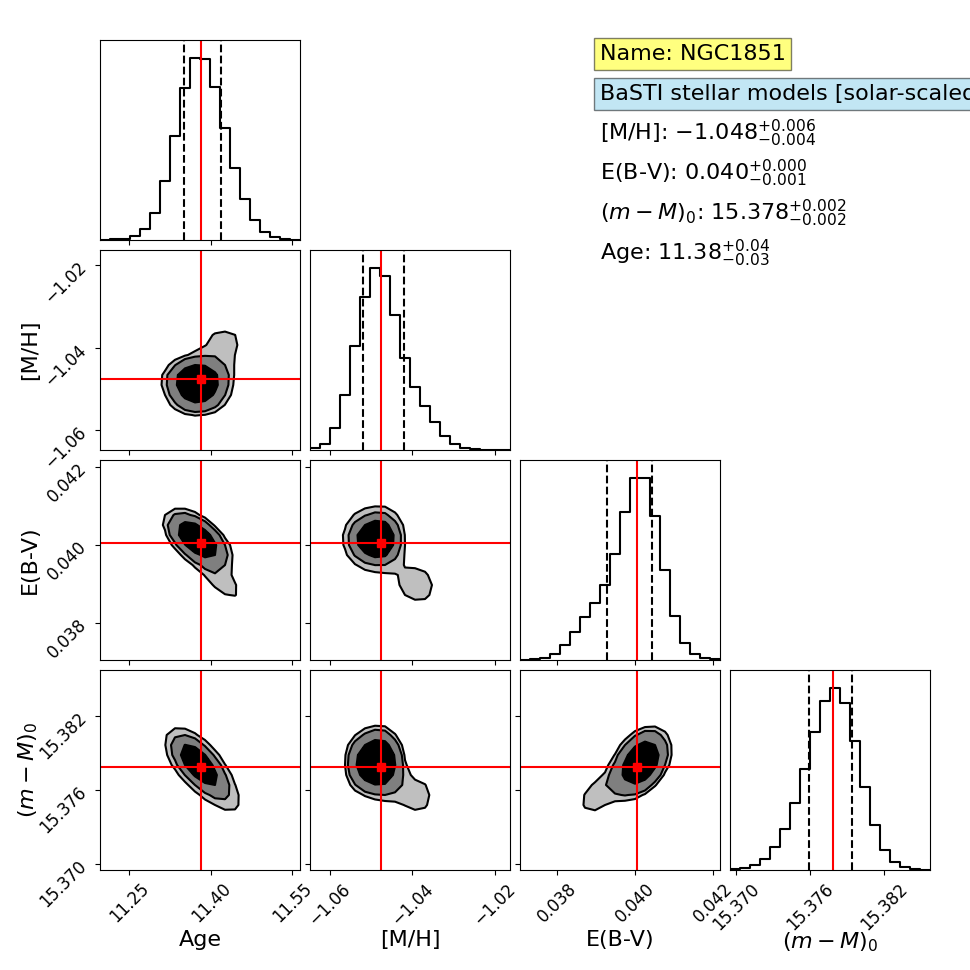}
            \caption[]
            {{\small }}    
        \end{subfigure}
        \caption[]
        {\small Results for \NGC. The meaning of the panels is the same as in Fig.~A.1} 
    \end{figure*}
    
\renewcommand{\NGC}{NGC2298}
\begin{figure*}
        \centering
        \begin{subfigure}{0.45\textwidth}
            \includegraphics[width=\textwidth]{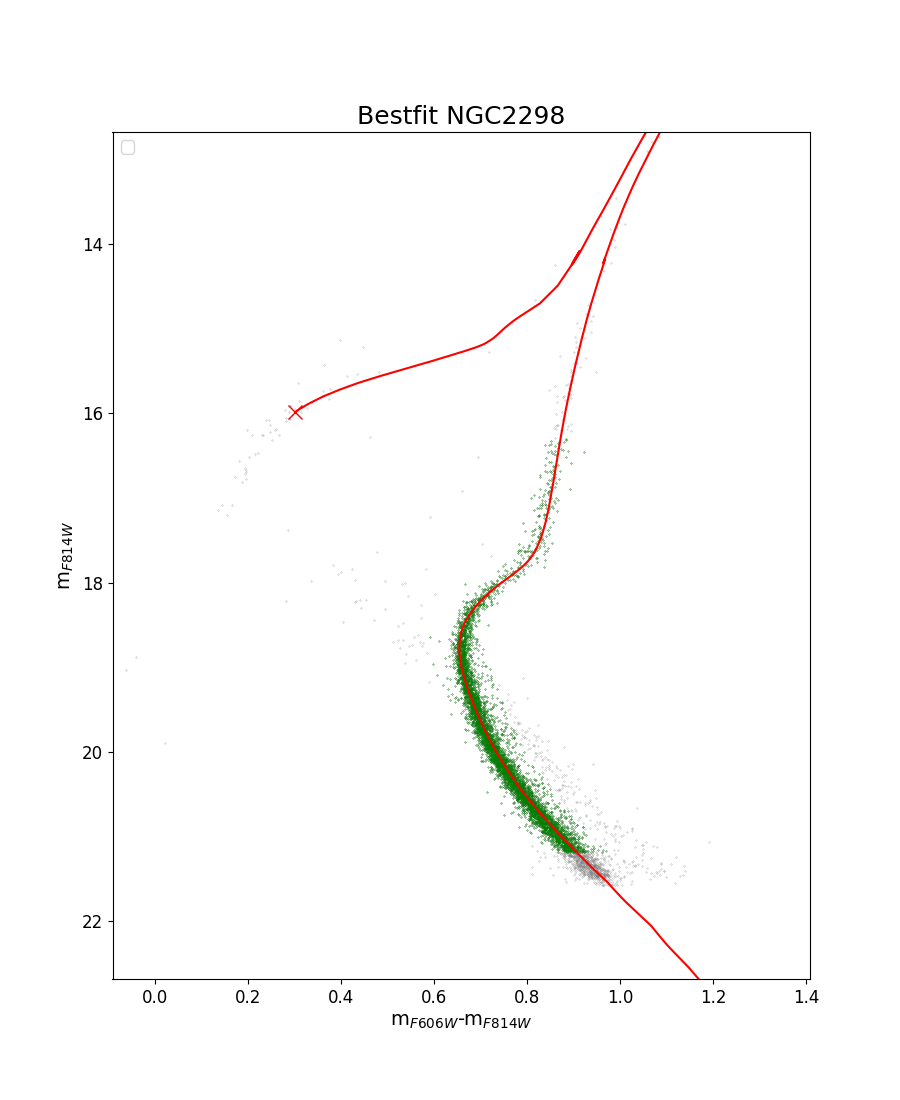}
            \caption[]
            {{\small }}    
        \end{subfigure}
        \begin{subfigure}{0.45\textwidth}  
            \includegraphics[width=\textwidth]{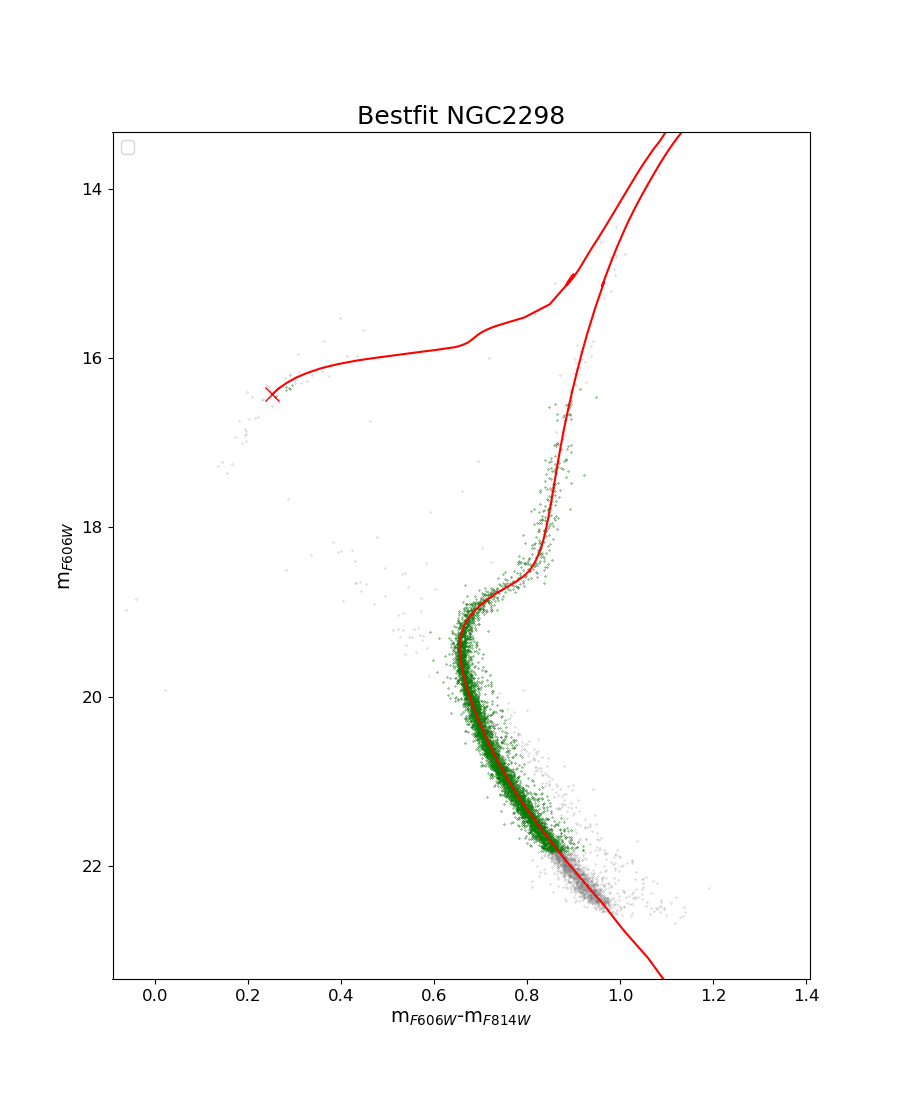}
            \caption[]
            {{\small }}    
        \end{subfigure}
        \vskip\baselineskip
        \begin{subfigure}{0.45\textwidth}   
            \includegraphics[width=\textwidth]{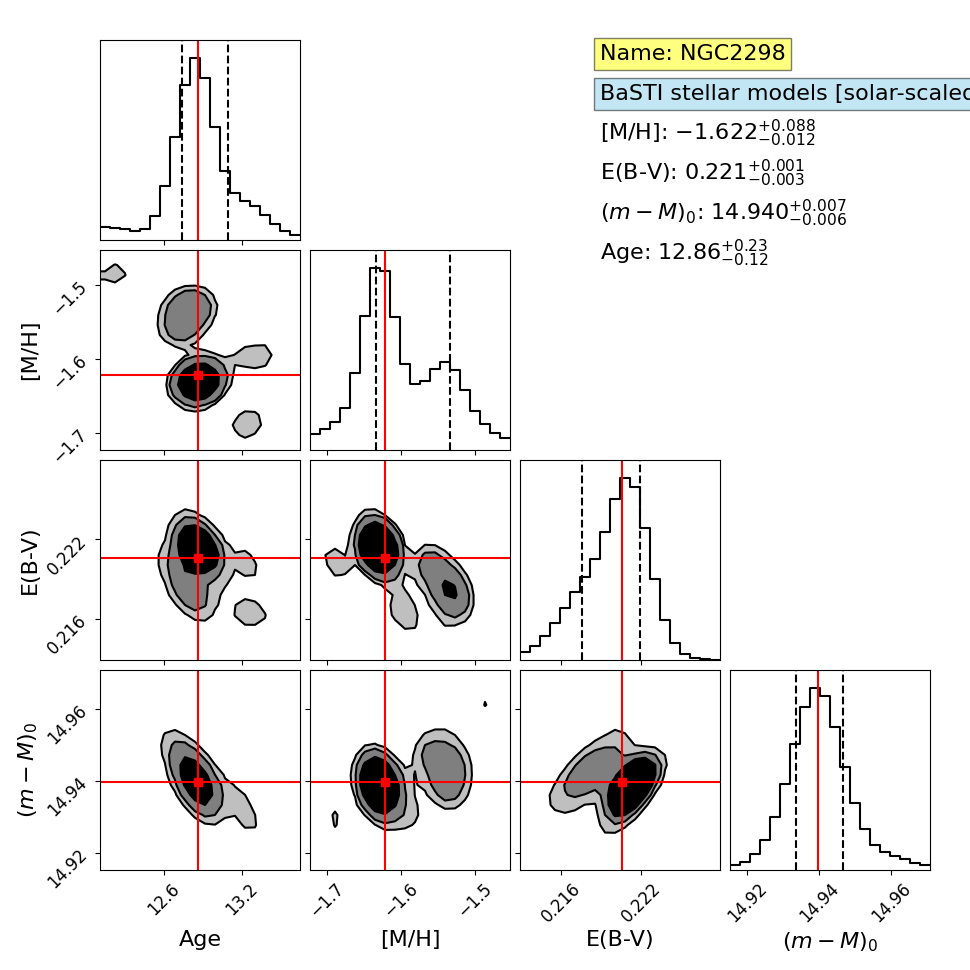}
            \caption[]
            {{\small }}    
        \end{subfigure}
        \begin{subfigure}{0.45\textwidth}   
            \includegraphics[width=\textwidth]{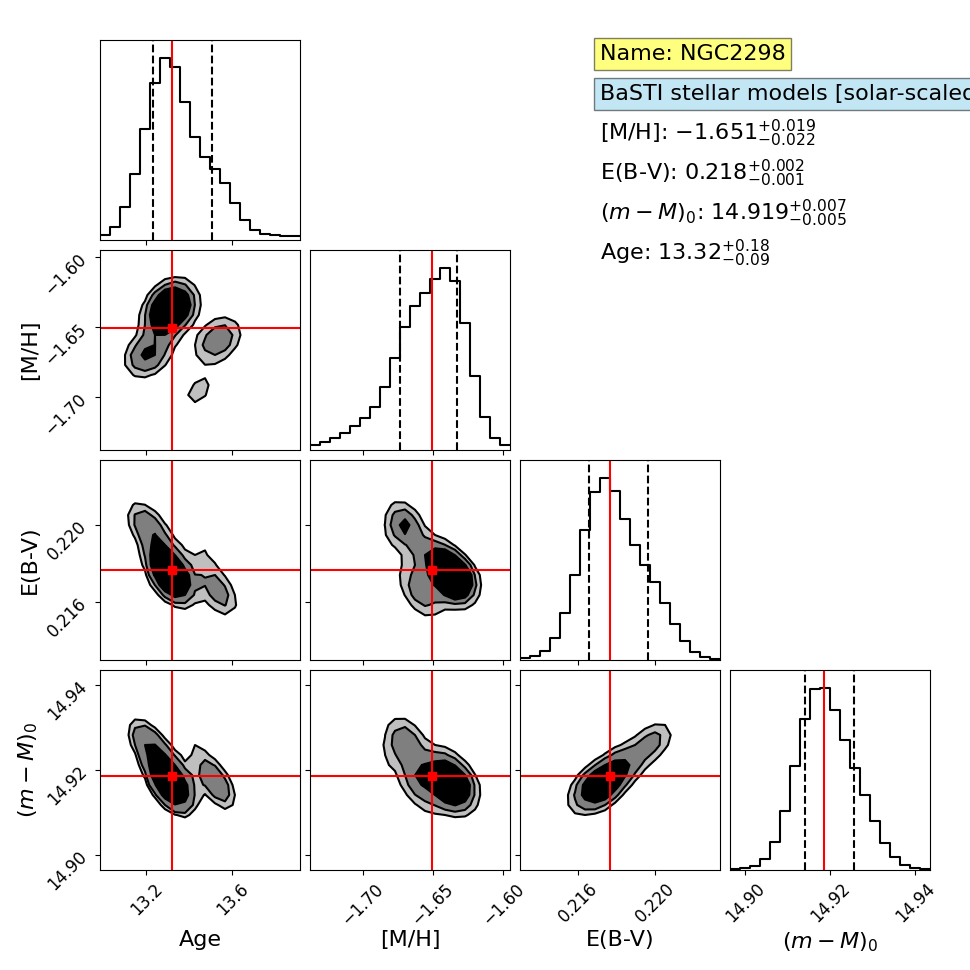}
            \caption[]
            {{\small }}    
        \end{subfigure}
        \caption[]
        {\small Results for \NGC. The meaning of the panels is the same as in Fig.~A.1} 
    \end{figure*}

\renewcommand{\NGC}{NGC2808}
\begin{figure*}
        \centering
        \begin{subfigure}{0.45\textwidth}
            \includegraphics[width=\textwidth]{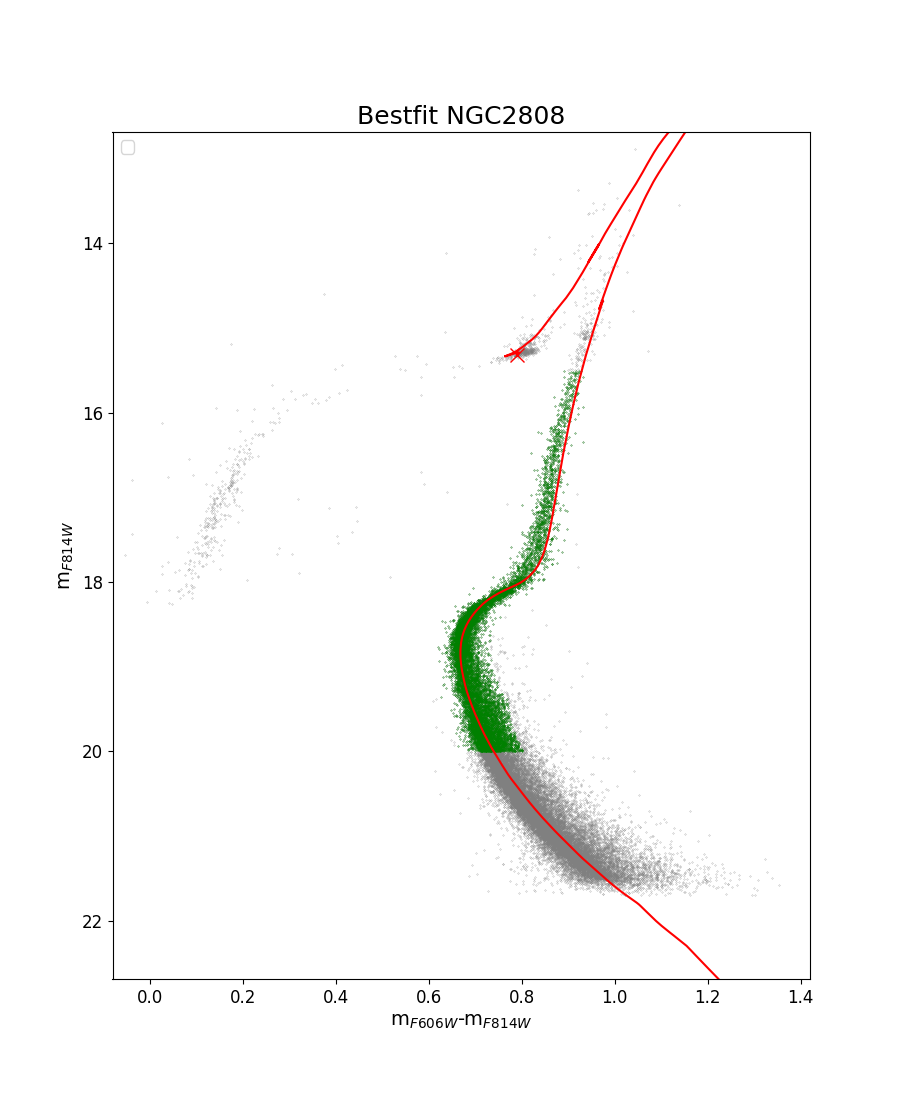}
            \caption[]
            {{\small }}    
        \end{subfigure}
        \begin{subfigure}{0.45\textwidth}  
            \includegraphics[width=\textwidth]{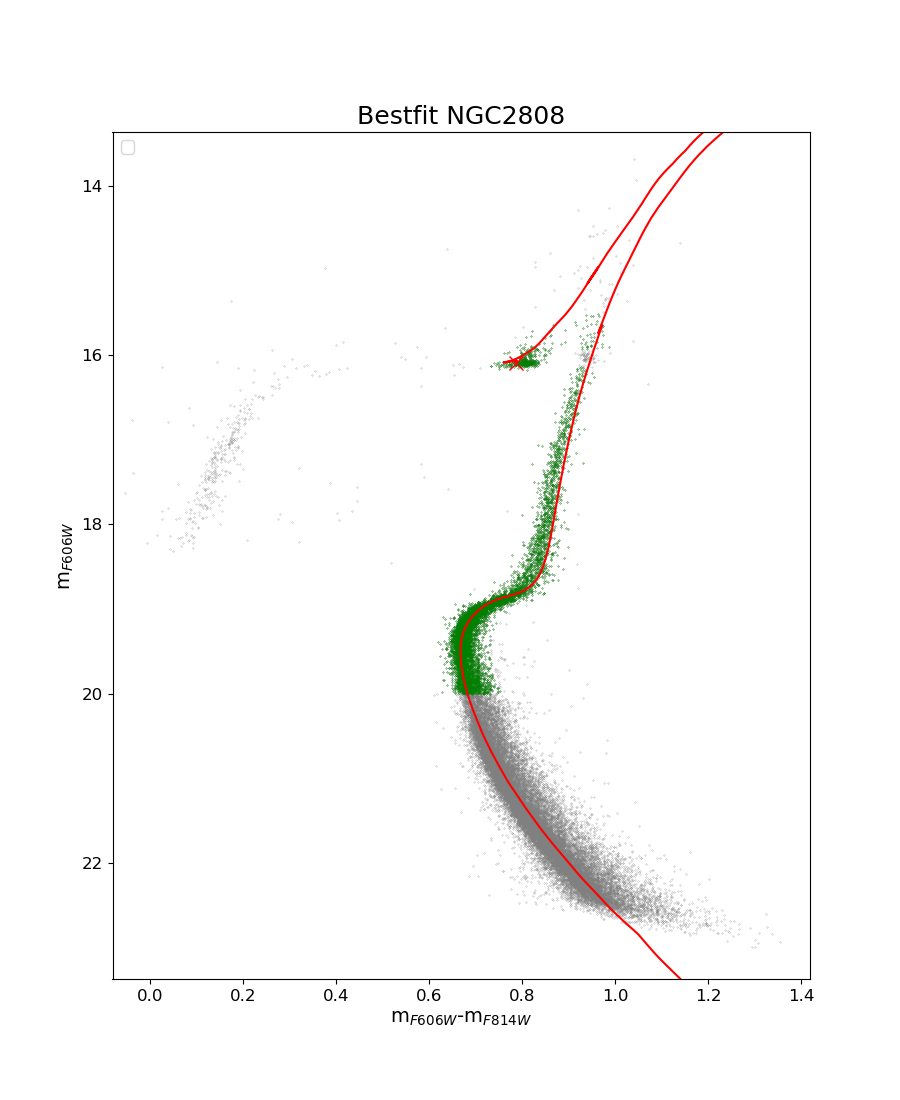}
            \caption[]
            {{\small }}    
        \end{subfigure}
        \vskip\baselineskip
        \begin{subfigure}{0.45\textwidth}   
            \includegraphics[width=\textwidth]{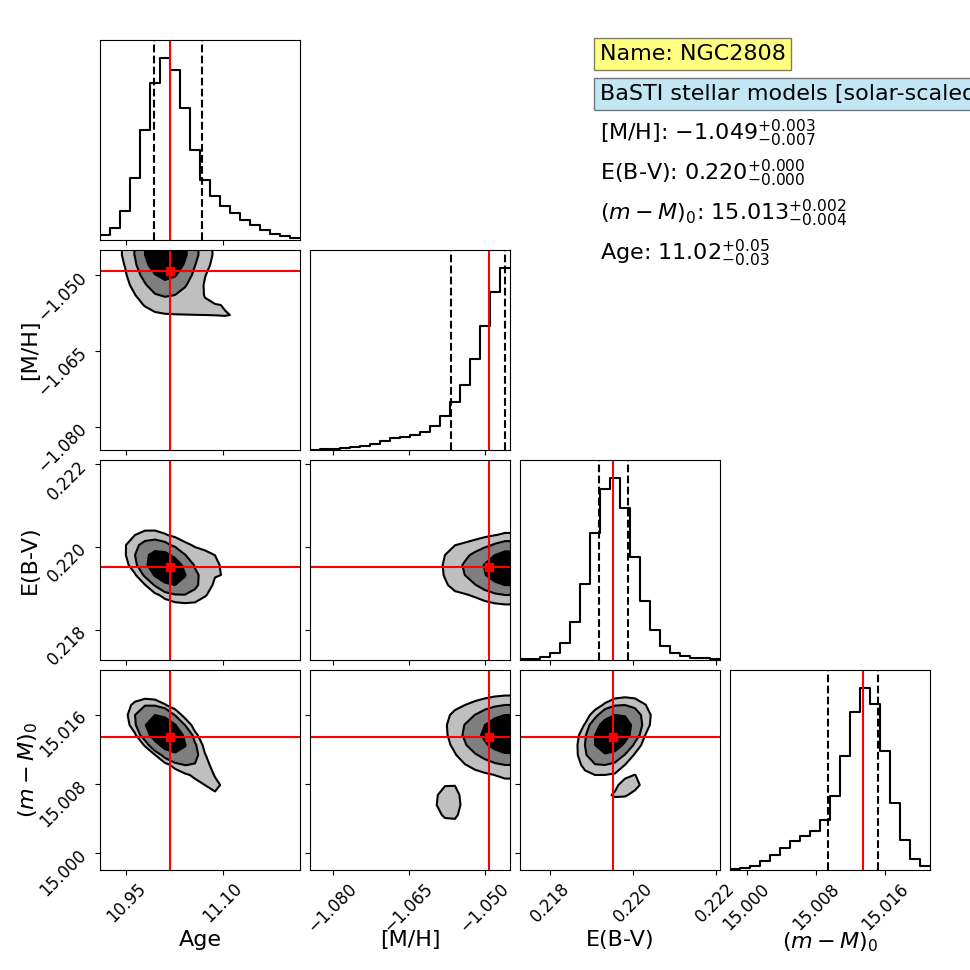}
            \caption[]
            {{\small }}    
        \end{subfigure}
        \begin{subfigure}{0.45\textwidth}   
            \includegraphics[width=\textwidth]{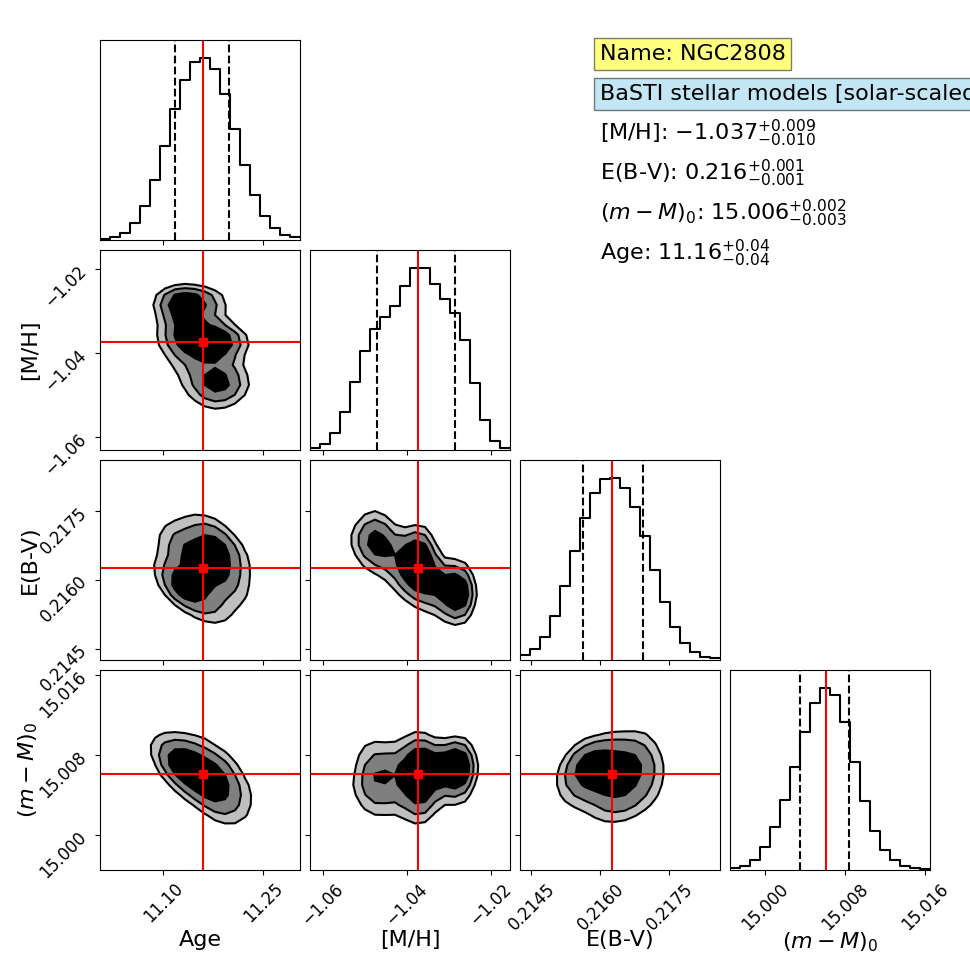}
            \caption[]
            {{\small }}    
        \end{subfigure}
        \caption[]
        {\small Results for \NGC. The meaning of the panels is the same as in Fig.~A.1} 
    \end{figure*}

\renewcommand{\NGC}{NGC5286}
\begin{figure*}
        \centering
        \begin{subfigure}{0.45\textwidth}
            \includegraphics[width=\textwidth]{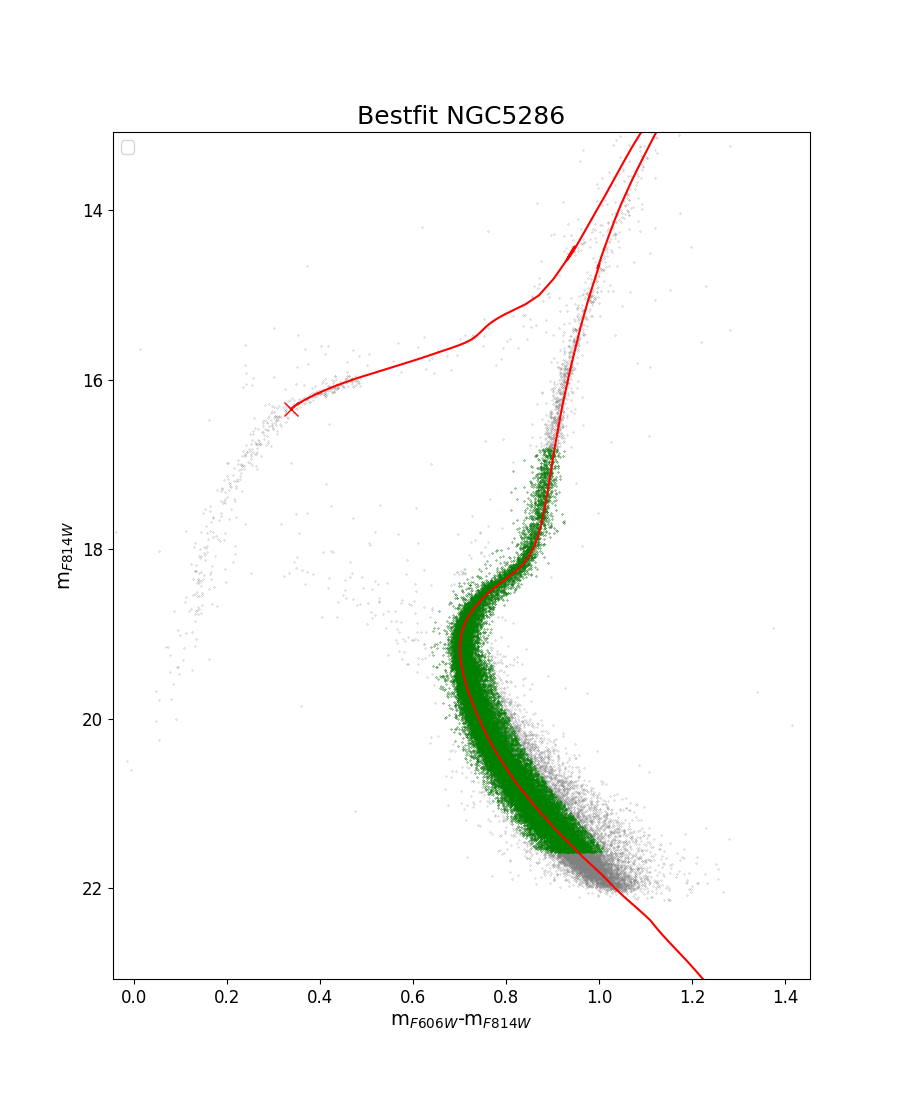}
            \caption[]
            {{\small }}    
        \end{subfigure}
        \begin{subfigure}{0.45\textwidth}  
            \includegraphics[width=\textwidth]{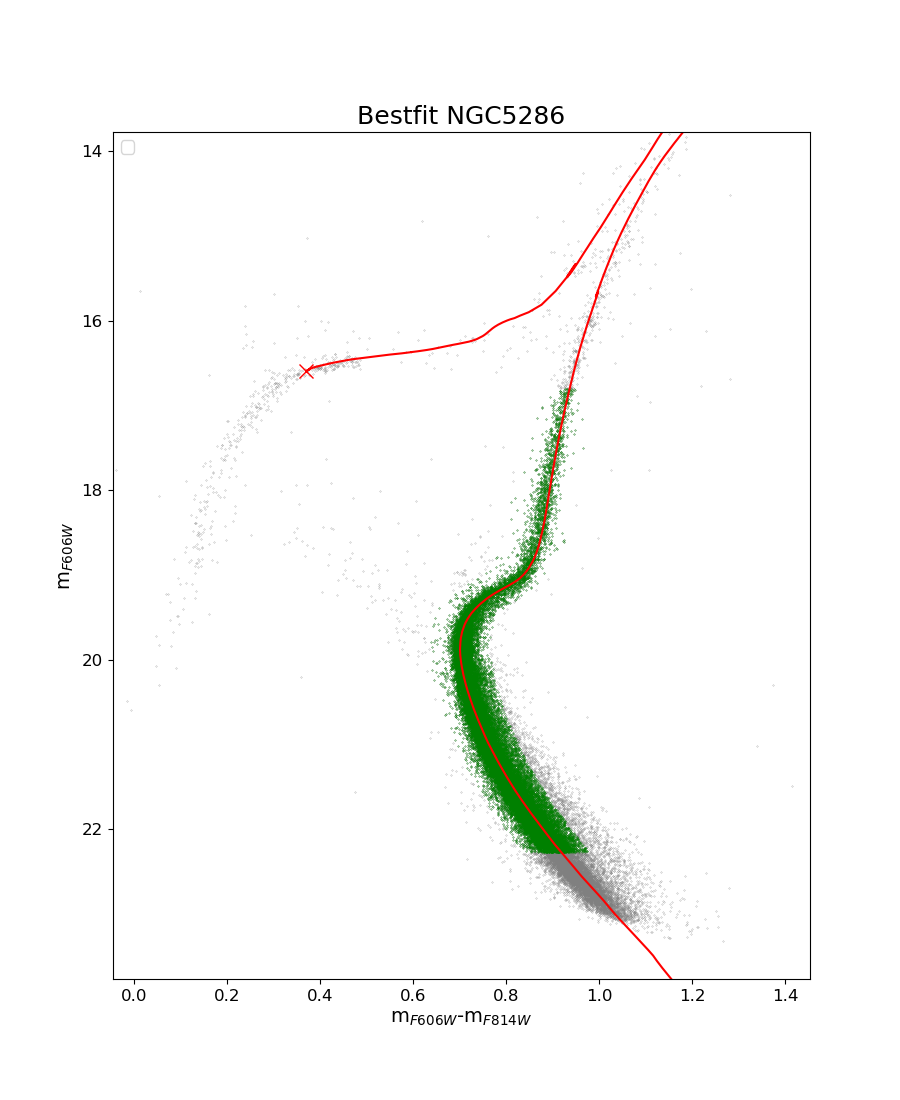}
            \caption[]
            {{\small }}    
        \end{subfigure}
        \vskip\baselineskip
        \begin{subfigure}{0.45\textwidth}   
            \includegraphics[width=\textwidth]{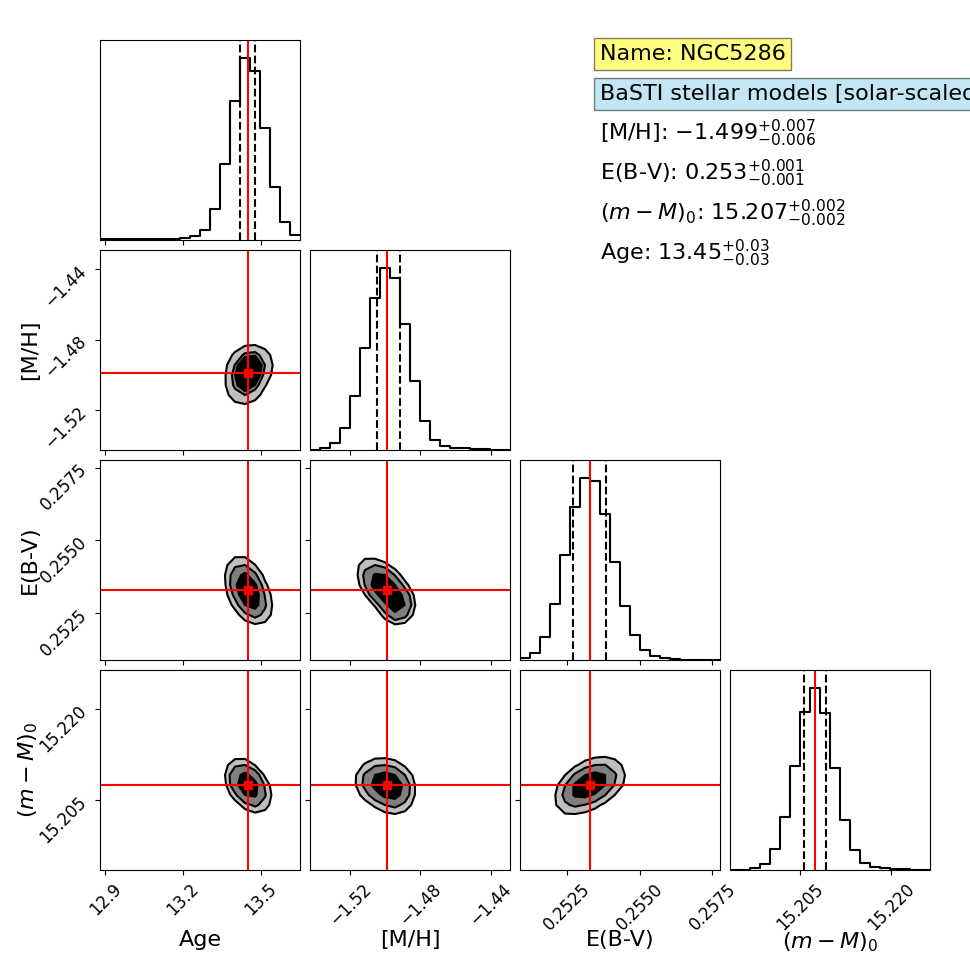}
            \caption[]
            {{\small }}    
        \end{subfigure}
        \begin{subfigure}{0.45\textwidth}   
            \includegraphics[width=\textwidth]{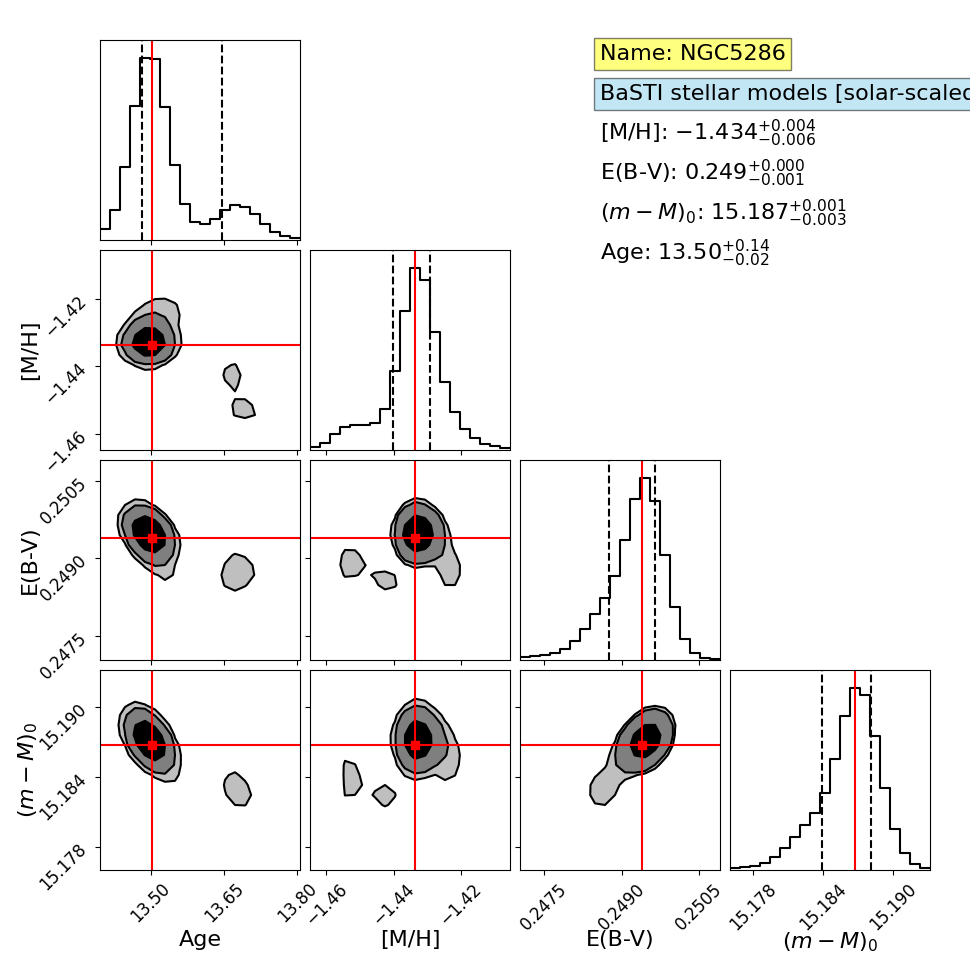}
            \caption[]
            {{\small }}    
        \end{subfigure}
        \caption[]
        {\small Results for \NGC. The meaning of the panels is the same as in Fig.~A.1} 
    \end{figure*}

\renewcommand{\NGC}{NGC5897}
\begin{figure*}
        \centering
        \begin{subfigure}{0.45\textwidth}
            \includegraphics[width=\textwidth]{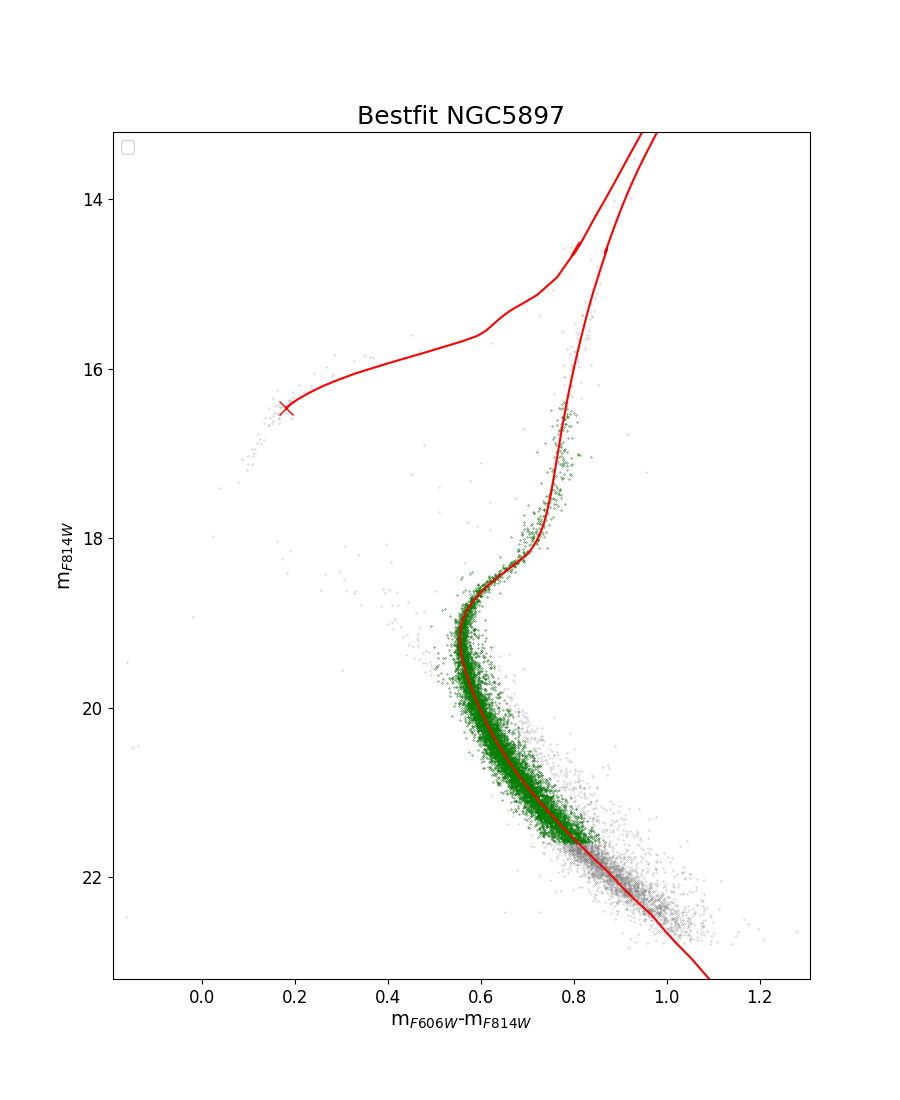}
            \caption[]
            {{\small }}    
        \end{subfigure}
        \begin{subfigure}{0.45\textwidth}  
            \includegraphics[width=\textwidth]{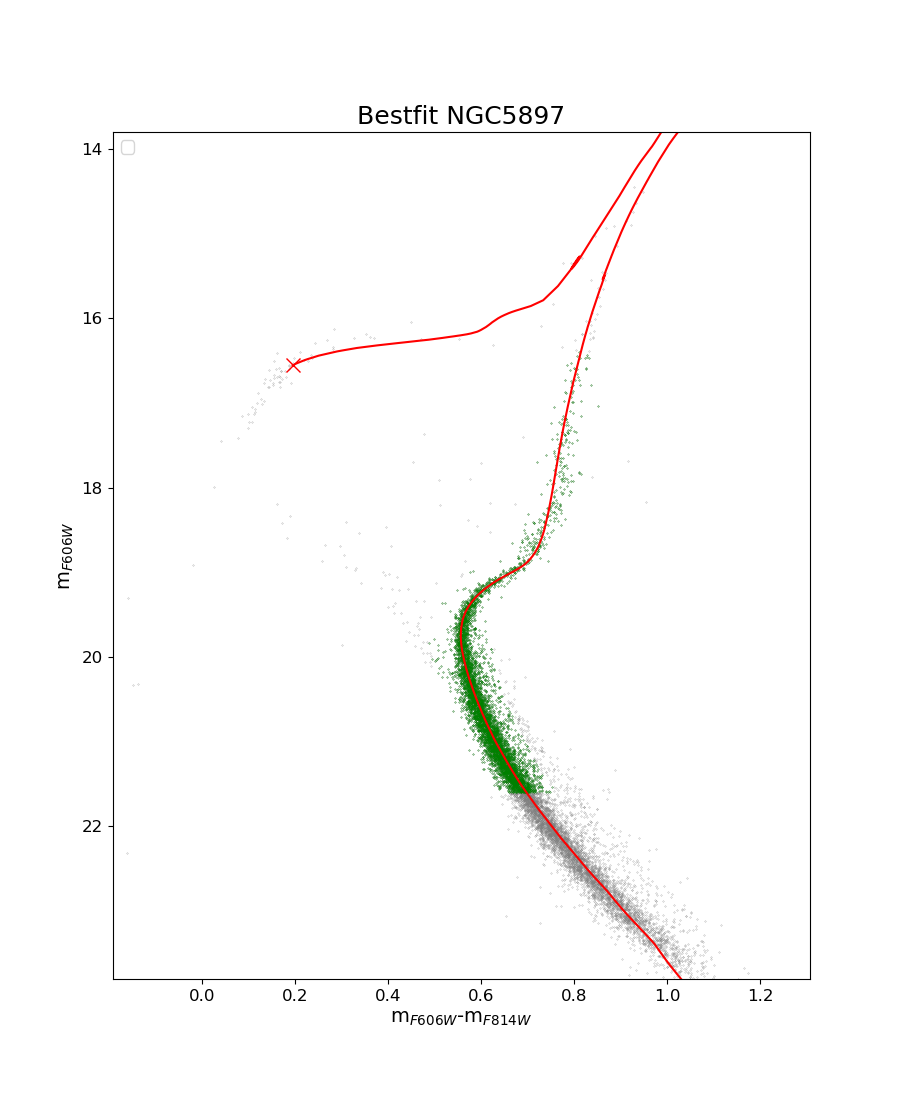}
            \caption[]
            {{\small }}    
        \end{subfigure}
        \vskip\baselineskip
        \begin{subfigure}{0.45\textwidth}   
            \includegraphics[width=\textwidth]{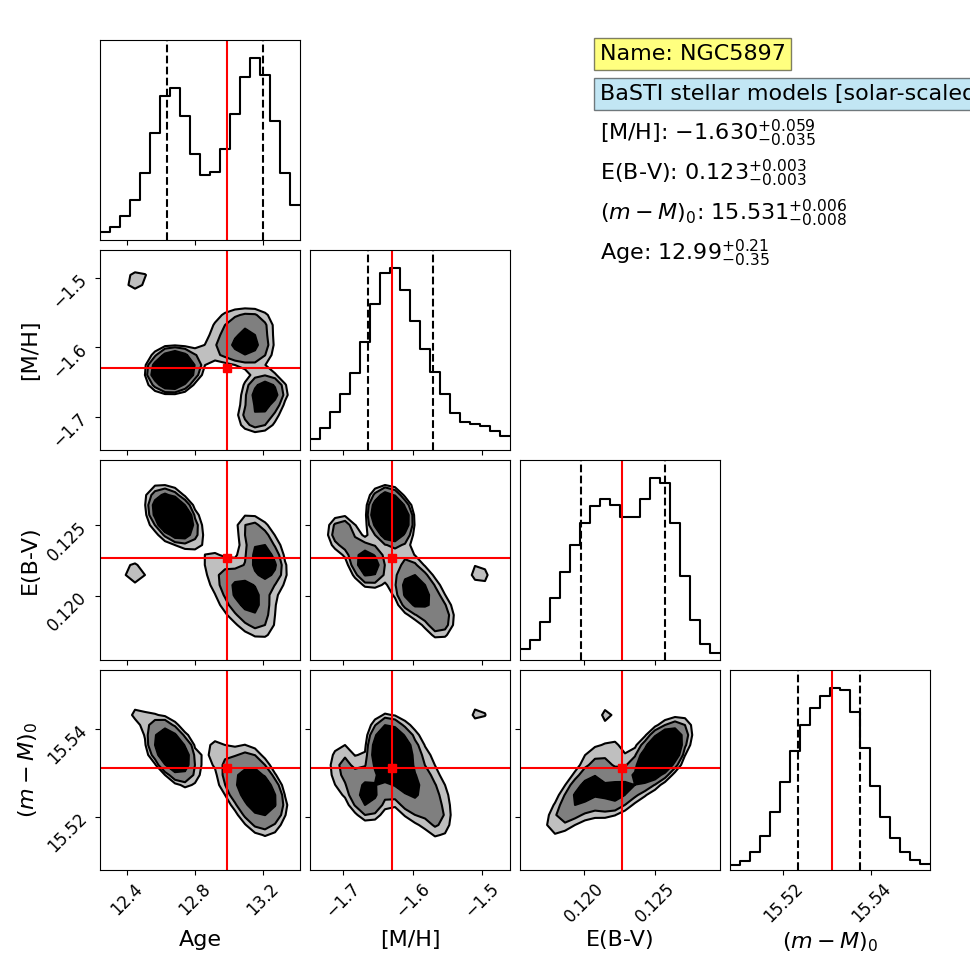}
            \caption[]
            {{\small }}    
        \end{subfigure}
        \begin{subfigure}{0.45\textwidth}   
            \includegraphics[width=\textwidth]{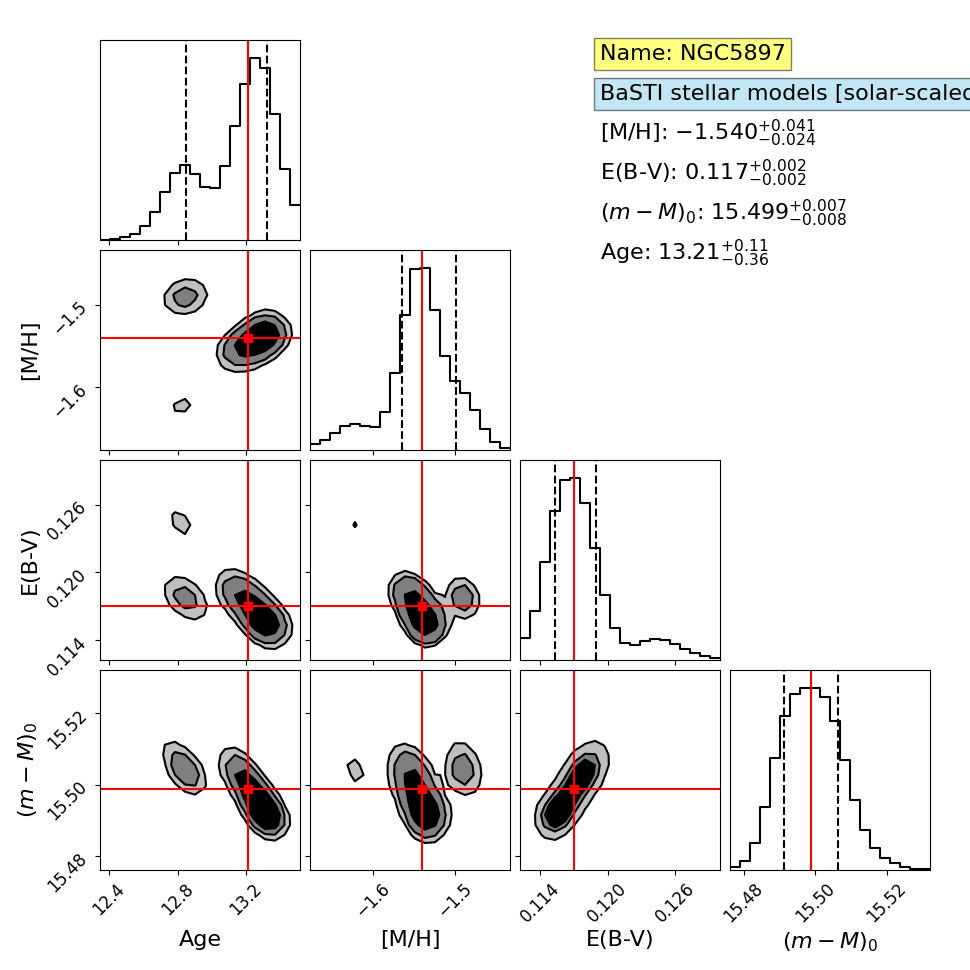}
            \caption[]
            {{\small }}    
        \end{subfigure}
        \caption[]
        {\small Results for \NGC. The meaning of the panels is the same as in Fig.~A.1} 
    \end{figure*}

\renewcommand{\NGC}{NGC6205}
\begin{figure*}
        \centering
        \begin{subfigure}{0.45\textwidth}
            \includegraphics[width=\textwidth]{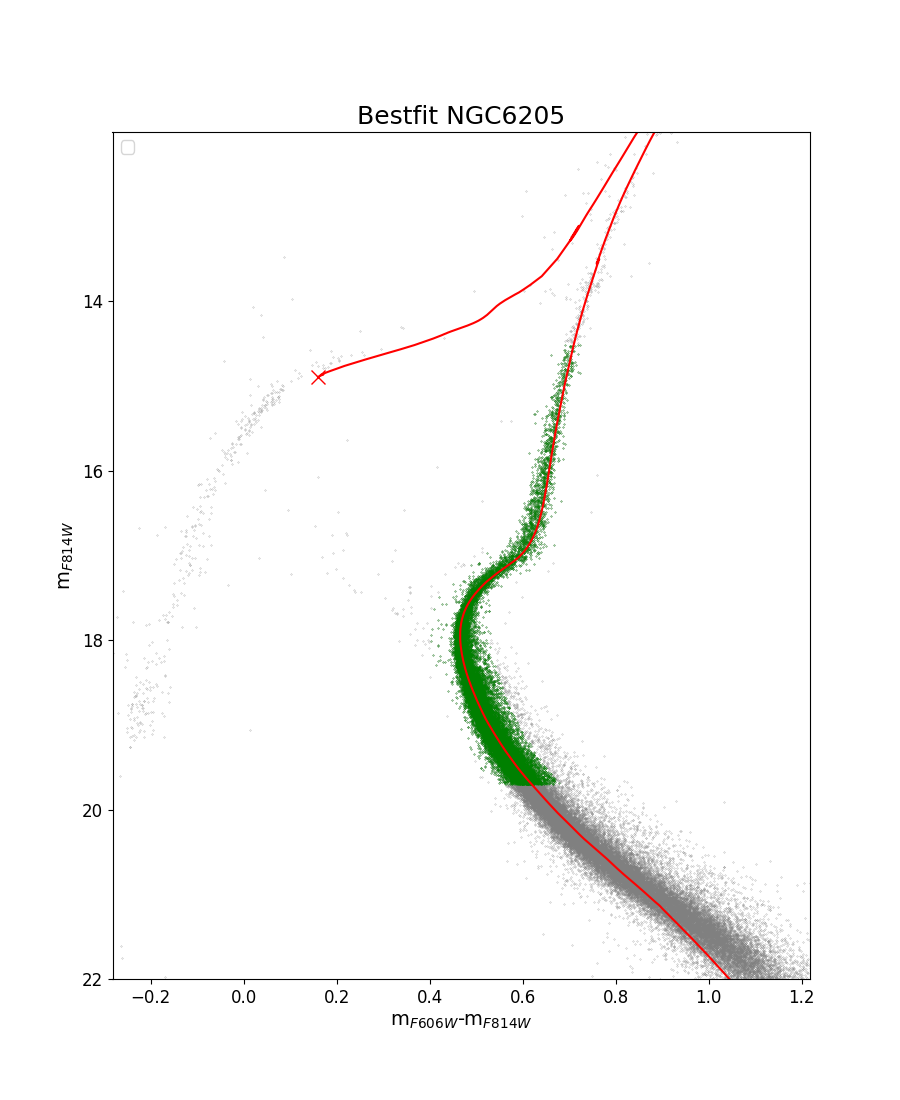}
            \caption[]
            {{\small }}    
        \end{subfigure}
        \begin{subfigure}{0.45\textwidth}  
            \includegraphics[width=\textwidth]{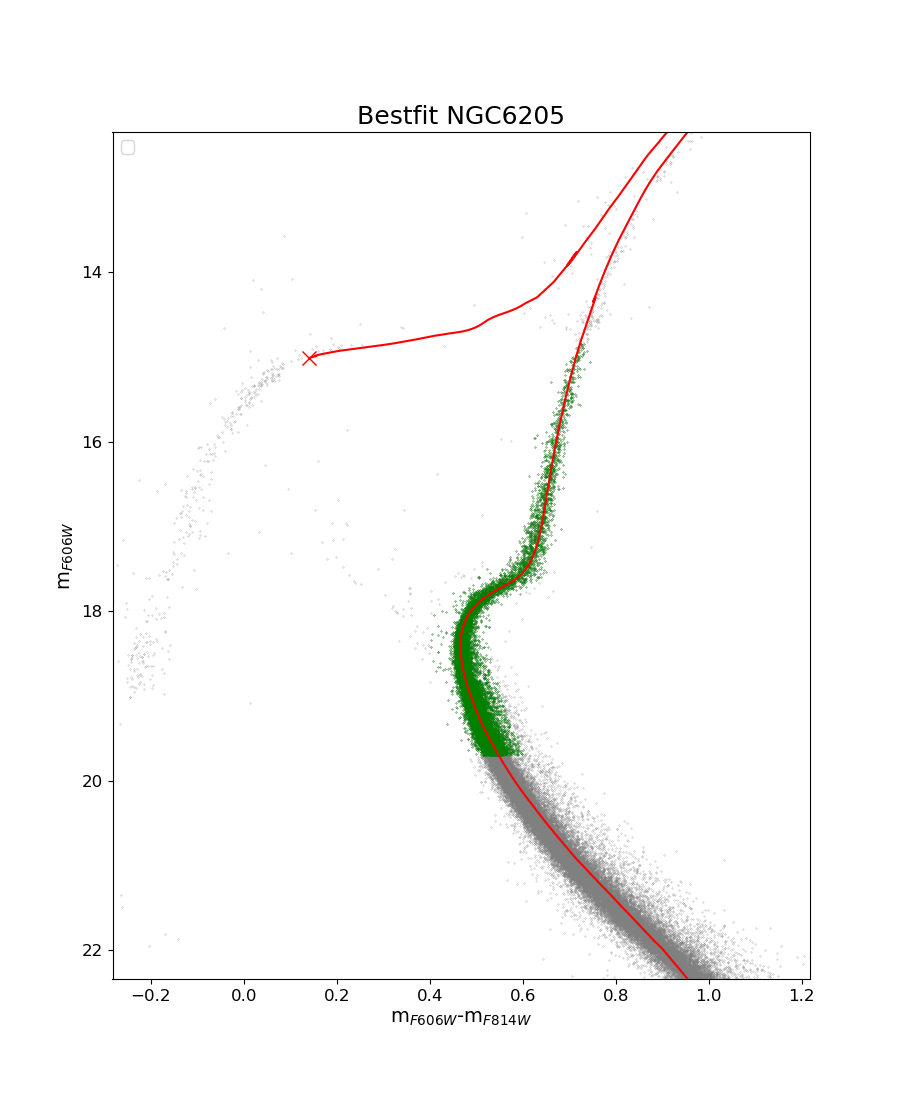}
            \caption[]
            {{\small }}    
        \end{subfigure}
        \vskip\baselineskip
        \begin{subfigure}{0.45\textwidth}   
            \includegraphics[width=\textwidth]{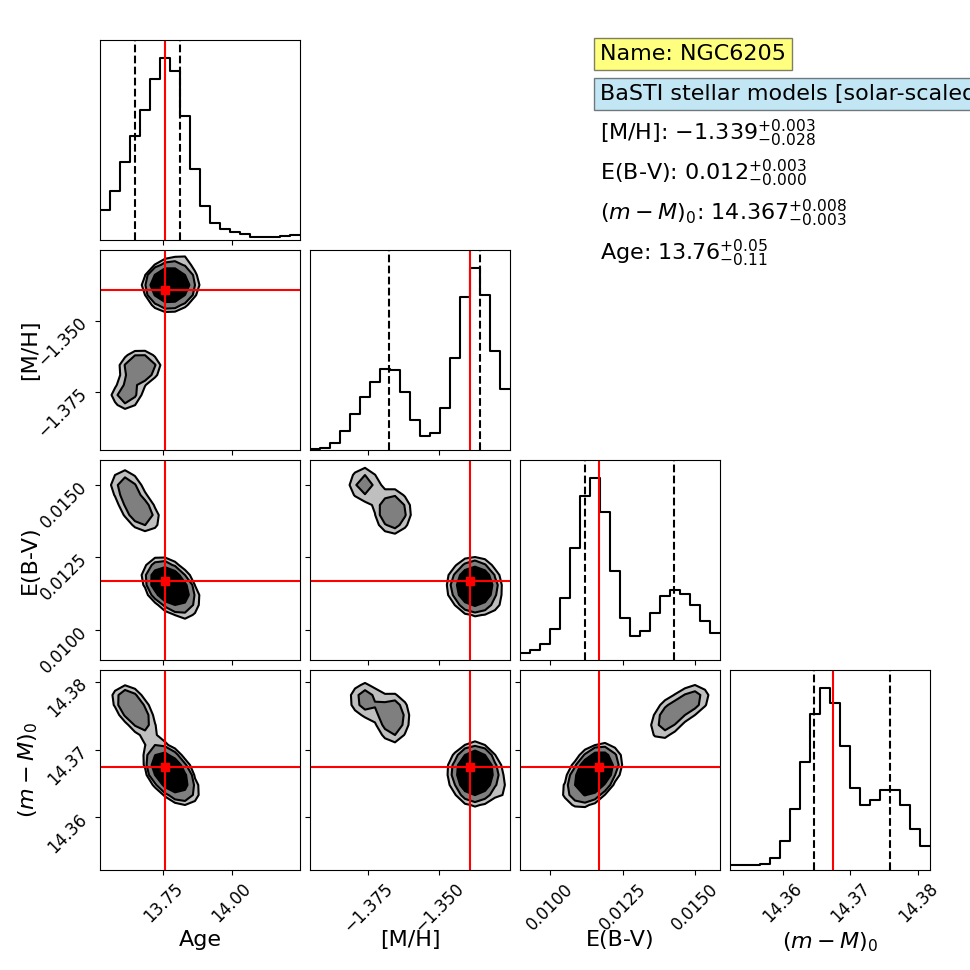}
            \caption[]
            {{\small }}    
        \end{subfigure}
        \begin{subfigure}{0.45\textwidth}   
            \includegraphics[width=\textwidth]{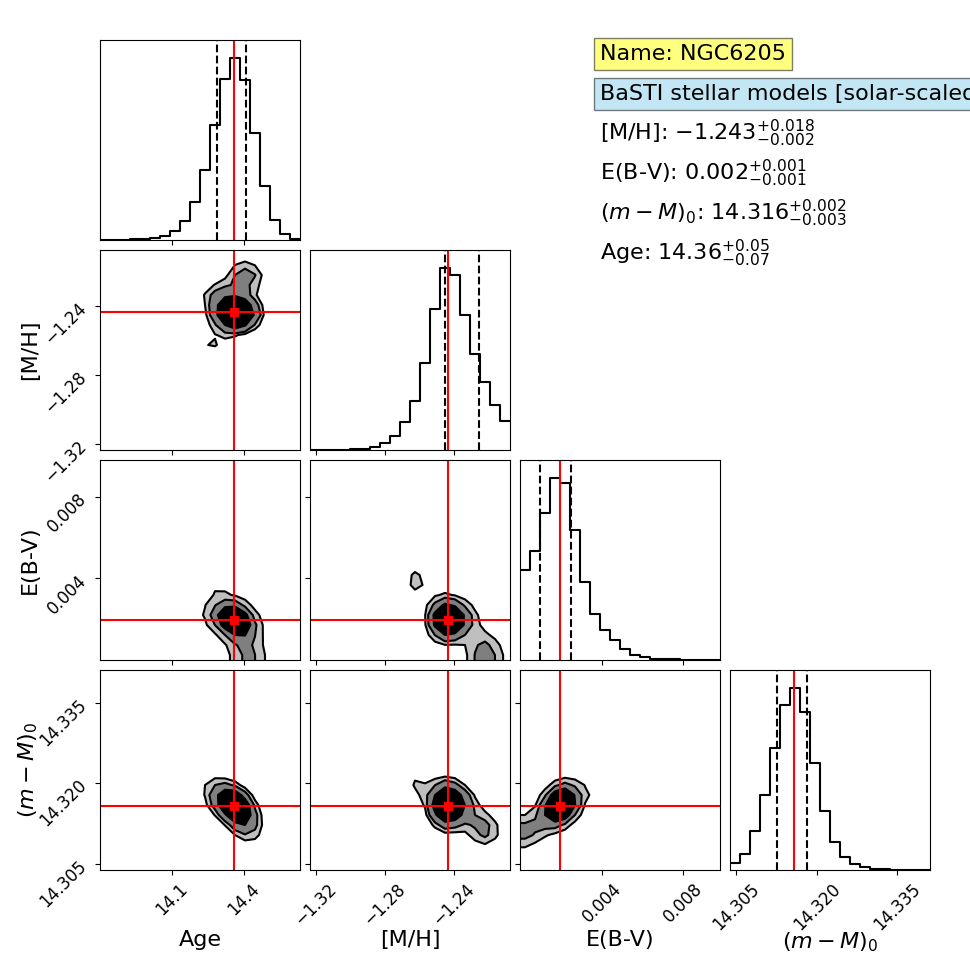}
            \caption[]
            {{\small }}    
        \end{subfigure}
        \caption[]
        {\small Results for \NGC. The meaning of the panels is the same as in Fig.~A.1} 
    \end{figure*}

\renewcommand{\NGC}{NGC6341}
\begin{figure*}
        \centering
        \begin{subfigure}{0.45\textwidth}
            \includegraphics[width=\textwidth]{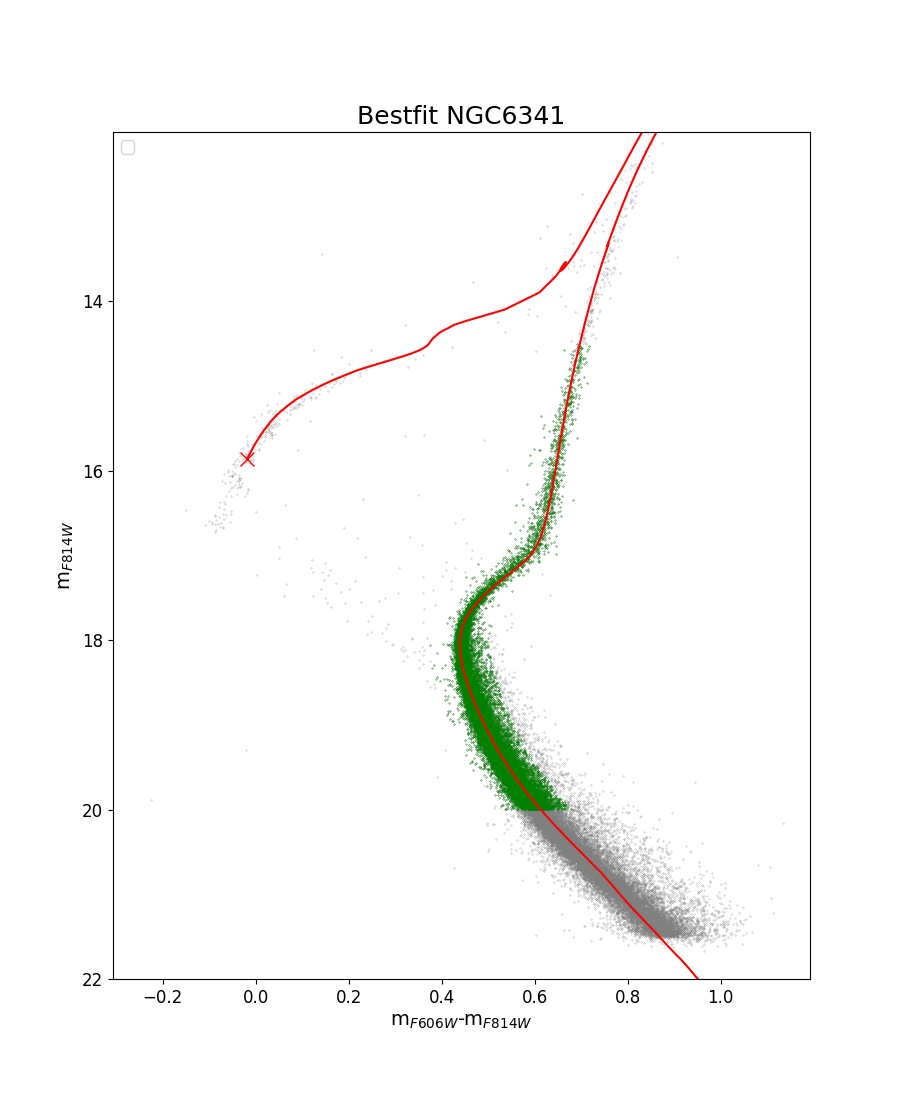}
            \caption[]
            {{\small }}    
        \end{subfigure}
        \begin{subfigure}{0.45\textwidth}  
            \includegraphics[width=\textwidth]{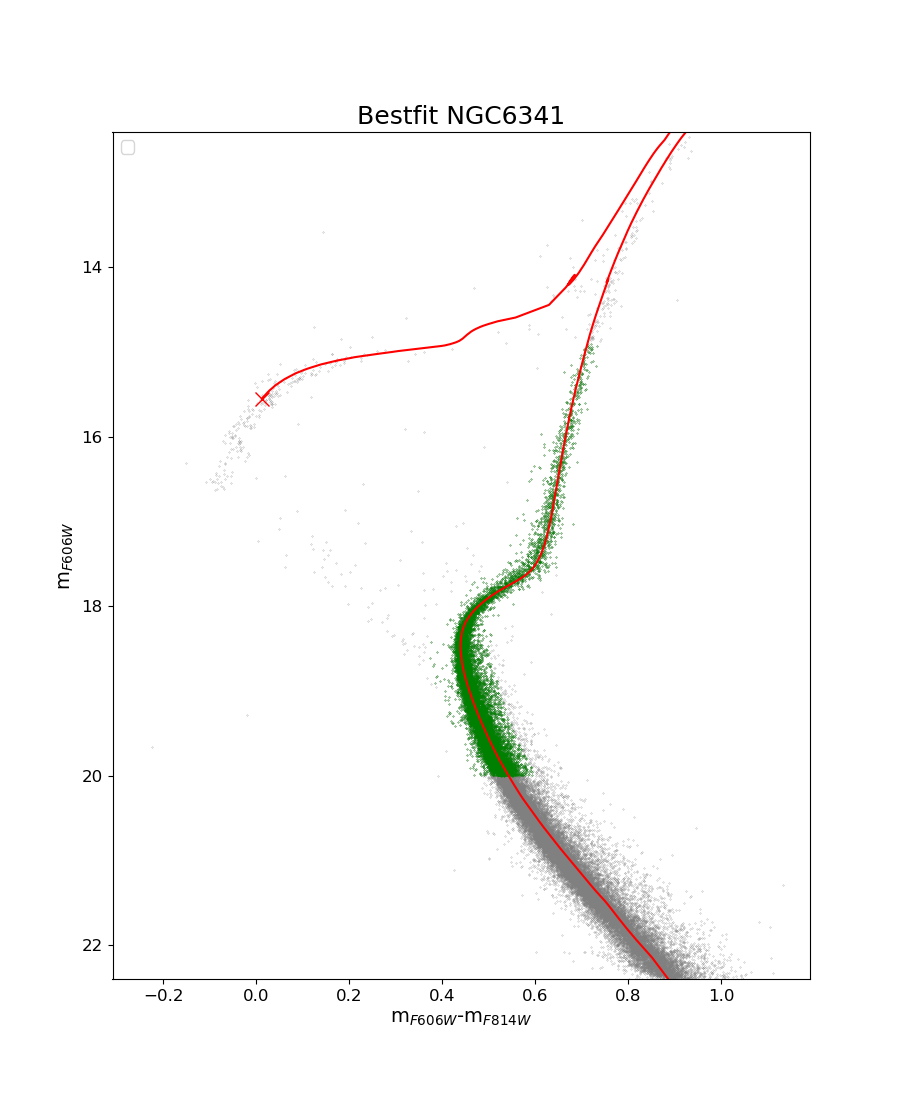}
            \caption[]
            {{\small }}    
        \end{subfigure}
        \vskip\baselineskip
        \begin{subfigure}{0.45\textwidth}   
            \includegraphics[width=\textwidth]{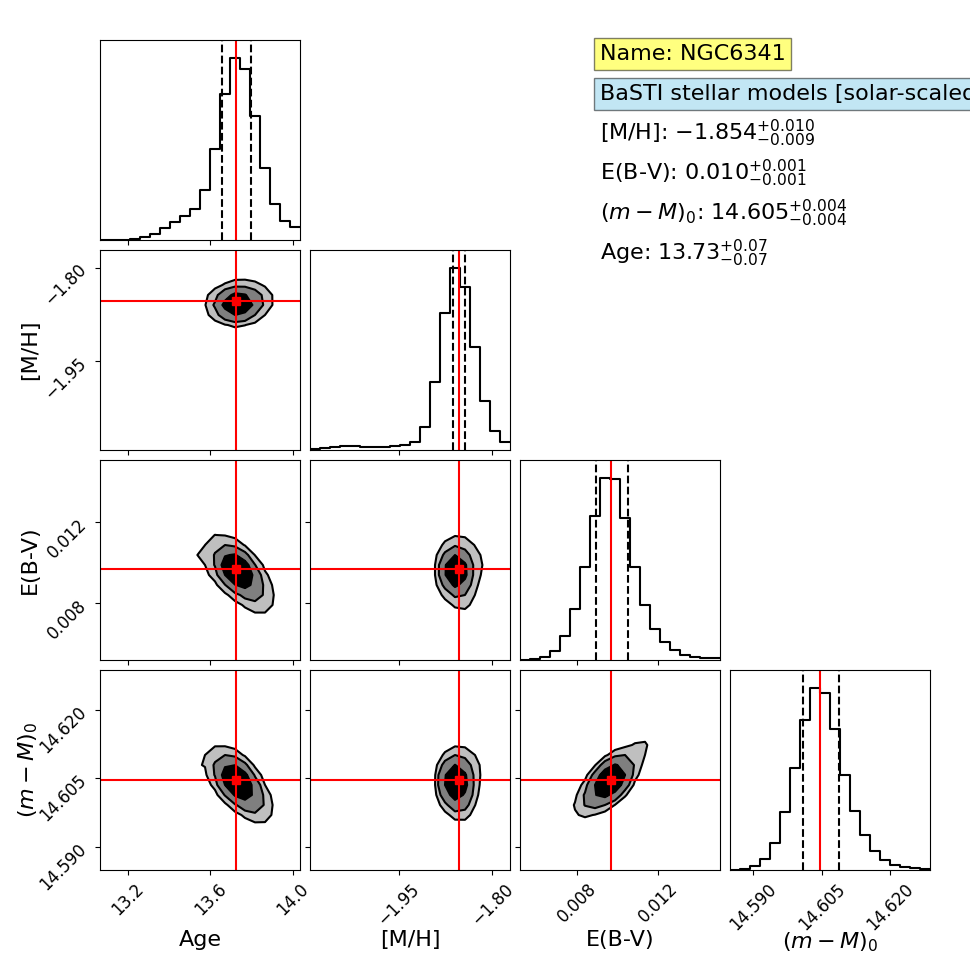}
            \caption[]
            {{\small }}    
        \end{subfigure}
        \begin{subfigure}{0.45\textwidth}   
            \includegraphics[width=\textwidth]{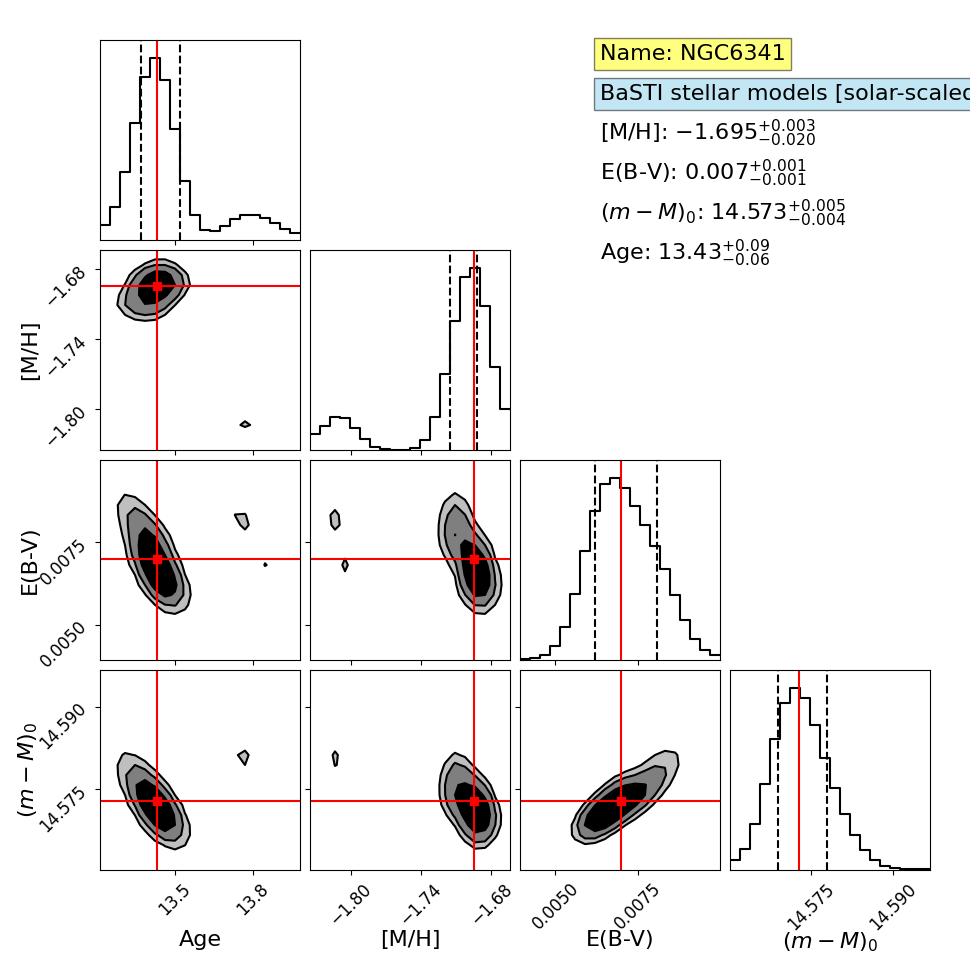}
            \caption[]
            {{\small }}    
        \end{subfigure}
        \caption[]
        {\small Results for \NGC. The meaning of the panels is the same as in Fig.~A.1} 
    \end{figure*}

\renewcommand{\NGC}{NGC6779}
\begin{figure*}
        \centering
        \begin{subfigure}{0.45\textwidth}
            \includegraphics[width=\textwidth]{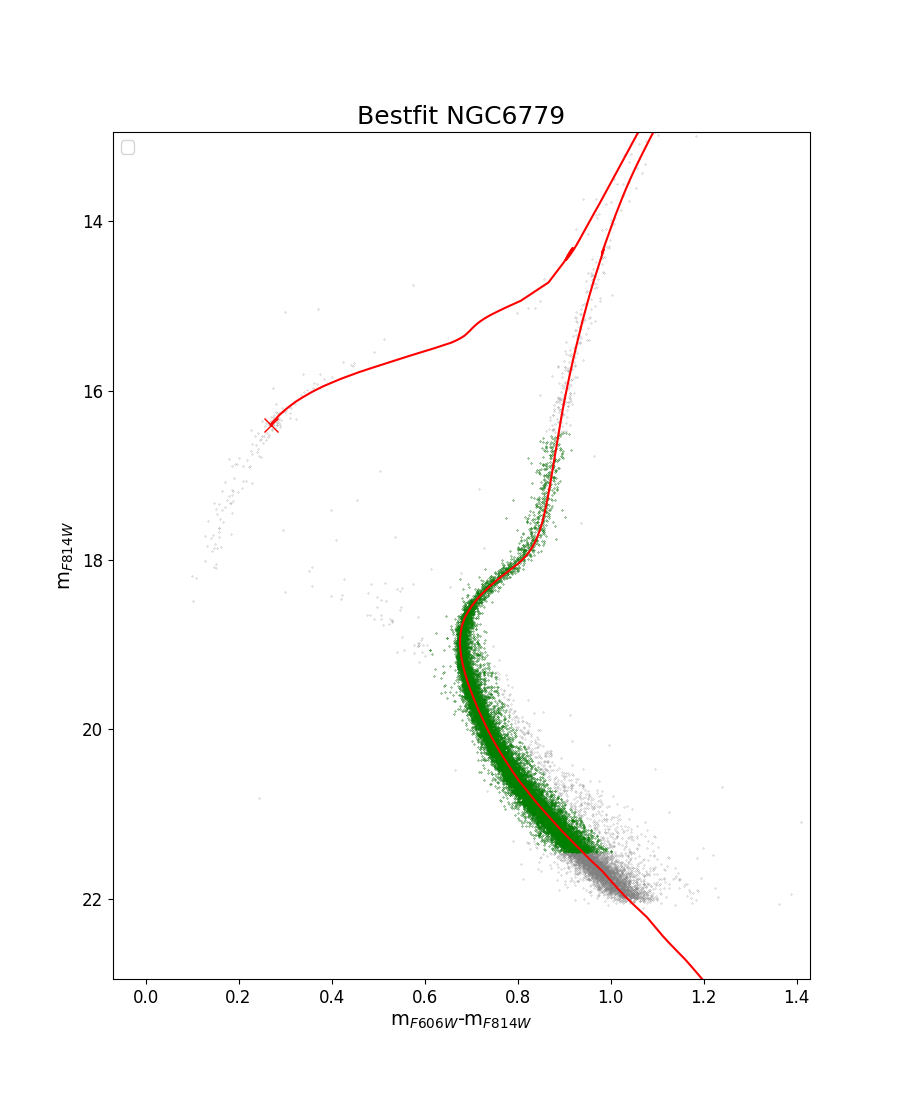}
            \caption[]
            {{\small }}    
        \end{subfigure}
        \begin{subfigure}{0.45\textwidth}  
            \includegraphics[width=\textwidth]{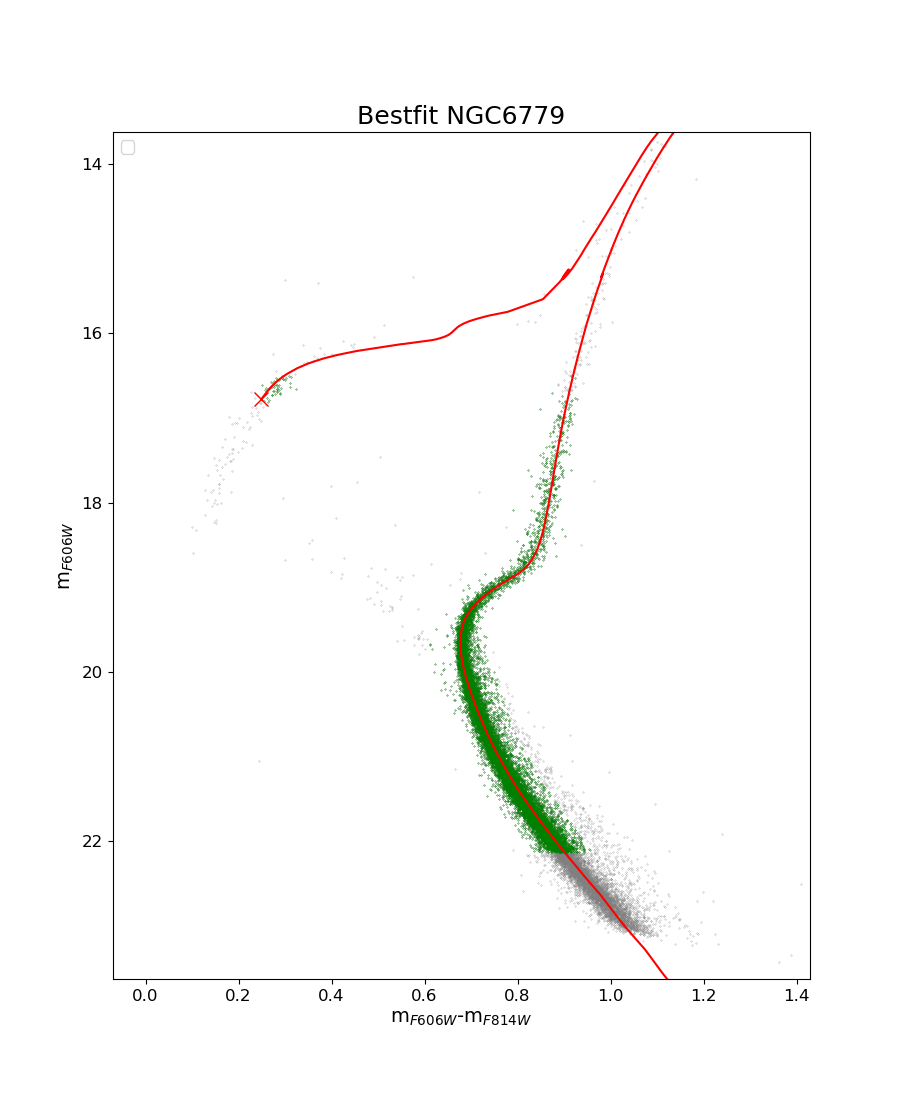}
            \caption[]
            {{\small }}    
        \end{subfigure}
        \vskip\baselineskip
        \begin{subfigure}{0.45\textwidth}   
            \includegraphics[width=\textwidth]{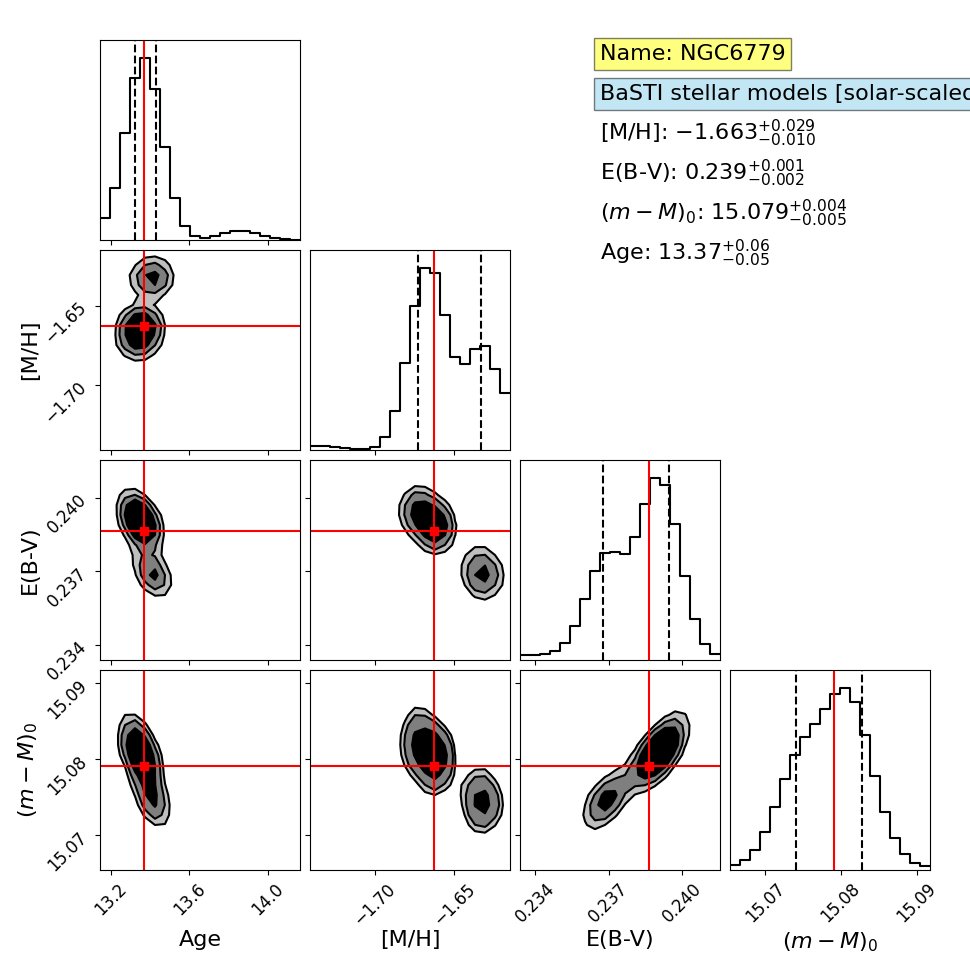}
            \caption[]
            {{\small }}    
        \end{subfigure}
m        \begin{subfigure}{0.45\textwidth}   
ov
            \includegraphics[width=\textwidth]{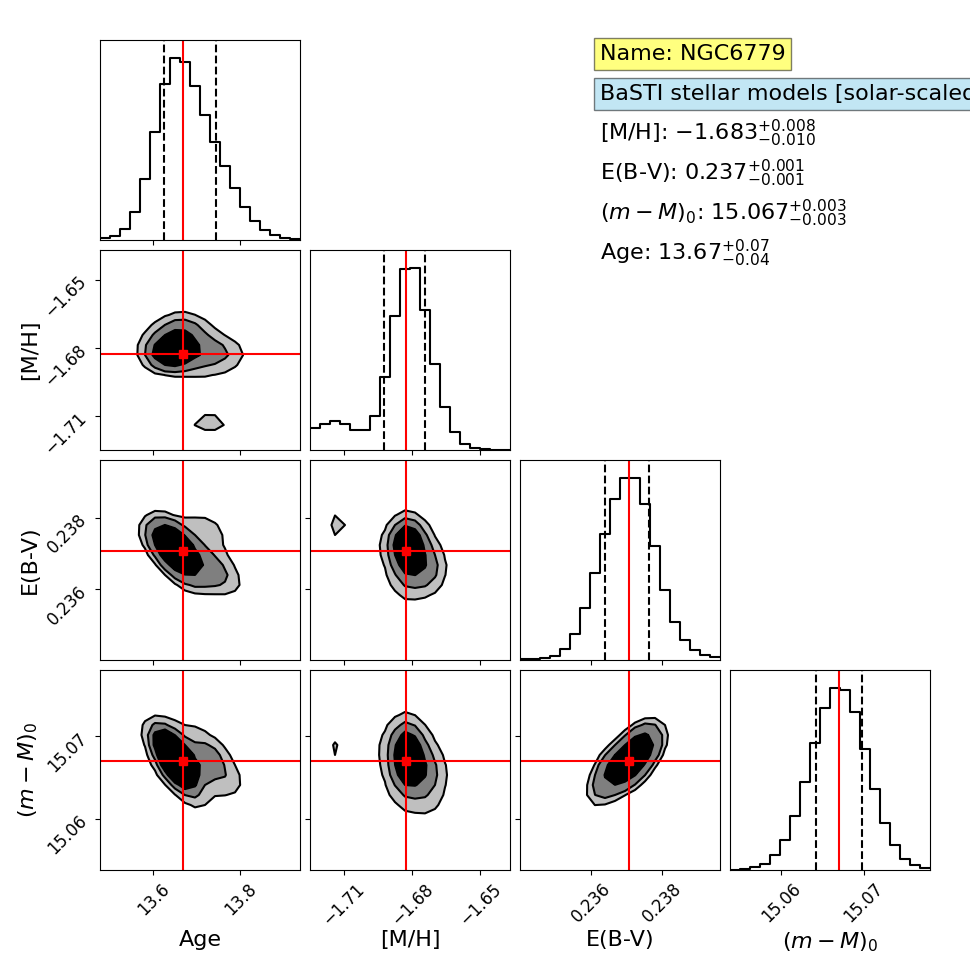}
            \caption[]
            {{\small }}    
        \end{subfigure}
        \caption[]
        {\small Results for \NGC. The meaning of the panels is the same as in Fig.~A.1} 
    \end{figure*}

\renewcommand{\NGC}{NGC7089}
\begin{figure*}
        \centering
        \begin{subfigure}{0.45\textwidth}
            \includegraphics[width=\textwidth]{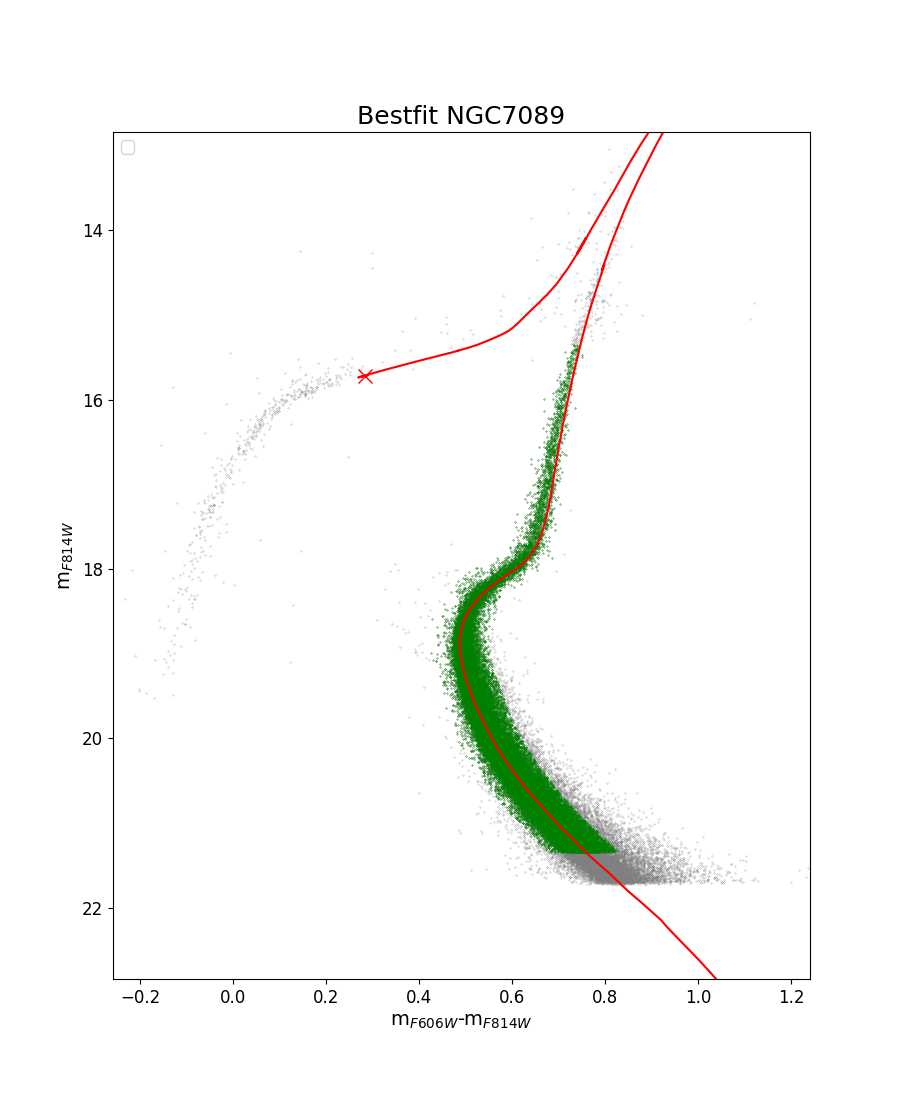}
            \caption[]
            {{\small }}    
        \end{subfigure}
        \begin{subfigure}{0.45\textwidth}  
            \includegraphics[width=\textwidth]{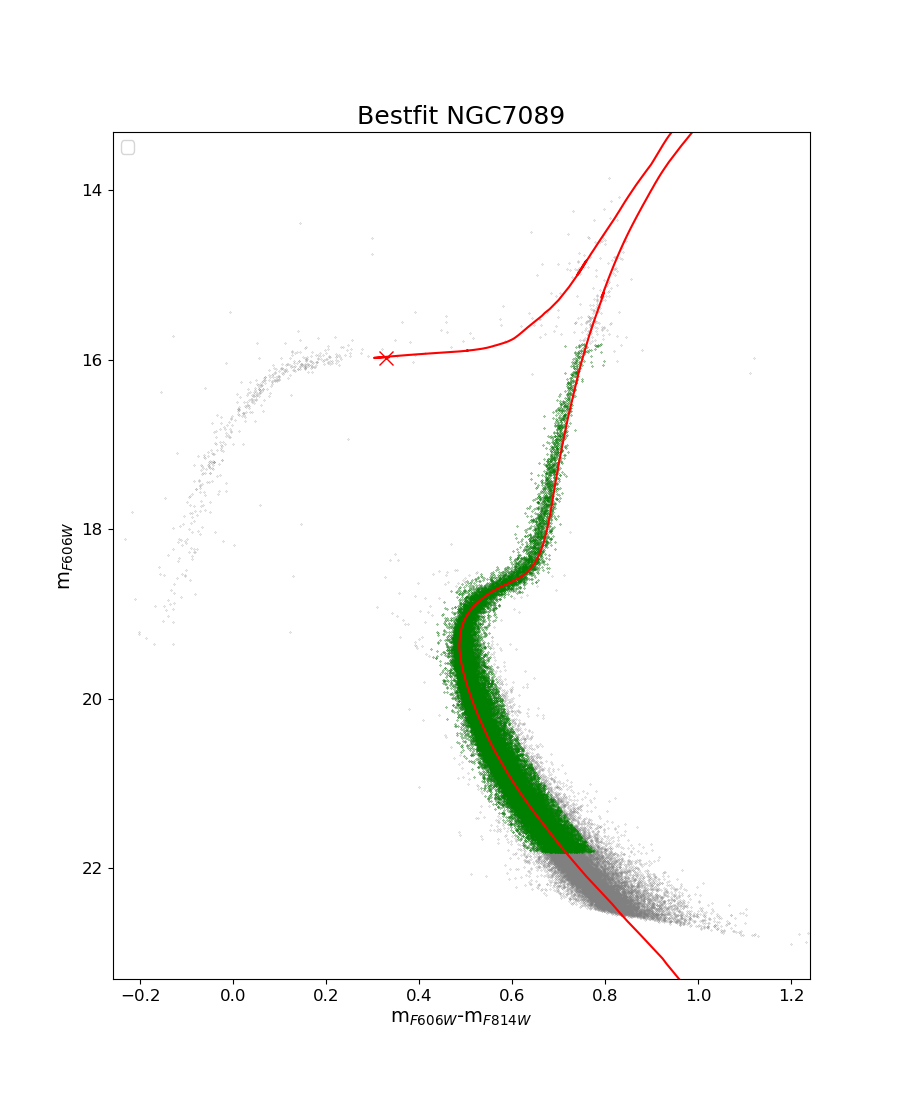}
            \caption[]
            {{\small }}    
        \end{subfigure}
        \vskip\baselineskip
        \begin{subfigure}{0.45\textwidth}   
            \includegraphics[width=\textwidth]{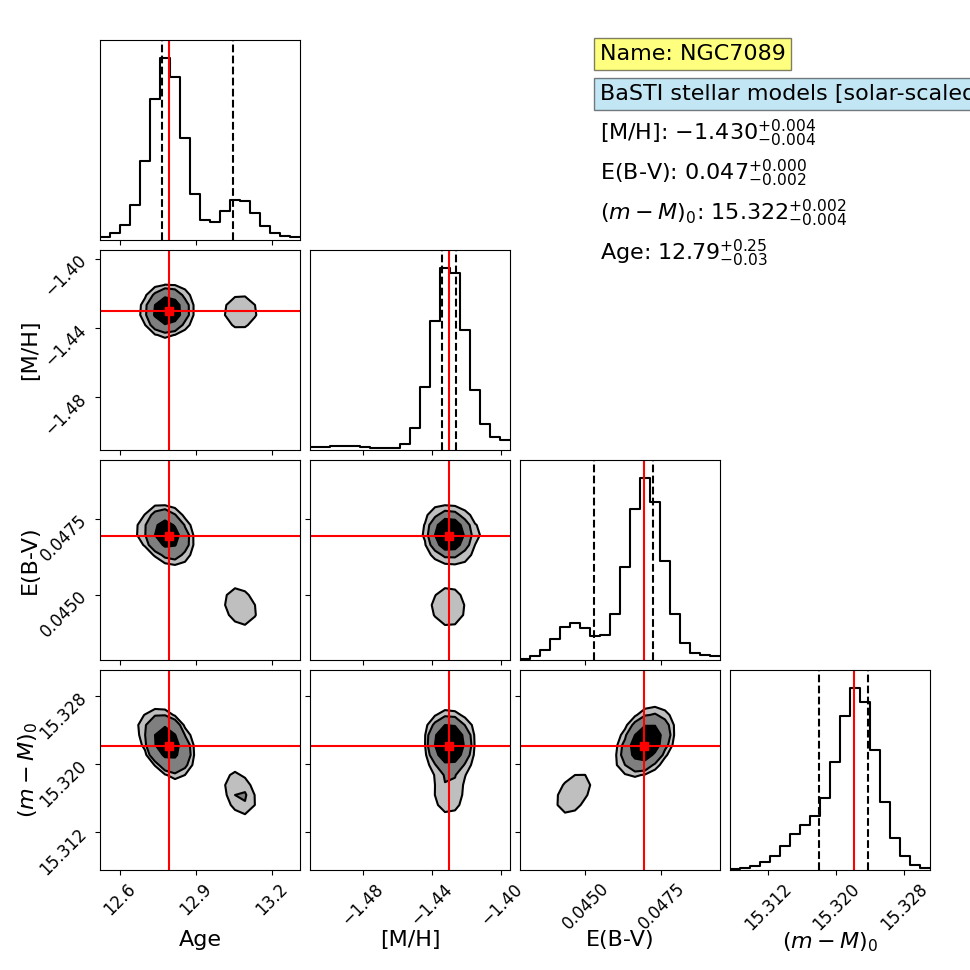}
            \caption[]
            {{\small }}    
        \end{subfigure}
        \begin{subfigure}{0.45\textwidth}   
            \includegraphics[width=\textwidth]{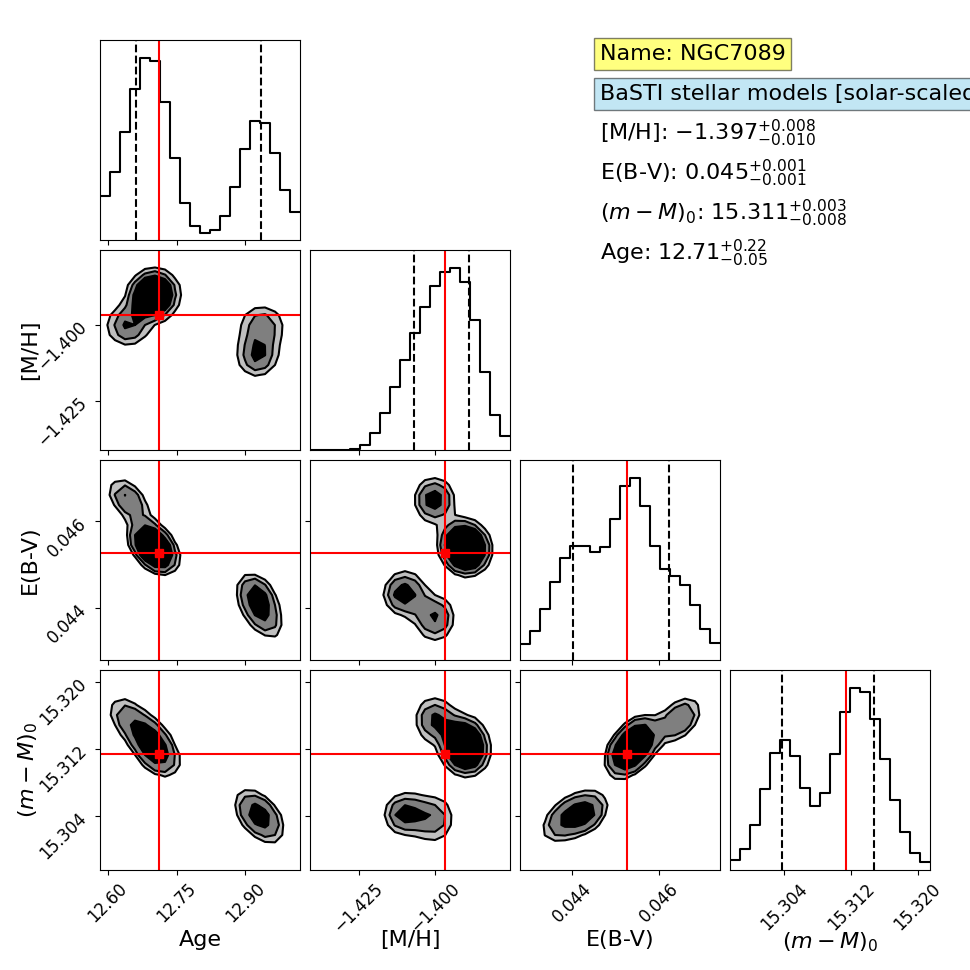}
            \caption[]
            {{\small }}    
        \end{subfigure}
        \caption[]
        {\small Results for \NGC. The meaning of the panels is the same as in Fig.~A.1} 
    \end{figure*}

\renewcommand{\NGC}{NGC7099}
\begin{figure*}
        \centering
        \begin{subfigure}{0.45\textwidth}
            \includegraphics[width=\textwidth]{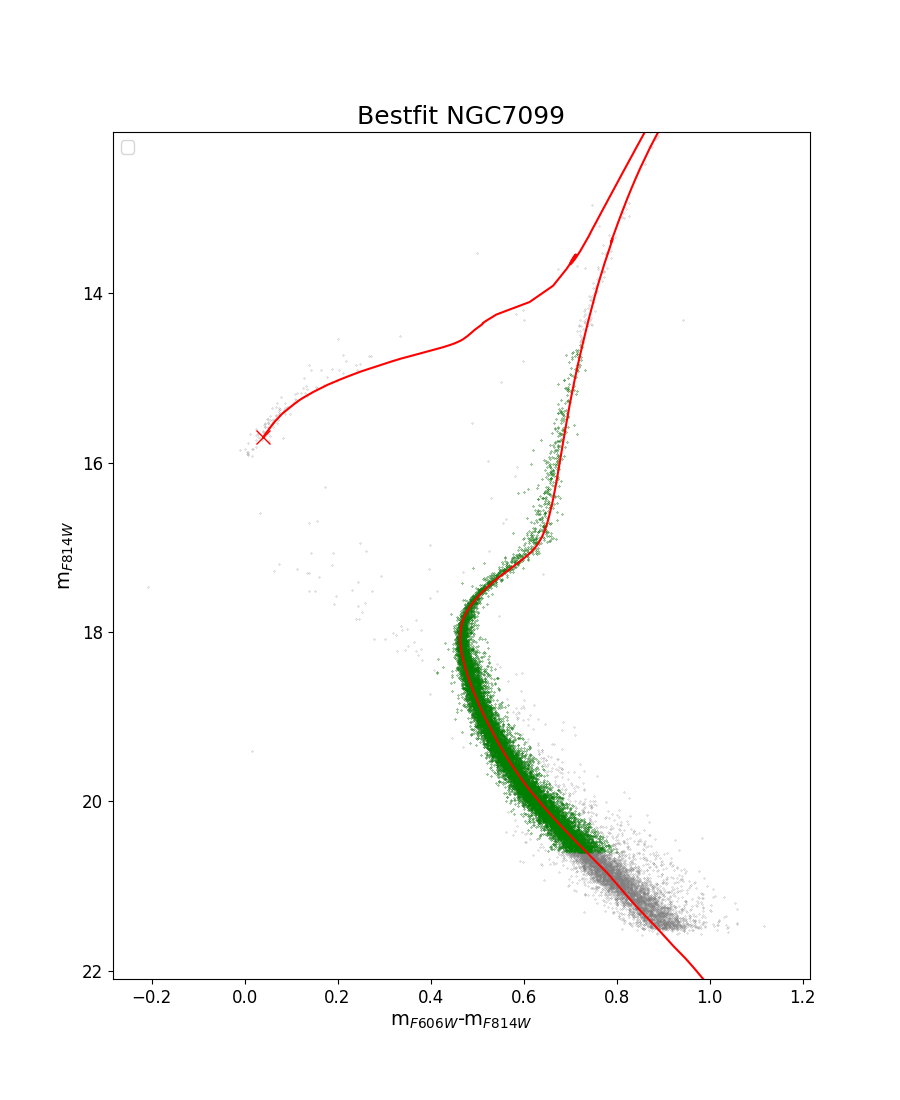}
            \caption[]
            {{\small }}    
        \end{subfigure}
        \begin{subfigure}{0.45\textwidth}  
            \includegraphics[width=\textwidth]{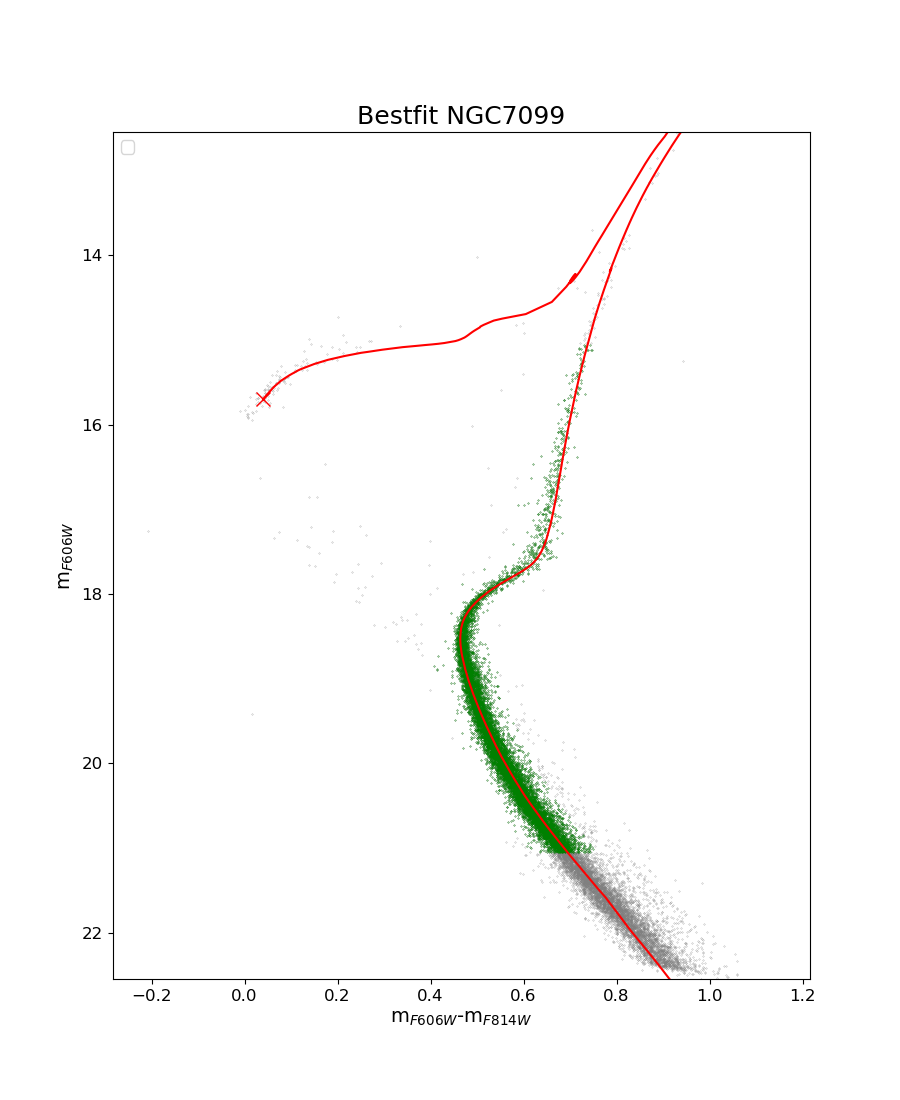}
            \caption[]
            {{\small }}    
        \end{subfigure}
        \vskip\baselineskip
        \begin{subfigure}{0.45\textwidth}   
            \includegraphics[width=\textwidth]{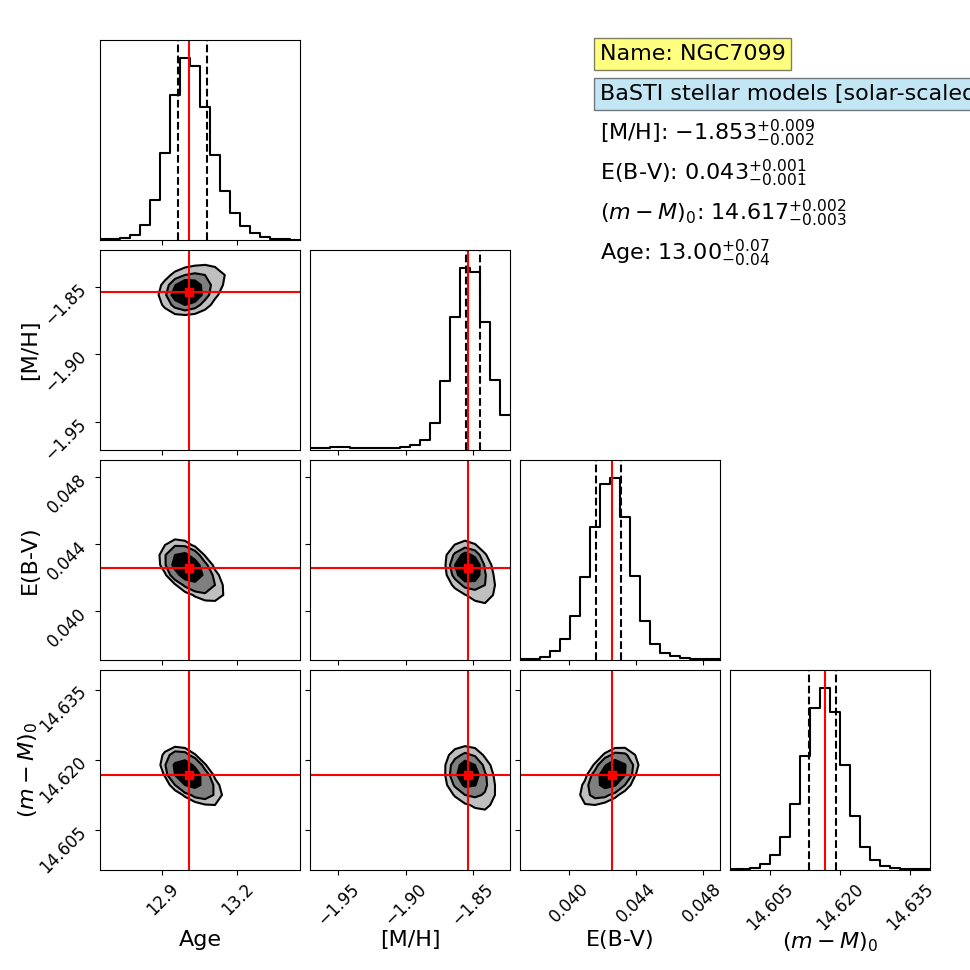}
            \caption[]
            {{\small }}    
        \end{subfigure}
        \begin{subfigure}{0.45\textwidth}   
            \includegraphics[width=\textwidth]{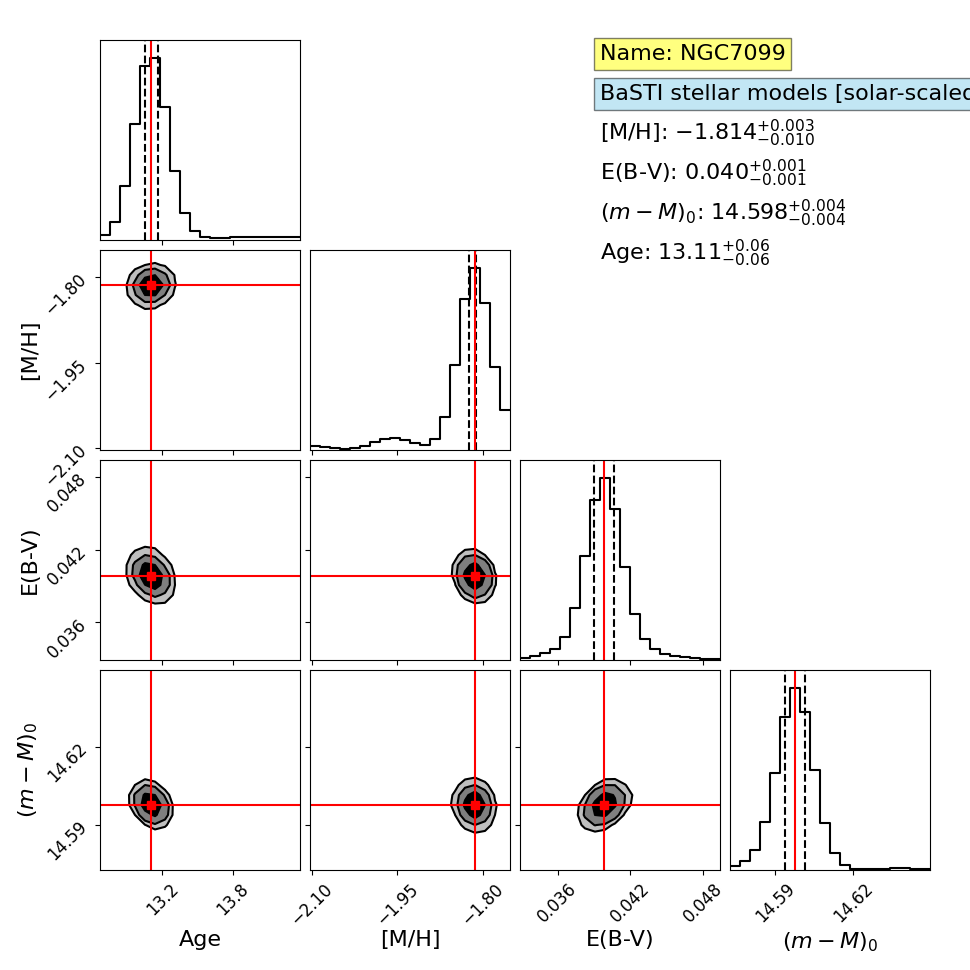}
            \caption[]
            {{\small }}    
        \end{subfigure}
        \caption[]
        {\small Results for \NGC. The meaning of the panels is the same as in Fig.~A.1} 
    \end{figure*}

\twocolumn

\end{appendix}
\end{document}